\documentclass[twocolumn,english,showpacs,superscriptaddress,prm]{revtex4-1}
\usepackage[colorlinks=true,citecolor=blue,linkcolor=blue,urlcolor=blue]{hyperref}
\usepackage[latin9]{inputenc}
\usepackage{color}
\usepackage{float}
\usepackage{textcomp}
\usepackage{amstext}
\usepackage{graphicx}
\usepackage{gensymb}
\makeatletter
\usepackage{graphics}\usepackage{subfigure}\usepackage{longtable}\usepackage{pstricks}\usepackage{dcolumn}\usepackage{bm}\usepackage{multirow}
\usepackage{amsmath}
\begin{document}

\title{Synthesis and physical properties of LaNiO$_2$ crystals}
\author{P. Puphal}
\email{puphal@fkf.mpg.de}
\affiliation{Max-Planck-Institute for Solid State Research, Heisenbergstra{\ss}e 1, D-70569 Stuttgart, Germany}
\author{B.~Wehinger}
\affiliation{European Synchrotron Radiation Facility, 71 Avenue des Martyrs, F-38043 Grenoble, France}
\author{J.~Nuss}
\affiliation{Max-Planck-Institute for Solid State Research, Heisenbergstra{\ss}e 1, D-70569 Stuttgart, Germany}
\author{K. K\"uster}
\affiliation{Max-Planck-Institute for Solid State Research, Heisenbergstra{\ss}e 1, D-70569 Stuttgart, Germany}
\author{U. Starke}
\affiliation{Max-Planck-Institute for Solid State Research, Heisenbergstra{\ss}e 1, D-70569 Stuttgart, Germany}
\author{G. Garbarino}
\affiliation{European Synchrotron Radiation Facility, 71 Avenue des Martyrs, F-38043 Grenoble, France} 
\author{B.~Keimer}
\affiliation{Max-Planck-Institute for Solid State Research, Heisenbergstra{\ss}e 1, D-70569 Stuttgart, Germany}
\author{M.~Isobe}
\affiliation{Max-Planck-Institute for Solid State Research, Heisenbergstra{\ss}e 1, D-70569 Stuttgart, Germany}
\author{M.~Hepting}
\email{hepting@fkf.mpg.de}
\affiliation{Max-Planck-Institute for Solid State Research, Heisenbergstra{\ss}e 1, D-70569 Stuttgart, Germany}

\date{\today}

\begin{abstract}
Infinite-layer (IL) nickelates are an emerging family of superconductors whose similarities and differences to cuprate superconductors are under intense debate. To date, the IL phase of nickelates can only be reached via topotactic oxygen reduction of the perovskite phase, using H$_2$ gas or reducing agents such as CaH$_2$. While the topotactic reduction method has been widely employed on thin film and polycrystalline powder samples, the reduction of La$_{1-x}$Ca$_x$NiO$_3$ single-crystals with lateral dimensions up to 150 $\mu$m was achieved only recently, using an indirect contact method with CaH$_2$. Here we report the topotactic transformation of much larger LaNiO$_3$ crystals with lateral dimensions of more than one millimeter, via direct contact with CaH$_2$. We characterize the crystalline, magnetic, and electronic properties of the obtained IL LaNiO$_{2}$ crystals by powder and single-crystal x-ray diffraction (XRD), magnetometry, electrical transport, and x-ray photoelectron spectroscopy (XPS) measurements. The amount of incorporated topotactic hydrogen due to the reduction process is determined by a gas extraction method. In addition, we investigate the evolution of the lattice parameters under hydrostatic pressure up to 12 GPa, using high-resolution synchrotron XRD. Furthermore, we provide a direct comparison of several physical properties of the LaNiO$_{2}$ crystals to their powder and thin film counterparts. 

\end{abstract}

\maketitle

\section{Introduction}

Superconductivity in the material class of nickelates was first discovered in Nd$_{0.8}$Sr$_{0.2}$NiO$_2$ thin films that were epitaxially grown on SrTiO$_3$ substrates \cite{Li2019}. In the meantime, the superconducting transition in Nd$_{0.8}$Sr$_{0.2}$NiO$_2$ films has been confirmed by other groups \cite{Zeng2020,Li2021MBE,Gao2021}, and was similarly observed for various choices of rare-earth ions ($R$ = La, Pr, Nd) \cite{Li20201,Osada2020,osada2021,Wang2022}, divalent dopant ions ($A$ = Sr, Ca) \cite{zeng2021}, and substrates \cite{Ren2021,Lee2022}. The reported transition temperatures $T_c$ are situated between 9-15 K, although it was found that hydrostatic pressure can enhance the $T_c$ of Pr$_{0.82}$Sr$_{0.18}$NiO$_2$ films to more than 30 K \cite{Wang2022b}. Moreover, while superconductivity in the infinite-layer compounds $R_{1-x}A_x$NiO$_2$ occurs only for substitution levels between $x \approx 0.1$ and 0.3 \cite{Lee2022}, it was recently reported that the quintuple-layer nickelate Nd$_6$Ni$_5$O$_{12}$ shows superconductivity already without rare-earth substitution \cite{Pan2021}.
Taken together, these findings suggest that low-valence nickelates are a novel class of superconductors. 

Nickelate superconductors exhibit several similarities but also differences to the cuprate superconductors. For instance, IL nickelates and cuprates share the same nominal $3d^9$ electronic configuration of the Ni (Cu) ion and are isostructural, with Ni (Cu) and O ions arranged in square planar coordination within NiO$_2$ (CuO$_2$) planes. On the other hand, the parent IL nickelates $R$NiO$_2$ lack long-range antiferromagnetic (AFM) order \cite{Hayward1999,Hayward2003,Lin202101,Fowlie2022,Ortiz2022}, which is a hallmark of parent cuprates \cite{Scalapino2012}, and show a distinct multiorbital electronic structure \cite{Hepting2020,Goodge2021}. Another notable distinction of nickelates is that to date only thin film samples were found to superconduct \cite{Bernardini2022a,Li2020,Wang20201,He2021}, whereas various types of cuprate samples, including single-crystals, polycrystalline powders, and films, can realize superconductivity. In fact, bulk single-crystalline cuprate samples often show superior properties over their thin film and polycrystalline counterparts, such as a higher crystalline quality and an elevated $T_c$ \cite{Rao1993}. Hence, the question arises whether epitaxy to the substrate and/or interface effects are critical for the emergence of superconductivity in nickelates \cite{Geisler2020,He2020,Bernardini2022a}, or the previously available nickelate bulk samples exhibited quality issues that prevented the emergence of superconductivity. For instance, in the case of powder samples, possible impediments for superconductivity include enhanced disorder, inclusions of secondary phases, and diminishing sizes of infinite-layer single domains in powder grains \cite{Hayward1999,Crespin2005,Li2020,Wang20201,He2021,Puphal2022}.

To shed new light on possible differences between bulk and thin film IL nickelates, some of us have recently synthesized high-quality perovskite La$_{1-x}$Ca$_x$NiO$_3$ single-crystals, which were subsequently reduced to the IL phase \cite{Puphal2021}. The samples in the perovskite phase were obtained via a perchlorate-chlorate flux mixture in an external temperature gradient growth under high pressure in a 1000-metric ton press, yielding crystals with typical lateral dimensions of 150 $\mu$m. For the topotactic reduction, which lasted 2-4 weeks, CaH$_2$ was used as a reducing agent in spatial separation from the crystals. The resulting La$_{1-x}$Ca$_x$NiO$_{2}$ crystals showed metallic behavior at high temperatures and a weakly insulating upturn at low temperatures, which is closely analogous to undoped and lightly-doped $R_{1-x}A_x$NiO$_2$ films \cite{Li2019,Zeng2020,Li2021MBE,Gao2021,Li20201,Osada2020,osada2021,Wang2022,zeng2021,Ren2021,Lee2022}, but in stark contrast to powder samples, which in turn show insulating behavior at all temperatures \cite{Li2020,Wang20201,He2021}. Yet, a superconducting transition was not observed in Ref.~\onlinecite{Puphal2021}, which is possibly due to the fact that electrical transport measurements were only realized on crystals with Ca-substitutions up to $x \approx 0.08$, \textit{i.e.}, at a hole-doping level below the onset of the superconducting dome of corresponding thin films \cite{zeng2021}. Single-crystals with higher substitution levels might thus allow to finally clarify whether superconductivity is an exclusive thin film property. However, the obtained sizes of high-pressure grown perovskite crystals \cite{Puphal2021} tend to decrease with increasing Ca-substitution level, which poses a challenge for electrical transport measurements and the characterization of other physical properties. 

In consequence, new strategies and synthesis methods for single-crystalline IL nickelates with enhanced sample sizes are highly desirable. Moreover, a considerable increase of the bulk sample sizes will facilitate the application of advanced spectroscopy techniques, including angle resolved photoemission, optical, x-ray and electron spectroscopy, tunneling microscopy, as well as inelastic x-ray and neutron scattering. These techniques have provided invaluable insights into the charge, spin, lattice, and orbital degrees of freedom of strongly correlated quantum materials such as cuprate superconductors \cite{Damascelli2003,Fink2001,Fischer2007,Basov2005,Devereaux2007,Ament2011,Fujita2012}, but were only occasionally applied to IL nickelates yet \cite{Hepting2021,Rossi2020,Gu2020,Ortiz2021,Rossi2022,Lu2021,Zeng2022n,Cervasio2022,Shen2022,Fursich2022,Krieger2022,Tam2021}, mostly because the sample mass of thin films is insufficient for techniques with a small scattering cross section and the highly invasive topotactic reduction deteriorates the film surface, which is problematic for surface sensitive techniques.

In the context of large single-crystalline samples, recent technical advances in the high-pressure floating zone growth have enabled the synthesis of centimeter-sized LaNiO$_{3}$ single-crystals under 30-130 bar oxygen partial pressure, using either a laser diode floating zone furnace \cite{Tomioka2021}, or a vertical optical-image floating-zone (OFZ) furnace that was designed for operation at elevated gas pressures \cite{Phelan2019,Zhang2017,Guo2018,Wang2018,Dey2019,Zheng2020}. Moreover, in a variant of the latter furnace, oxygen gas pressures of up to 300 bar can be achieved, providing the required oxidizing conditions to stabilize the Ni$^{3+}$ valence state not only in perovskite LaNiO$_{3}$, but even in PrNiO$_{3}$ \cite{Zheng2019}. 

In this work, we utilize the OFZ method to synthesize LaNiO$_{3}$ single-crystals, from which we obtain millimeter-sized LaNiO$_{2}$ crystals via topotactic reduction using CaH$_{2}$ in direct contact with the crystals. For insufficiently reduced crystals, powder x-ray diffraction (PXRD) measurements reveal the presence of residual contributions of the LaNiO$_{2.5}$ phase, whereas excessively reduced crystals decompose into elemental Ni, La$_2$O$_3$, and lanthanum oxide hydride LaOH, with the latter compound emerging due to a reaction with the reducing agent, as evidenced by hydrogen gas extraction analysis. For selected LaNiO$_{2}$ crystals, we carry out single-crystal x-ray diffraction (XRD) under hydrostatic pressure to determine the lattice constants and the bulk modulus. Furthermore, we compare the magnetic susceptibility and electrical resistivity of reduced crystals with corresponding polycrystalline powder and thin film samples, suggesting that our crystals are closely analogous to thin films. Finally, we perform x-ray photoelectron spectroscopy
(XPS) measurements, which show a satellite peak around the Ni 3$p$ core level peak, which we interpret as a signature of the $3d^9$ electronic configuration of LaNiO$_{2}$.



\section{Methods}
\subsection*{High gas pressure optical floating zone growth}

For the high gas pressure OFZ growth, La$_{2}$O$_{3}$ powder was dried at 1000$^{\circ}$C for one day. Subsequently, we  mixed and ground the dried La$_{2}$O$_{3}$ powder (10.284 g, 31.6 mmol, Alfa  Aesar 99.99\%)  in a 0.5:1 molar ratio with NiO (4.7155 g, 31.6 mmol, Alfa  Aesar 99.998\%) to obtain by oxidization in a high O$_{2}$ partial pressure stoichiometric LaNiO$_{3}$ single crystals. The powder mixture was transferred to an alumina crucible and annealed for 2 days at 900$^{\circ}$C in air followed by subsequent grinding and repeated annealing. The powder was then transferred to rubber forms, evacuated, and pressed at 700 bar in a Riken CD-10 press and subsequently sintered at 1000$^{\circ}$C in air. The rods were centered and installed in a high-pressure high-temperature optical floating zone furnace (Model HKZ, SciDre GmbH, Dresden), followed by a growth at 150 bar oxygen pressure with a flow of 0.1 l/min. The floating zone was moved with 4 mm/h, where due to the low viscosity of the melt at these conditions a constant feeding of 2 mm/h was necessary since a wider diameter of the seed developed. The obtained boule was analyzed with x-ray Laue diffraction, and crystals with lateral dimensions of several millimeters were extracted for subsequent reduction treatment.

\subsection*{Topotactic reduction via direct and indirect contact}

For the topotactic reduction with direct contact between the LaNiO$_{3}$ single-crystals and the reducing agent CaH$_{2}$, a number of as-grown single-crystals with lateral dimensions between $\sim$1 mm and $\sim$500 $\mu$m were selected, such that a total mass of $\sim$100 mg was reached. Subsequently, $\sim$250 mg CaH$_{2}$ powder (Sigma Aldrich 97\%) was loaded in a glass crucible and put into a DURAN glass tube, together with the LaNiO$_{3}$ crystals. The glass tube was then evacuated ($\sim 10^{-7}$ mbar) and sealed to ampules with dimensions of $\phi_{\text{out}} = 1.7$ mm, $\phi_{\text{in}} = 1.5$ mm, and $h = 10$ cm. The reduction was carried out by heating slowly to 300$^\circ$C in an optimized low temperature furnace and held for 2-3 weeks before slowly cooling down to room temperature. Subsequently, the ampules were opened and the obtained crystals were handled in air. 

In addition, several OFZ grown LaNiO$_{3}$ crystals were reduced in spatial separation to the reducing agent CaH$_{2}$. This indirect contact method and the used parameters are closely similar to the conditions in previous reports for the reduction of LaNiO$_{3}$ powder samples \cite{Ortiz2022} and small La$_{1-x}$Ca$_x$NiO$_{3}$ crystals \cite{Puphal2021}. Here, several OFZ grown crystals with lateral dimensions of a few hundred micrometers and a total mass of $\sim$100 mg were wrapped in aluminum foil and transferred to a glove box. A glass crucible was loaded with 400-600 mg CaH$_{2}$ powder and put into a DURAN glass tube. The wrapped crystals were inserted into the glass tube and placed above the crucible with CaH$_{2}$. Subsequently, the glass tube was evacuated and sealed to ampules. The reduction was carried out by heating the specimen slowly to 300$^\circ$C and holding it for 8 weeks before slowly cooling down to room temperature. 


\subsection*{Hydrogen gas extraction}
The hydrogen content of the samples was determined with an Eltra ONH-2000 analyzer, for which pulverized crystals were placed in a Ni crucible, clipped and heated. A  carrier  gas  takes  the hydrogen out of the sample and the hydrogen is detected by a thermal-conductivity-cell. Each measurement is repeated three times and compared to a standard. The quoted error bars give the statistical error. We remark that the investigated LaNiO$_{2.5}$ sample was obtained from an indirect contact reduction after 8 weeks. The LaNiO$_{2}$ sample and the sample containing $\alpha$-La$_{2}$O$_{3}$ and elemental Ni were obtained from a direct contact reduction after 2 weeks.

\subsection*{Single-crystal x-ray diffraction}

The single-crystal XRD data at ambient pressure were collected at room temperature with a SMART APEXI CCD X-ray diffractometer (Bruker AXS, Karlsruhe, Germany), using graphite-monochromated Mo-K$_{\alpha}$ radiation ($\lambda=0.71073\,$\AA). Since very small crystal pieces are generally more suitable for single-crystal XRD, all crystals investigated in this study are fragments from larger crystals that were broken off in high viscosity oil. A piece with lateral dimensions of approximately 50 $\mu$m was broken off from a larger LaNiO$_{2.5}$ crystal that was obtained from an indirect contact reduction after 8 weeks. A similar piece was broken off from a LaNiO$_{2}$ crystal that was obtained from a direct contact reduction after 2 weeks. The pieces were mounted with some grease on a loop made of Kapton foil (Micromounts$^{TM}$, MiTeGen, Ithaca, NY). The reflection intensities were integrated with the SAINT subprogram in the Bruker Suite software package. The investigated crystals with tetragonal symmetry showed systematic twinning by reticular merohedry. The threefold axes of the (pseudo)cubic perovskite structure become twinning elements, and the three twin domains are related by transformation matrices: (1 0 0) (0 1 0) (0 0 1); (0 1 0) (0 0 1) (1 0 0); (0 0 1) (1 0 0) (0 1 0). To handle this, the reflection intensities were integrated with the help of the orientation matrices of all three twin-domains, and a multi-scan absorption correction was applied using TWINABS. The structure was solved by direct methods and refined by full-matrix least-square fitting with the SHELXTL software package.

The single-crystal XRD data at high pressure were collected at the beamline ID27 at the European Synchrotron Radiation Facility (ESRF). We used membrane driven diamond anvil cells with an angular opening of 70 degrees. These were equipped with single crystalline diamonds with culet sizes of 500 $\mu$m. Helium was used as pressure transmitting medium and the applied pressure was determined by ruby fluorescence. Monochromatic x-rays with a wavelength of 0.3738\,\AA~ and a spot size of 1.5 $\times$ 1.8 $\mu$m$^2$ were scattered from the LaNiO$_2$ crystal. Scattering intensities were collected in continuous rotation and shutterless mode using an Eiger9M detector (Dectris, Switzerland) equipped with 750 $\mu$m thick CdTe sensors with a pixel size of 75  $\times$ 75 $\mu$m$^2$. The angular steps were 0.25 degrees, spanning 64 degrees in total. The scattering geometry was calibrated using a CeO$_2$ powder standard (NIST, USA) and a vanadinite single-crystal. The detector distance was 216.445 mm. For the measurement of the XRD at ambient pressure, a piece with a size of approximately 40 $\times$ 40 $\times$ 30 $\mu$m$^3$ was broken off from a larger LaNiO$_{2}$ crystal that was obtained from a direct contact reduction after 2 weeks. For the high-pressure XRD another piece was broken off from the same crystal, with a size of approximately 50 $\times$ 50 $\times$ 30 $\mu$m$^3$.

\subsection*{X-ray photoelectron spectroscopy}
Prior to the XPS measurements, a centimeter-sized LaNiO$_{3}$ single crystals was cleaved and mounted on carbon tape. LaNiO$_{2}$ crystals with sizes between 1 mm and 500 $\mu$m were transferred into the glove box rapidly after the reduction process was finished. In the glove box, they were broken into slightly smaller pieces to expose freshly cleaved surfaces and distributed on an indium foil. All samples were transferred under inert gas into the XPS chamber.
The XPS data were collected using a commercial Kratos AXIS Ultra spectrometer and a monochromatized Al $K_\alpha$ source (1486.6 eV). The base pressure during XPS was below $10^{-9}$ mbar. Survey spectra were acquired with a pass energy of 80 eV and detailed spectra with a pass energy of 20 eV. Since the LaNiO$_{2}$ spectra showed some charging the binding energy was calibrated to the La 3$d_{5/2}$ peak at the lowest binding energy of LaNiO$_{3}$. The intensity of the spectra was normalized to the most intense La 3$d$ peak.

\subsection*{Physical properties measurements}
Magnetic susceptibility measurements were carried out in a range of 1.8 - 350\,K and 0 - 7\,T using a Quantum Design Magnetic Property Measurements System (MPMS).  The resistivity measurements were carried out in a four-probe geometry with Au wires attached on the surface of the crystals by silver paint, using the standard option of a Quantum Design Physical Property Measurements System (PPMS).

\begin{figure*}[tb]
 \begin{centering}
\includegraphics[width=2\columnwidth]{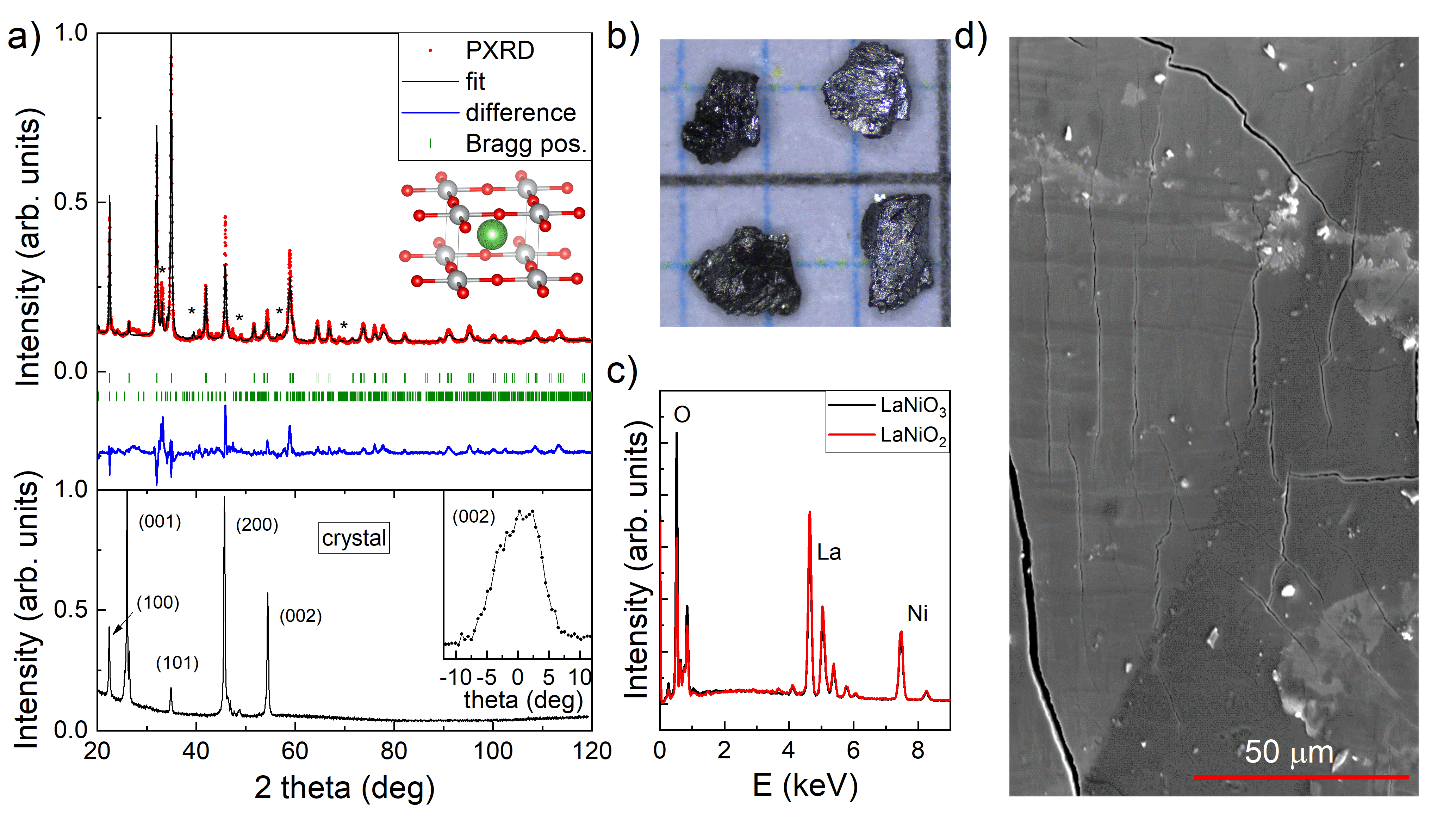}
\par\end{centering}
\caption{a) The top panel shows a PXRD pattern of pulverized LaNiO$_{2}$ crystals taken with Cu K$_\alpha$ radiation. The solid red line in the top panel is the fit from a Rietveld refinement, which includes the LaNiO$_{2}$ phase (upper green symbols) and a minority LaNiO$_{2.5}$ phase (lower green symbols). The most prominent Bragg peaks of the latter phase are indicated by asterisks. The inset shows the IL nickelate structure with the green, grey, and red spheres corresponding to La, Ni, and O respectively. The bottom panel shows a XRD pattern from an out-of-plane scan of a crystal prior to pulverization. The peaks of the (0 0 $L$), ($H$ 0 0), and ($H$ 0 $L$) families of the LaNiO$_{2}$ phase with the tetragonal $P4/mmm$ unit cell are indicated. The inset displays a rocking curve of the (0 0 2) peak. b) Photograph of four representative reduced crystals on millimeter paper, where one crystal was used for the diffraction measurements in panel a. c) EDX spectra from cleaved surfaces of an as-grown LaNiO$_{3}$ (black) and a reduced (red) crystal. Note the reduced intensity of the O $K_{\alpha_1}$ peak after reduction, whereas the La and Ni lines remain essentially unchanged, as expected. d) SEM image of a crystal surface after reduction (not cleaved). Note that the cracks and furrows visible on the surface are mostly arranged in rectangular patterns and likely correspond to domain boundaries between tetragonal LaNiO$_{2}$ twin domains. A few cracks are arranged in angles of $\sim120^\circ$, which is possibly a consequence of the presence of rhombohedral domain boundaries in the pristine LaNiO$_{3}$ phase prior to the reduction. The same crystal was used for the resistivity measurements in Fig.~\ref{resistivity}.}
\label{powder}
\end{figure*}

\section{Results}
\subsection{Crystal growth and topotactic reduction}

For the synthesis of sizable perovskite LaNiO$_{3}$ crystals, we employ a high gas pressure OFZ furnace with 150 bar oxygen pressure. In the obtained boule with a diameter of 6-8 mm and a length of 8 cm, a single grain develops during the growth after a few mm. For the subsequent topotoactic reduction, crystals from the boule are broken into smaller fragments, from which we select pieces with lateral dimensions between 1.5 mm and 500 $\mu$m.

In a previous reduction study, an indirect contact method with up to 4 weeks reduction time was applied to La$_{1-x}$Ca$_x$NiO$_3$ crystals with typical dimensions of 150 $\times$ 150 $\times$ 150 $\mu$m$^3$  \cite{Puphal2021}. Notably, we find for millimeter sized LaNiO$_{3}$ crystals that even 8 weeks of indirect contact reduction only lead to the intermediate LaNiO$_{2.5}$ phase, suggesting that for sizable crystals the reduction beyond the intermediate phase progresses exponentially slow \cite{Puphal2022} and a tremendous reduction time would be required to reach the IL phase via the indirect contact method. In the present study, we thus apply direct contact between the millimeter-sized LaNiO$_{3}$ crystals and the reducing agent CaH$_2$ for 2-3 weeks. This yields a transformation of the majority of our crystals to the IL phase (Fig.~\ref{powder}a)), although a minority phase of incompletely reduced LaNiO$_{2.5}$ can still be detected in PXRD (top panel in Fig.~\ref{powder}a)).

Figure~\ref{powder}b) shows four representative crystals with lateral dimensions of approximately 1 $\times$ 1 mm$^2$, which were obtained after two weeks of direct contact reduction. The weight of each crystal is about 2 mg.  Indications for the reduced oxygen content in the crystals after the topotactic treatment can be recognized from scanning electron microscopy (SEM) with energy-dispersive x-ray spectroscopy (EDX) analysis (Fig.~\ref{powder}c)), although a quantitative determination of the LaNiO$_{2}$ and LaNiO$_{2.5}$ fractions is challenging due to the low sensitivity of the technique to light elements such as oxygen. On the other hand, a Rietveld refinement of PXRD data can allow for a more robust analysis, in particular due to the distinct lattice constants of the different phases. In more detail, as the entire crystal is pulverized for the PXRD, all constituent phases of the crystal can be exposed (top panel in Fig.~\ref{powder}a)). The PXRD data can be refined reasonably well when assuming that the crystal consists of  LaNiO$_{2}$ in space group $P4/mmm$ \cite{Hayward1999,Lin202101,Puphal2021} and the intermediate LaNiO$_{2.5}$ phase in space group $C2/c$ \cite{Alonso1997}, with the former contributing 83.7 wt\% and the latter 16.3 wt\%. The lattice parameters obtained from the refinement are presented in Tab.~\ref{tab}. While the fraction of the incompletely reduced minority phase is considerable in the PXRD of the crystal of Fig.~\ref{powder}a), 
a XRD out-of-plane scan ($\omega-2\Theta$ scan) taken on a crystal prior to pulverization shows virtually no traces of the LaNiO$_{2.5}$ phase (bottom panel in Fig.~\ref{powder}a)). Instead, only specific LaNiO$_{2}$ Bragg reflections are present, which are associated with the (0 0 $L$), ($H$ 0 0), and ($H$ 0 $L$) peak families of the IL structure in space group $P4/mmm$. Note that in the XRD pattern the relative intensities between the peak families differ from the PXRD pattern, which is due to a dominance of the (0 0 $L$) and ($H$ 0 0) oriented domains in the measured crystal in the bottom panel of Fig.~\ref{powder}a). The inset in Fig.~\ref{powder}a) shows a rocking curve of the (0 0 2) peak of the LaNiO$_{2}$ crystal, which exhibits a relatively broad width. This indicates that the mosaic spread of the crystallographic planes is large when averaging over the entire crystal. Nonetheless, it is likely that on a local scale the mosaic spread is substantially smaller and the planes of the IL crystal lattice are highly aligned, similarly to the local crystal structure that was revealed by scanning
transmission electron microscopy (STEM) within the domains of La$_{1-x}$Ca$_x$NiO$_2$ crystals \cite{Puphal2021}.

Notably, it was found previously that La$_{1-x}$Ca$_x$NiO$_2$ crystals tend to break along the boundaries between the three orthogonal domains of the tetragonal space group $P4/mmm$ \cite{Puphal2021}. This separation of crystallographic domains occurs especially upon excessive reduction, but precursors of the separation can be observed already earlier in the form of cracks and furrows on the crystal surface. A representative SEM image of a LaNiO$_{2}$ crystal surface is shown in Fig.~\ref{powder}d), where several adjacent cracks are oriented either in a parallel or orthogonal fashion. The average distance between these cracks is much larger than the typical domain size of 10 $\times$ 10 $\times$ 10 $\mu$m$^3$ in La$_{1-x}$Ca$_x$NiO$_2$ crystals \cite{Puphal2021}, suggesting an increased domain size for the OFZ grown and directly reduced crystals in the present study. This notion is supported 
by the fact that the 50 $\times$ 50 $\times$ 30 $\mu$m$^3$ crystal used in the high-pressure measurements (see below) shows an XRD signal that originates almost exclusively from one domain. The occurrence of numerous cracks on larger lengths scales is likely responsible for the broad width of the rocking curve in the inset in Fig.~\ref{powder}a), as the spatial separation of the domains can lead to a misalignment of the crystallographic planes in the macroscopic average. Nevertheless, we note that in spite of the onset of domain separation, the crystals in Fig.~\ref{powder}b) were robust and not brittle when they were handled during our study.

\begin{figure}[tb]
 \begin{centering}
\includegraphics[width=1\columnwidth]{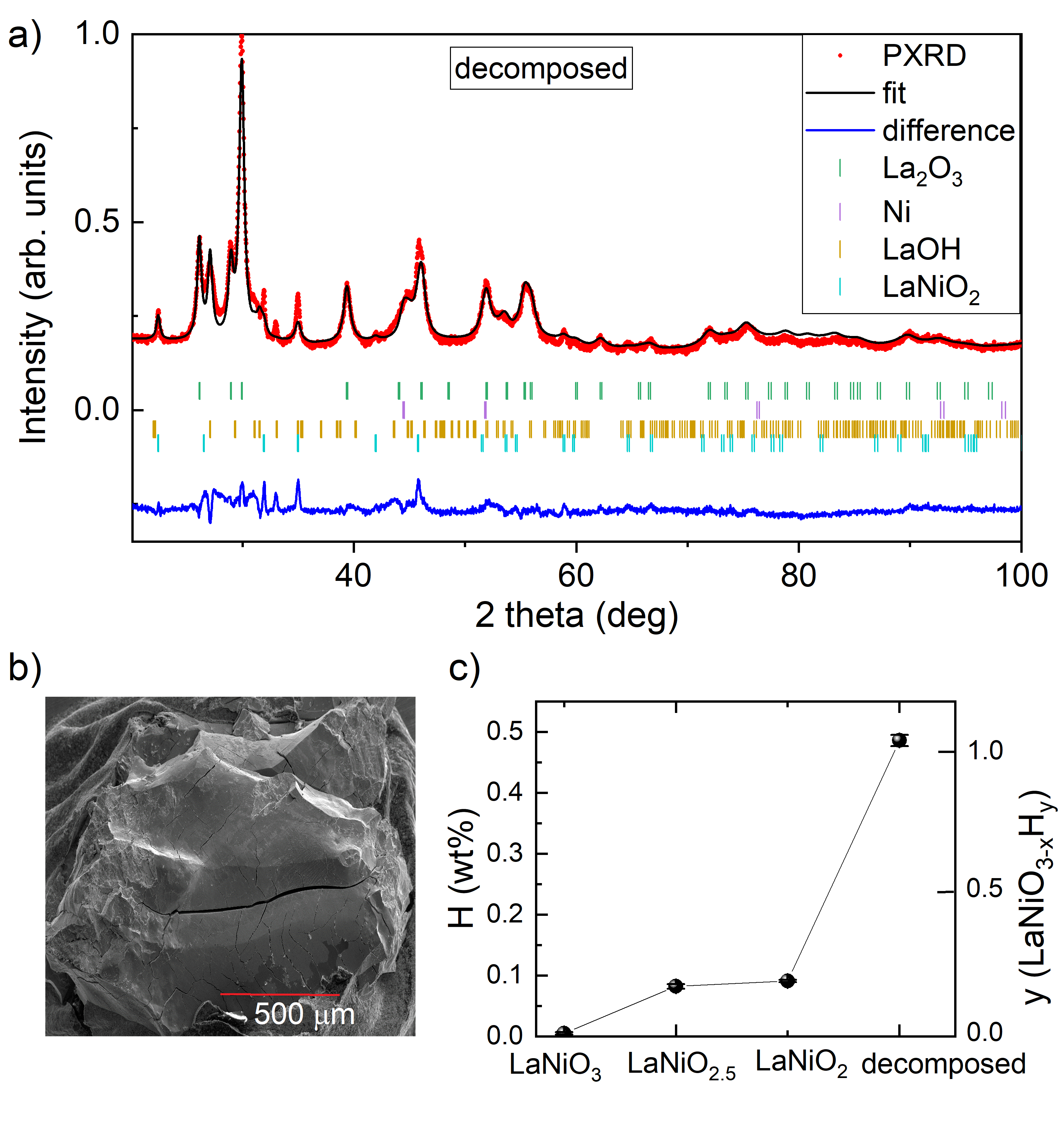}
\par\end{centering}
\caption{a) PXRD pattern of a pulverized crystal that experienced excessive reduction (decomposition), taken with Cu K$_\alpha$ radiation. The Rietveld refinement includes various phases, as indicated in the legend. b) SEM image of the decomposed crystal. The image shows a cleaved surface of the crystal, which was exposed after the topotactic reduction. c) Hydrogen content determined by gas extraction from a pristine sample (LaNiO$_{3}$), indirect contact reduced sample (LaNiO$_{2.5}$), direct contact reduced sample (LaNiO$_{2}$), as well as the decomposed direct contact sample. The hydrogen is given in weight-\% (left axis) and for the assumption of a hypothetical LaNiO$_{3-x}$H$_y$ composition of each sample, where $x$ is 0, 0.5, and 1, respectively (right axis).}
\label{overred}
\end{figure}

\begin{figure*}[tb]
 \begin{centering}
\includegraphics[width=1.5\columnwidth]{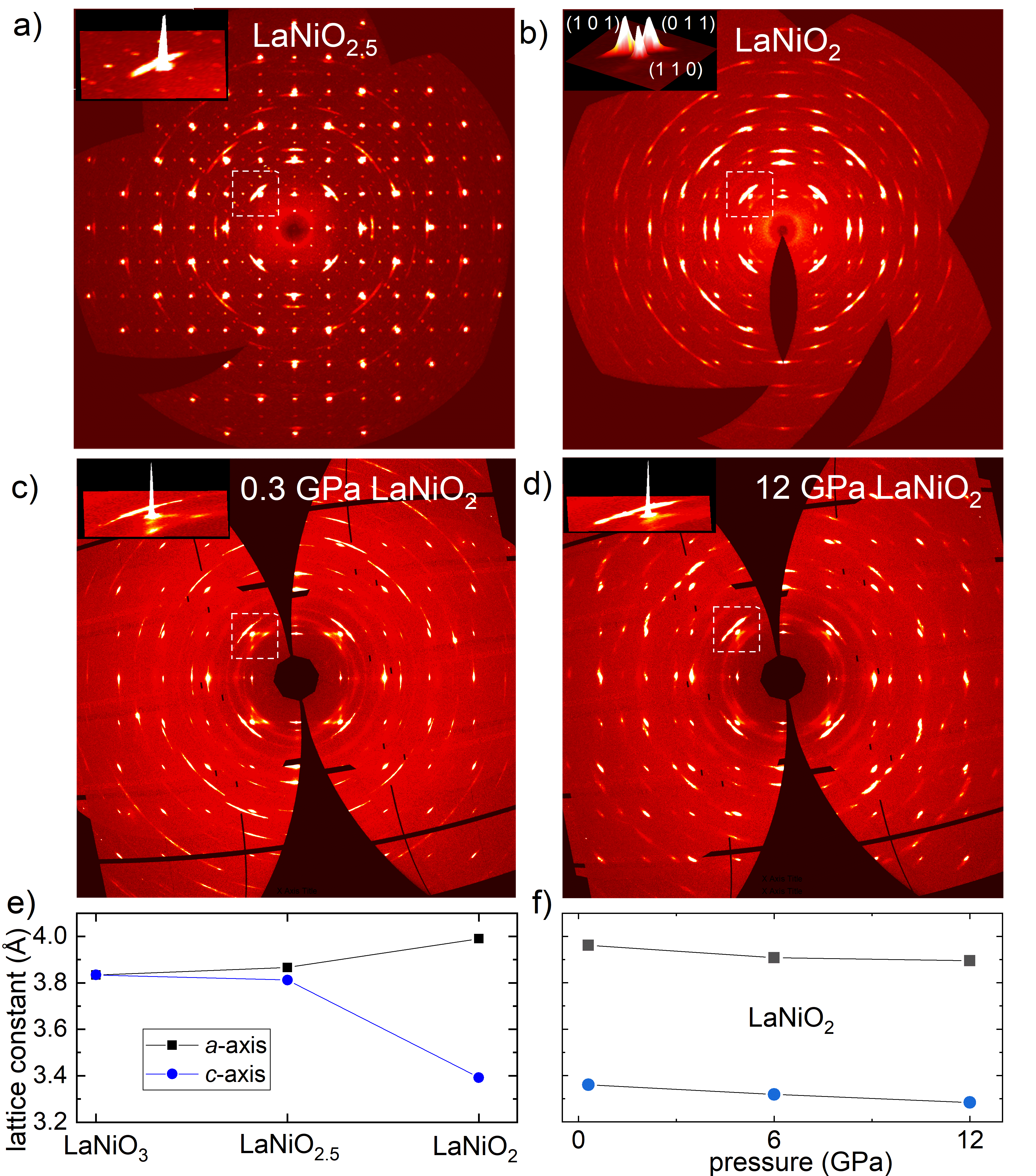}
\par\end{centering}
\caption{Maps of XRD intensities in the ($H$ $K$ 0) planes of the reduced crystals. a), b) Maps acquired at ambient conditions with Mo K$_{\alpha}$ radiation ($\lambda=0.71073\,$\AA) from a LaNiO$_{2.5}$ and LaNiO$_{2}$ crystal, respectively. The LaNiO$_{2.5}$ map shows superstructure reflections between the main Bragg reflections, which are not present in the LaNiO$_{2}$ map. The inset in each panel magnifies the region around a three-fold split Bragg reflection (see dashed white box), with the out-of-plane dimension corresponding to the XRD intensity. In the inset in panel a) the main reflection can be indexed as (1 1 0), while the two weaker reflections originate from the (1 0 1) and (0 1 1) twin domain, respectively. Very weak superstructure reflections can also be seen in the inset. In the inset in panel b), the three reflections of the twin domains exhibit similar intensities, indicating an equal population of the domains in the investigated crystal. c), d) High-resolution XRD maps of a different LaNiO$_{2}$ crystal acquired with synchrotron radiation ($\lambda=0.3738\,$\AA) for applied pressures of 0.3 and 12 GPa, respectively. The insets highlights the dominant intensity of the of the (1 1 0) domain in the investigated crystal. e) Evolution of the lattice constants as a function of reduced oxygen content. For LaNiO$_{3}$, the lattice constants $a=c$ refer to the pseudocubic unit cell (Ni-Ni distance, see text). For LaNiO$_{2.5}$ and LaNiO$_{2}$, $a$ and $c$ are derived from their tetragonal $P4/mmm$ unit cells (see text). f) Lattice constants of LaNiO$_{2}$ as a function of hydrostatic pressure. }
\label{crys}
\end{figure*}

\subsection{Excessive reduction and hydrogen incorporation}

Besides the presence of the insufficiently reduced LaNiO$_{2.5}$ phase, also an excessive reduction is detrimental for the realization of high-quality LaNiO$_{2}$ crystals. Fig.~\ref{overred}a) shows the PXRD pattern of a crystal that was reduced far beyond the ideal IL LaNiO$_{2}$ composition, revealing that the sample contains elemental Ni, crystalline $\alpha$-La$_{2}$O$_{3}$ and LaOH, as well as remnants of LaNiO$_{2}$. In more detail, the composition in weight-\% obtained from Rietveld refinement is 62.3 wt\% La$_{2}$O$_{3}$, 15.15 wt\% Ni, 13.22 wt\% LaOH, and 9.33 wt\% LaNiO$_{2}$. Note that lanthanum oxide hydride LaOH is a relatively uncommon compound, but it is typically obtained as a byproduct when treating La$_{2}$O$_{3}$ with hydrogen \cite{Malaman1984}.  Also note that generally a small amount of Ni inclusions can be expected even in the non-excessively reduced samples in Fig.~\ref{powder} due to NiO impurities in the parent LaNiO$_{3}$ crystals. A refinement of the data in Fig.~\ref{powder}a with a forced third phase (not shown here) yields a weight percentage of Ni of 1.60(1) wt\%, although the third phase does not improve the quality of the fit appreciably. In the following, we will refer to excessively reduced samples as decomposed crystals, although we note that the sample retains some crystalline integrity, while exhibiting an increased number of cracks and furrows (Fig.~\ref{overred}b)). In other words, the LaNiO$_{2}$ phase in the sample is mostly decomposed, whereas a crystalline structure of other phases is still present to a certain degree. 

Remarkably, the decomposed crystals (Figs.~\ref{overred}a,b)) and the LaNiO$_{2}$ crystals in Fig.~\ref{powder} were obtained from the same reduction batch, \textit{i.e.}, the crystals were exposed to closely similar reduction conditions. However, one notable difference between the crystals are the dimensions, with the decomposed crystal exhibiting a much larger size of approximately 2 $\times$ 1.5 $\times$ 1.5 mm$^3$. Moreover, we find that crystals with somewhat smaller dimensions than the crystals from Fig.~\ref{powder} show a trend of enhanced LaNiO$_{2.5}$ contributions, \textit{i.e.}, the reduction appears to have not progressed as much. We attribute these observations to an exothermal nature of the topotactic reduction process, where additional heat generated within large crystals expedites the reduction, which is very sensitive to temperature \cite{Hayward2003}. Hence, we conclude that the reduction parameters of the present study (see Methods) are best suited for the reduction of LaNiO$_{3}$ crystals with dimensions of approximately 1 mm$^3$, whereas modifications would be required for an optimal reduction of larger or smaller crystals.  

One vividly debated aspect of the topotactic reduction of nickelates is the possible incorporation of hydrogen in the crystal lattice \cite{Onozuka2016,Si2020,Malyi2022,Alvarez2022,Bernardini2022,Si2022}, which could lead to the formation of oxyhydride nickelates, in analogy to topotactically reduced vanadates of composition SrVO$_2$H \cite{DenisRomero2014}. In fact, some of us recently reported the presence of a significant amount of hydrogen that accumulates in powder samples during the reduction from LaNiO$_{3}$ to LaNiO$_{2}$ when using CaH$_2$ in indirect contact \cite{Puphal2022}. However, as the key result, the analysis of neutron diffraction data revealed that hydrogen or OH were likely not incorporated at proper lattice positions in the LaNiO$_{2}$ unit cell. Instead, we proposed that the hydrogen, which was detected by a gas extraction method, is trapped at Ni impurities and/or at grain surfaces and boundaries \cite{Puphal2022}.

Here, we also use gas extraction to investigate the occurrence of possible hydrogen incorporation during the reduction of our OFZ grown LaNiO$_{3}$ single-crystals. Figure~\ref{overred}c) shows the extracted hydrogen content in weight-\% for LaNiO$_{3}$, LaNiO$_{2.5}$, LaNiO$_{2}$, as well as the decomposed crystal from Figs.~\ref{overred}a,b). For the former three cases, we find a steady increase of the detected amount of hydrogen with increasing reduction progress, quantitatively similar to the powder reduction in Ref.~\onlinecite{Puphal2022}. In the case of the decomposed crystal, however, the amount of hydrogen is strongly increased by more than a factor of five, although this crystal was exposed to the reduction process for the same duration as the LaNiO$_{2}$ crystal. This observation supports the notion that LaNiO$_{2}$ is not prone to topotactic hydrogen incorporation \cite{Puphal2022}, but instead that hydrogen in topotactic nickelate samples is closely associated with impurities and/or decomposed phases. In particular, while the decomposition into La$_{2}$O$_{3}$ and Ni as a source for hydrogen trapping was already suggested in Ref.~\onlinecite{Puphal2022}, the Rietveld refinement of the significantly decomposed crystal in Fig.~\ref{overred}a) can only be performed when also including LaOH. Thus, while we do not find indications for the formation of any oxyhydride nickelate, the observed La-based oxyhydride likely explains the steep increase of detected hydrogen gas for the decomposed sample (Fig.~\ref{overred}c)). The small amount of hydrogen gas detected for the LaNiO$_{2.5}$ and LaNiO$_{2}$ samples can be due to subtle NiO impurities that are readily incorporated in as-grown LaNiO$_{3}$ crystals and gradually reduce to elemental Ni, acting as a hydrogen trap. Alternatively, local parts of the LaNiO$_{2.5}$ and LaNiO$_{2}$ crystals might have overheated in the exothermal reaction, readily yielding minor amounts of LaOH, although we note that this phase was not identified in the corresponding PXRD patterns (see for instance Fig.~\ref{powder}a)).

In conclusion, the reducing agent CaH$_2$ provides not only the hydrogen atmosphere facilitating the topotactic reduction, but can also lead to a solid-solid (oxide-hydride) reaction and the formation of LaOH. Future studies are required to clarify the onset of the reaction and its influence on the progression of the topotactic reduction, similarly to previous thin film studies comparing the application of reducing agents in direct contact and the reduction in a hydrogen gas atmosphere \cite{Ikeda2014}.

\subsection{Crystal structure at ambient and high pressures}

Detailed information on the structure of the reduced crystals can be obtained from reconstructed intensity maps of single-crystal XRD data (Fig.~\ref{crys}). As a first step, we analyze the XRD data of a crystal that was reduced until the intermediate LaNiO$_{2.5}$ phase. In the reconstructed map of the ($H$ $K$ 0) planes (Fig.~\ref{crys}a)), sharp superstructure reflections with minor intensity appear in addition to the main Bragg reflections. These superstructure reflections cannot be captured by a simple tetragonal unit cell. Nonetheless, for better comparability with the fully reduced samples, we perform the Rietveld refinement in the tetragonal space group $P4/mmm$, \textit{i.e.}, the same space group that is commonly employed for the IL nickelates \cite{Hayward1999,Hayward2003,Puphal2022,Ortiz2022}. However, in contrast to the IL crystal structure, we allow for the occupation of apical oxygen sites in the unit cell, reflecting the oxygen stoichiometry of the LaNiO$_{2.5}$ phase. Note that our simplified refinement of LaNiO$_{2.5}$ in $P4/mmm$ for comparative purposes is different from previous powder studies on the material, which employed the larger monoclinic $C2/c$ unit cell \cite{Alonso1997}. The refined lattice parameters are presented in Tab.~\ref{tab}.

Next, we acquire single-crystal XRD maps from a crystal after completed reduction. This crystal is a fragment with approximate dimensions of 50 $\times$ 50 $\times$ 50 $\mu$m$^3$, which was broken off from a much larger reduced crystal. The obtained reconstructed maps are similar to those of La$_{1-x}$Ca$_x$NiO$_{2}$ crystals \cite{Puphal2021}. In particular, a three-fold splitting of the XRD reflexes is present (see  Fig.~\ref{crys}b)), which is indicative of the domain formation upon reduction that was already recognized in the SEM image (Fig.~\ref{powder}d)). The onset of a three-fold splitting is also observable in the XRD map of the intermediate LaNiO$_{2.5}$ phase (see Fig.~\ref{crys}a). The refinement of the LaNiO$_{2}$ crystal is carried out taking into account a statistical distribution of three orthogonally oriented twin domains of the tetragonal $P4/mmm$ unit cell, with the lattice constants $a,b=3.9901(11)$\,\AA~ and $c=3.3918(14)$\,\AA. These lattice constants are slightly larger than the values reported for LaNiO$_{2}$ powder and La$_{1-x}$Ca$_x$NiO$_{2}$ crystals \cite{Hayward1999,Puphal2022,Ortiz2022,Puphal2021}, which might point to a small amount of remaining apical oxygen ions or residual domains of the LaNiO$_{2.5}$ phase in the investigated crystal fragment. 

Figure~\ref{crys}e) focuses on the evolution of the lattice parameters as a function of reduced  oxygen content. For better comparability, we convert the rhombohedral lattice constants of LaNiO$_{3}$ into a pseudocubic setting, where the pseudocubic lattice constants $a=b=c$ are equivalent to the Ni-Ni distance. The rhombohedral lattice constants (see Tab.~\ref{tab}) were obtained from Rietveld refinement of a pulverized LaNiO$_{3}$ single-crystal. As a key result we find that the topotactic reduction from LaNiO$_{3}$ to the intermediate LaNiO$_{2.5}$ phase leaves the Ni-Ni distance almost unchanged, whereas a clear variation occurs for the transition from LaNiO$_{2.5}$ to LaNiO$_{2}$, with a moderate increase of the in-plane lattice constant and a dramatic collapse of the $c$-axis lattice constant. This behavior can be rationalized when considering that the partial occupation of the apical oxygen sites in LaNiO$_{2.5}$ can still stabilize the dimensions of the original unit cell, whereas the complete lack of apical oxygen in the IL structure leads to the $c$-axis collapse. Notably, while the $c$-axis shrinks, the $a/b$ lattice constant expands such that the unit cell volume changes only slightly between LaNiO$_{3}$ and LaNiO$_{2}$. This approximate conservation of the volume might facilitate the topotactic transformation of the macroscopic crystals in our study, whereas crystals of materials with strong volume changes during reduction might build up high strain gradients that can yield brittle and fractionalized crystals.

\begin{table}[tb]
\caption{Refined lattice constants of unreduced and reduced nickelate crystals. The powder samples were a pulverized pristine LaNiO$_{3}$ crystal that was refined in the rhombohedral space group $R\bar3c$ (hexagonal axes) and a pulverized reduced crystal that was refined with a LaNiO$_{2.5}$ minority phase in the space group $C2/c$ (monoclinic angle $\alpha=93.54(2)^{\circ}$) and a LaNiO$_{2}$ majority phase in the tetragonal space group $P4/mmm$. The crystalline samples were a LaNiO$_{2.5}$ and a LaNiO$_{2}$ crystal. Unlike the LaNiO$_{2.5}$ powder with space group  $C2/c$, the LaNiO$_{2.5}$ crystal was considered in a simplified approach and refined in $P4/mmm$ (see text). For crystalline LaNiO$_{2}$, high-resolution synchrotron XRD data was refined in $P4/mmm$ for the cases of 0, 0.3, 6, and 12 GPa pressure.}
\begin{tabular}{c|ccccc}
\hline \hline 
 & type & symmetry & p {[}GPa{]} & $a$ {[}\,\AA~{]} & $c$ {[}\,\AA~{]}\tabularnewline
\hline 
LaNiO$_{3}$ & powder & $R\bar3c$ & - & 5.45409(7) & 13.12546(19)\tabularnewline
LaNi$_{2.5}$ & powder  &  $C2/c$ & - & 7.914(2) & 7.443(2) \tabularnewline
  & crystal  & $P4/mmm$ & - & 3.8664(17) & 3.812(2)\tabularnewline
LaNiO$_{2}$ & powder & $P4/mmm$ & - & 3.9540(3) & 3.3744(3) \tabularnewline
 & crystal & $P4/mmm$ & - & 3.9901(11) & 3.3918(14)\tabularnewline
 & crystal & $P4/mmm$ & 0.3 & 3.9619(3) & 3.360(3)\tabularnewline
 & crystal & $P4/mmm$ & 6.0 & 3.9082(7) & 3.318(7)\tabularnewline
 & crystal & $P4/mmm$ & 12 & 3.8953(8) & 3.28(10)\tabularnewline
 \hline \hline 
\end{tabular}
\label{tab}
\end{table}

To gain insights into the stability and elasticity of the IL structure obtained after the $c$-axis collapse, we carry out high-resolution synchrotron XRD measurements under high pressure in a diamond anvil cell. Figure~\ref{crys}f) displays the dependence of the lattice parameters on the applied pressure, which was measured on a piece broken off from the same LaNiO$_{2}$ crystal that was measured in ambient conditions. We find that under the application of 12 GPa, the lattice parameters compress to $a=3.8953(8)$\,\AA~and $c=3.284(10)$\,\AA. 
The calculated bulk modulus via the unit cell volume $V$ extracted from refinement yields a large value of $K=-V\cdot dp/dV=155.39$ GPa, which is similar to what is reported for LaNiO$_{2}$ via first-principles calculations \cite{Bernardini2022} as well as for cuprates ranging from 60-180 GPa \cite{Qin2005,Hyatt2001}, highest for the related IL cuprates. In this context, we note that the application of pressure is a powerful knob to tune the properties of cuprate superconductors \cite{Mark2022}, yielding the record transition temperature $T_{c}$ of more than 150 K in mercury-based cuprates at 30 GPa \cite{Chu1993}. Similarly, a recent pressure study on IL nickelates demonstrated that the $T_{c}$ of Pr$_{0.82}$Sr$_{0.18}$NiO$_2$ thin films can be increased from 17 K to $\sim 31$ K under 12.1 GPa of hydrostatic pressure, suggesting a $T_{c}$ vs. pressure enhancement law with a slope of 0.96 K/GPa \cite{Wang2022b}. Along these lines, our analysis reveals that 12 GPa pressure shrinks the $c$-axis lattice constant of LaNiO$_{2}$ by 0.11\,\AA. This change is much larger than the reported effect of chemical pressure induced in LaNiO$_2$ films via Ca-doping \cite{zeng2021}. For instance, a Ca-substitution of $x$=0.35 changes the $c$-axis of LaNiO$_2$ films grown on SrTiO$_3$ substrates from 3.405 \AA\, to 3.368 \AA, \textit{i.e.}, the lattice constant shrinks only by 0.037 \AA. The strong change due to 12 GPa pressure presumably increases the hybridization between the Ni $3d$ and rare-earth $5d$ orbitals considerably. The presence of this hybridization is a hallmark of IL nickelates \cite{Hepting2021}, whereas it is absent in the IL cuprate CaCuO$_{2}$ \cite{Botana2020}. On the other hand, the decrease of the in-plane lattice constant under pressure leads to a decrease of the Ni-O-Ni bond angle in the NiO$_2$ plane and therewith a decrease of the hybridization between the planar Ni $3d_{x^2-y^2}$ and O $2p$ orbitals can be expected \cite{Wang2022b,Kang2022}. While this hybridization is already small in unstrained IL nickelates \cite{Lee2004}, the application of hydrostatic pressure apparently provides a tool to impede it further.

\subsection{Magnetic susceptibility}

\begin{figure}
 \begin{centering}
\includegraphics[width=1.0\columnwidth]{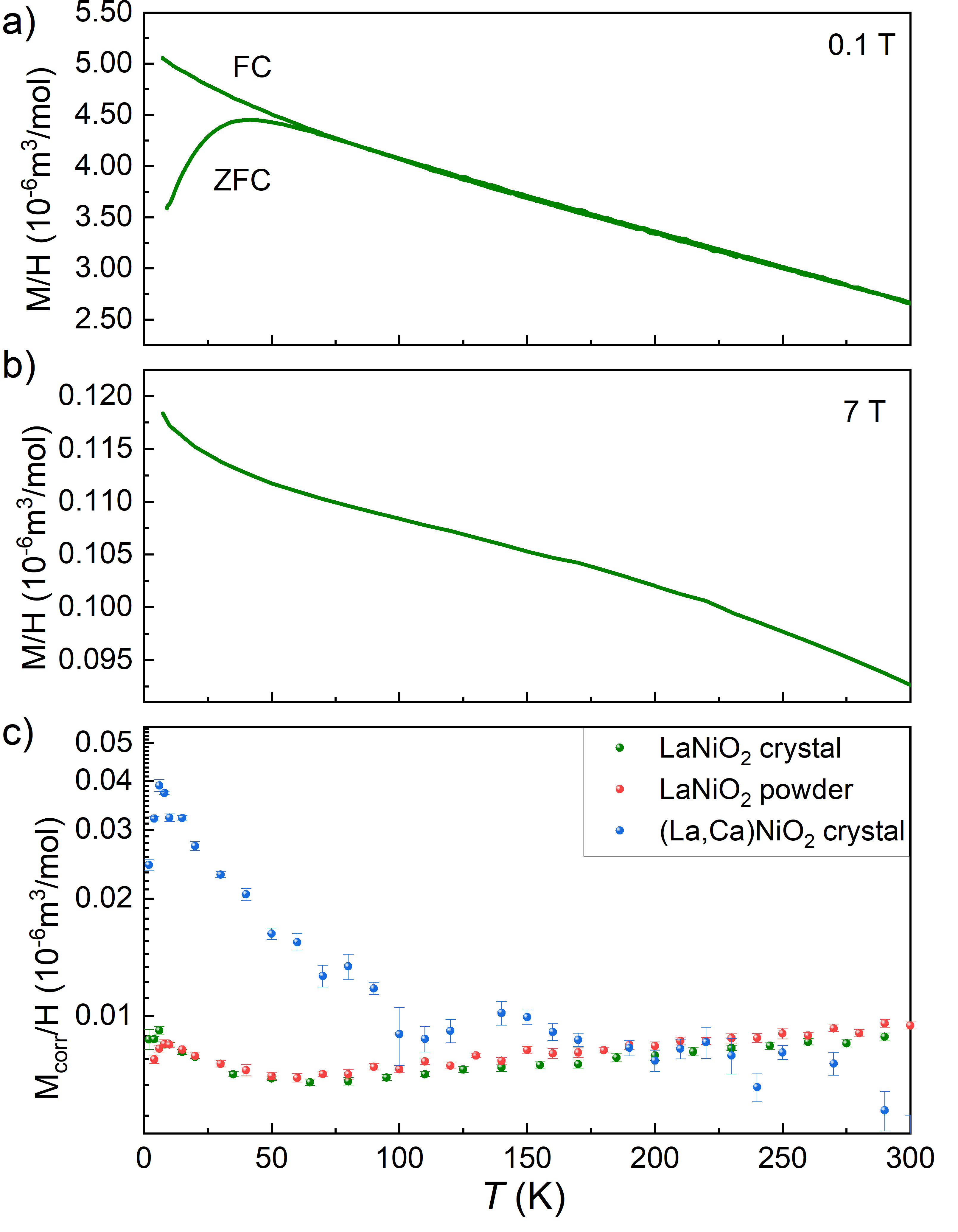}
\par\end{centering}
\caption{a) Susceptibility of a LaNiO$_2$ crystal, measured upon zero field cooling (ZFC) and field cooling (FC) in a small external field of 0.1 T. b) Susceptibility of the same crystal measured in a strong field of 7 T.
c) Comparison of the corrected magnetic susceptibilities of the LaNiO$_2$ crystal (green symbols), LaNiO$_2$ powder (red symbols), and a La$_{0.93}$Ca$_{0.07}$NiO$_2$ crystal (blue symbols). The LaNiO$_2$ powder was obtained from an indirect contact reduction of LaNiO$_3$ powder synthesized via the citrate-nitrate method \cite{Ortiz2022}. The La$_{0.93}$Ca$_{0.07}$NiO$_2$ crystal was obtained from an indirect contact reduction of a La$_{0.93}$Ca$_{0.07}$NiO$_3$ crystal synthesized via salt flux growth with oxidizer in a multi-anvil press \cite{Puphal2021}. The corrected susceptibilities were extracted via the Honda-Owen method (see text) from magnetization versus field curves.}
\label{magnetization}
\end{figure}

Figure~\ref{magnetization}a) shows the susceptibility of a LaNiO$_2$ crystal measured in a small external field. Prior thin film studies focused on the paramagnetic response at the superconducting transition of infinite-layer nickelates \cite{Zeng2022n}, whereas the detailed evolution of the susceptibility across a wide temperature range remains unreported.  Yet, our susceptibility data of LaNiO$_2$ are reminiscent of prior works on polycrystalline samples of layered nickelates \cite{Huangfu2020,Ortiz2022,Lin202101,Wissel2022}, with a splitting of the curves measured upon zero field cooling (ZFC) and field cooling (FC), as well as a cusp of the ZFC curve around $\sim$30\,K. For stronger applied fields the bifurcation between the ZFC and FC curve decreases gradually, and for a field of 7 T we find the gap to be completely closed and both curves merged (Fig.~\ref{magnetization}b)). Complementary ac susceptibility measurements in Refs.~\onlinecite{Huangfu2020,Ortiz2022,Lin202101} have identified a spin freezing as the origin of this behavior. Specifically in Ref.~\onlinecite{Ortiz2022}, these glassy dynamics observed in a polycrystalline LaNiO$_2$ sample were attributed to the presence of subtle local oxygen disorder in the form of remaining apical oxygen. Such excess oxygen might result in clusters with effective Ni$^{2+}$ moments that dominate both the dc susceptibility signal and the dynamics in the ac signal. In principle, the glassy dynamics can be suppressed in strong fields. However, especially the high-field susceptibility (Fig.~\ref{magnetization}b)) can be influenced by a ferromagnetic signal from elemental Ni impurities \cite{Ortiz2022}, which are likely present in our LaNiO$_2$ crystals, although they are estimated to correspond to less than 2 wt\%.

In order to discern between the ferromagnetic Ni background and intrinsic paramagnetic behavior of the majority phase, we apply the Honda-Owen method \cite{Honda1910,Owen1912} to isotherms measured between $4$ and $7$\,T. This method extrapolates the measured susceptibility $M/H = M_{\text corr}/H + C_{\text{sat}}M_{\text{sat}}/H$ for $1/H \rightarrow 0$, where $M/H$ is the measured susceptibility, $M_{\text corr}/H$ the corrected (intrinsic) susceptibility, $C_{\text{sat}}$ the presumed ferromagnetic impurity content and $M_{\text{sat}}$ its saturation magnetization. The resulting corrected susceptibility of a LaNiO$_2$ crystal is presented in Fig.~\ref{magnetization}c) (green symbols). In addition, the corrected susceptibility from a powder sample of LaNiO$_2$ \cite{Ortiz2022} (red symbols) and a batch of several La$_{1-x}$Ca$_x$NiO$_2$ crystals (blue symbols) \cite{Puphal2021} are shown. Whereas the corrected susceptibilities of the former two samples exhibit a subtle minimum around $\sim$65 K and increase almost linearly towards room temperature, the susceptibility of the Ca-substituted sample decreases in this temperature range. The increase of the susceptibility of LaNiO$_2$ indicates non-Curie-Weiss behavior and can be ascribed to enhanced AFM fluctuations \cite{Ortiz2022}. We note that the corrected  susceptibilities of the LaNiO$_2$ crystal and the LaNiO$_2$ powder are almost identical across the entire measured temperature range, in spite of the fact that the LaNiO$_3$ precursor phases were synthesized by different methods and presumably contained a different amount of NiO and other secondary phases. Hence, the closely similar susceptibility values extracted by the Honda-Owen method, corroborate that the method is viable to determine the intrinsic susceptibility of the majority IL phase in the samples. Notably, the non-Curie-Weiss behavior of LaNiO$_2$ resembles the susceptibility of lightly doped cuprates, such as La$_{2-x}$Sr$_x$CuO$_4$ with $x \geq 0.04$ \cite{Nakano1994}, hinting towards an analogy between magnetic correlations in parent IL nickelates and doped cuprates \cite{Ortiz2022}. In this context, the effects of self-doping due to rare-earth electron pockets on the magnetic response are an interesting topic for future studies on parent IL nickelates. 

In contrast, the corrected susceptibility of La$_{1-x}$Ca$_x$NiO$_2$, which is hole-doped due to the Ca-substitution, is clearly distinct from LaNiO$_2$. Remarkably, overdoped cuprates, such as La$_{2-x}$Sr$_x$CuO$_4$ with $x \geq 0.22$ \cite{Nakano1994}, show a similar susceptibility with a  strong upturn towards lower temperatures, thus underscoring the analogy between the magnetic correlations in IL nickelates and cuprates, although with a substantial offset with respect to the nominal hole-doping levels. For completeness, we note that the corrected susceptibility of La$_{1-x}$Ca$_x$NiO$_2$ exhibits a sharp drop below $\sim$6\,K, which is a similar temperature as the $T_c$ of La$_{1-x}$Ca$_x$NiO$_2$ films \cite{zeng2021}. However, we emphasize that the data in Fig.~\ref{magnetization}c) were extracted from isotherms measured in high fields between $4$ and $7$\,T where any superconductivity should be (at least partially) suppressed. Instead, the sharp downturn---which more subtly occurs also in the LaNiO$_2$ data--- is likely related to the glassy state with spins that are frozen in arbitrary directions at these low temperatures, although with underlying AFM correlations, which is an effect that is not properly filtered out by the Honda-Owen method that is suitable for ferromagnetic impurities.

\begin{figure*}
 \begin{centering}
\includegraphics[width=2\columnwidth]{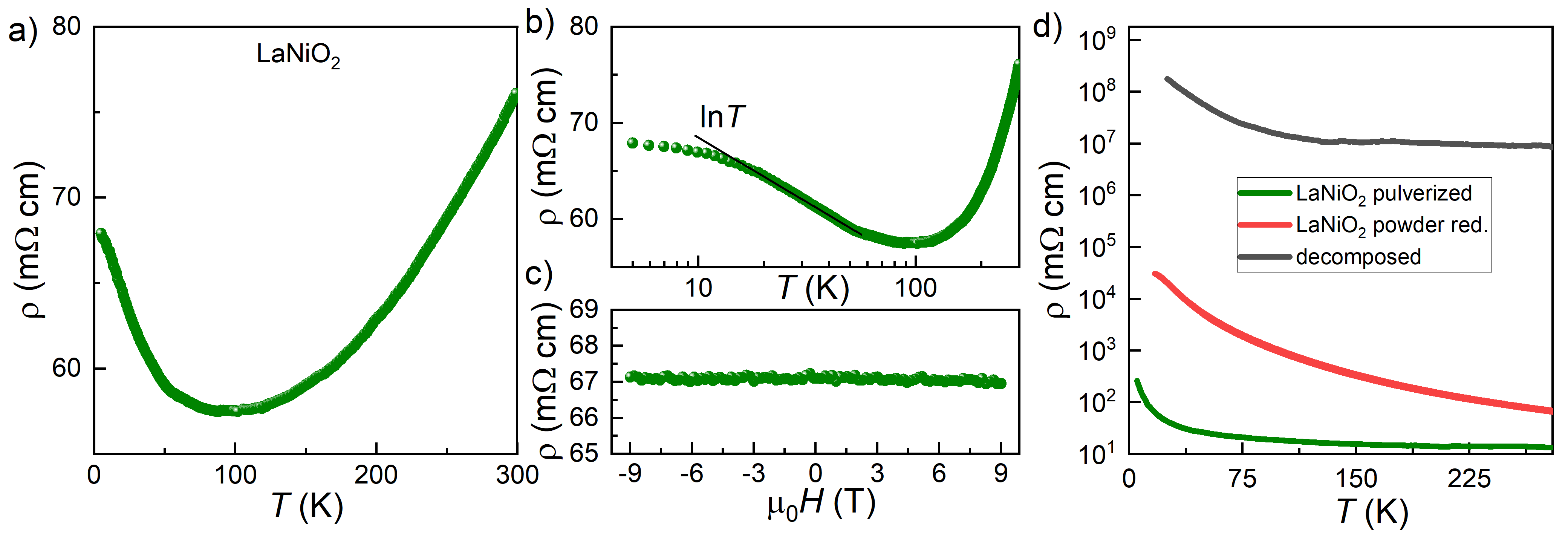}
\par\end{centering}
\caption{{Electronic transport of reduced samples. a) Resistivity of a LaNiO$_2$ crystal. b) The same resistivity data, but on a linear-log scale. The solid black line is a guide to the eye, highlighting the temperature range with a logarithmic dependence. c) Resistivity at 4 K measured as a function of the applied magnetic field from -9 to 9 T. No magnetoresistance effect is observed within the experimental error. d) Resistivities of pressed powder pellets. The green curve corresponds to a pulverized LaNiO$_2$ crystal, the red curve to a LaNiO$_2$ powder samples that was readily reduced in powder form (see Ref.~\onlinecite{Ortiz2022}), and the black curve is from a decomposed crystal.}}
\label{resistivity}
\end{figure*}

\subsection{Electronic transport}

Figure~\ref{resistivity} displays the resistivity of various reduced samples. As expected, the measurement of a LaNiO$_2$ crystal reveals metallic behavior at high temperatures and a weakly insulating upturn at low temperatures (Fig.~\ref{resistivity}a)). Qualitatively, this behavior is analogous to the resistivity of small La$_{0.92}$Ca$_{0.08}$NiO$_{2}$ crystals \cite{Puphal2021} and the in-plane resistivity of $R_{1-x}A_x$NiO$_2$ films with low doping \cite{Osada2020,Zeng2020,Li2021MBE,Gao2021,Li20201,osada2021,Wang2022,Ren2021,Lee2022}. Moreover, metallic behavior with a low-temperature upturn is reminiscent of underdoped cuprates \cite{Takagi1992}. The magnitude of our resistivity is comparable to previous reports on LaNiO$_2$ thin films \cite{Kawai2009,Kaneko2009,zeng2021}, although we note that some studies achieved films with much lower residual resistivities \cite{osada2021,Ikeda2016}. Since our data were acquired from a reduced crystal with dimensions as large as 1.4 $\times$ 0.8 $\times$ 0.8 mm$^3$, it can be assumed that a large quantity of IL domains contributes to the resistivity in Fig.~\ref{resistivity}a), with each domain likely exhibiting an anisotropy between the in-plane and out-of-plane transport, which, however, is not yet established quantitatively for IL nickelates. Furthermore, it is possible that residual domains in the LaNiO$_{2.5}$ phase and/or the presence of insulating La$_2$O$_3$ influence the transport and lead to an increased value of our measured resistivity. 

Nevertheless, we emphasize that on a linear-log scale, our data closely follow a logarithmic temperature (ln $T$) dependence below the resistivity minimum at $\sim$100 K (Fig.~\ref{resistivity}b)). This temperature dependence has been identified as a hallmark of the resistivity of the parent compounds of the IL nickelates \cite{Hsu2022} and was discussed in the context of a Mott-Kondo scenario \cite{Zhang20201,Yang2022}, or strong electron correlations \cite{Hsu2021}. Moreover, below 10 K, high-quality LaNiO$_2$ films showed a deviation from the logarithmic temperature dependence and tend towards a plateau \cite{Hsu2022,Zhang20201}, while arguably even further optimized films showed a downturn beyond the plateau at lowest temperatures, which was interpreted as a possible onset of superconductivity \cite{osada2021}. In Fig.~\ref{resistivity}b), the emergence of a plateau-like behavior is also visible, yet no downturn occurs until the lowest measured temperature ($\sim$4 K). 
 
Insights into the origin of the upturn of the resistivity below $\sim$100 K can be obtained from magnetoresistance measurements. Figure~\ref{resistivity}c) presents the resistivity of the LaNiO$_2$ crystal at 4 K as a function of an applied magnetic field. Remarkably, the resistivity shows virtually no dependence on the field, which makes a Mott-Kondo scenario implausible and also suggests that the upturn is not dominated by disorder and weak localization effects. The absence of a field dependence was similarly observed in lightly hole-doped nickelate films \cite{Lee2022} and is compatible with an effective freezing of mobile charge carriers due to intrinsic strong correlations \cite{Hsu2021}. Nonetheless, we remark that the presence of diffuse scattering intensity around the Bragg peaks in the XRD maps (Fig.~\ref{crys}) and spin glass behavior (Fig.~\ref{magnetization}) are clear indications for the presence of disorder, although it is still possible that the correlation effect is predominant in the low-temperature transport.

In order to further explore the effects of disorder and an anticipated transport anisotropy, we probe the resistivity of a LaNiO$_2$ crystal that was pulverized in air and pressed into a pellet form (Fig.~\ref{resistivity}d)). The PXRD of the same powder is given in Fig.~\ref{powder}a). While at high temperatures the magnitude of the resistivity of the crystal (Fig.~\ref{resistivity}a)) and the pressed powder are similar, the latter samples lack a decrease of the resistivity with decreasing temperature between 300 and $\sim$100 K, where the resistivity minimum of the former sample occurred. The continuous increase of the resistivity of the powder, which signals semi-conducting behavior, is presumably a consequence of the hampered transport across boundaries between powder grains. Moreover, since the powder resistivity measurement corresponds to an average of the different crystallographic directions of LaNiO$_2$, it reflects the transport anisotropy and also suggests that the data in Figs.~\ref{resistivity}a,b) were mostly influenced by the transport within the more conducting NiO$_2$ planes. 

In addition, we compare the resistivity of the pulverized crystal and LaNiO$_2$ powder that was obtained via the topotactic reduction of LaNiO$_3$ powder. This powder exhibits a very high purity, with secondary phases below 2~wt\% (details of the synthesis and characterization are given in Refs.~\onlinecite{Ortiz2022,Puphal2022}). Nevertheless, we find that the resistivity of the powder obtained readily from powder reduction is strongly enhanced across the entire temperature range and reaches even the semi-conducting/insulating regime (Fig.~\ref{resistivity}d)). We ascribe this behavior to the exposure of an increased surface area of the powder grains during the topotactic reduction. In particular, the reduction of powder likely facilitates the formation of local defects in the surface regions of grains and might also lead to an enhanced water adsorption \cite{Puphal2022}. Regarding a future realization of superconducting bulk IL nickelates, these findings suggest that for achieving superconductivity it might be detrimental to reduce powder samples. On the other hand, if bulk crystals with sufficient hole-doping host superconductivity, a crystal pulverized after the topotactic reduction might show superconductivity, provided that the superconducting coherence length in the $c$-axis direction is sufficiently long. 

For comprehensiveness, we also report the resistivity of a decomposed crystal. As can be seen in Fig.~\ref{resistivity}d), the resistivity evolves in a strongly insulating regime at all temperatures, \textit{i.e.}, the presence of metallic elemental Ni in such a sample is not relevant for the macroscopic transport properties.    

\subsection{Electronic structure}

\begin{figure}[tb]
 \begin{centering}
\includegraphics[width=1\columnwidth]{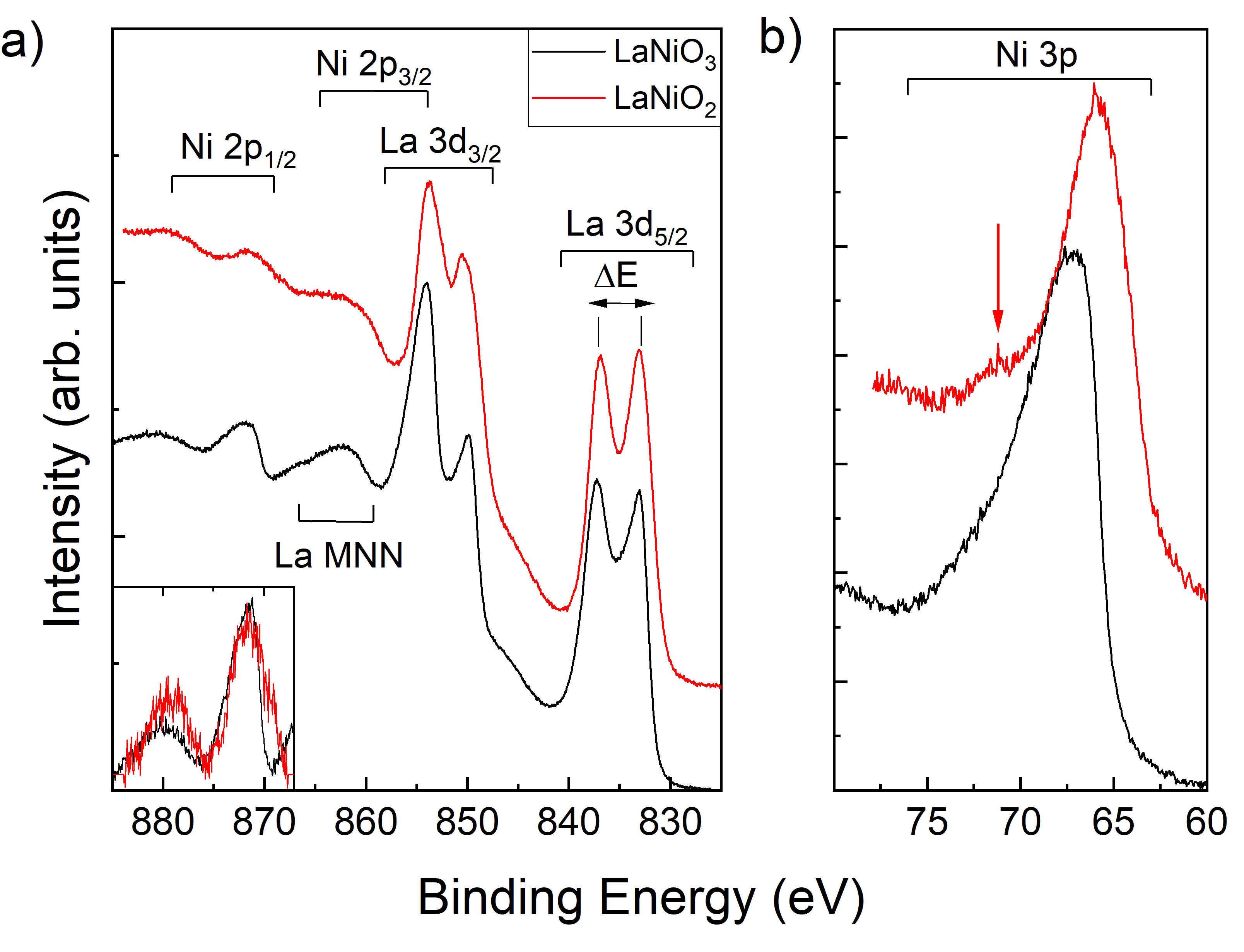}
\par\end{centering}
\caption{XPS spectra of LaNiO$_3$ (black) and LaNiO$_2$ crystals (red) for a) binding energies between 885 and 825 eV and b) energies between 80 and 60 eV. The spectra are normalized to the maximum intensity and offset in vertical direction for clarity. The LaNiO$_2$ spectra are shifted horizontally by -2.1\,eV to compensate for a charging effect. The inset in a) zooms into the Ni $2p_{1/2}$ region, after subtraction of a fitted Shirley background. The red arrow in b) indicates the satellite peak for LaNiO$_2$. }
\label{XPS}
\end{figure}

XPS can provide valuable information on the electronic structure of materials, but is highly surface sensitive as photoelectrons are emitted only from regions in close proximity to the surface. As a consequence, previous photoemission studies on IL nickelates are sparse \cite{Chen2022,Fu2020,Higashi2021}, because the invasive topochemical reduction typically deteriorates the surface region of thin film and powder samples. In principle, a few unit cell thick SrTiO$_3$ capping layer on top of thin films can prevent an impairment of the topmost nickelate layers, however, at the same time, the capping shields the nickelate film, which greatly hampers the realization of photoemission experiments. Along these lines, large LaNiO$_2$ crystals provide opportunities, as they can be cleaved. Thus, fresh surfaces can be obtained that were not directly exposed during the topotactic reduction process.

For our XPS measurements, we prepared samples in a glove box under Ar atmosphere, which were subsequently transferred under inert gas into the XPS chamber. Specifically, we fractured a large as-grown LaNiO$_3$ crystal in the glove box and unveiled a $\sim$16 mm$^2$ freshly cleaved surface. In the case of LaNiO$_2$, several $\sim$1 mm$^2$ pieces were transferred to the glove box rapidly after completion of the topotactic reduction and broken into smaller pieces that were distributed on an indium foil. Hence, the XPS signal from this sample stems from a mixture of fresh surfaces and surfaces that were exposed to the topotactic process. Nonetheless, it can be expected that the hallmarks of the transformation of Ni$^{3+}$ (nominal $3d^7$ configuration) to Ni$^{1+}$ (nominal $3d^9$ configuration) are discernible in the XPS spectra. We note that the presence of elemental and/or hydrogenated Ni can in principle influence the XPS spectra, but due to their estimated content of less than 2 wt\% this spectral contribution is expected to be minor or negligible.

Figure~\ref{XPS}a) compares the XPS of our LaNiO$_3$ and LaNiO$_2$ samples across a wide range of binding energies, including the La $3d$ and Ni $2p$ edges. As expected, the La $3d_{5/2}$ peaks \cite{Mickevicius2006,Qiao2011}, which are separated by the multiplet splitting $\Delta E$, remain similarly before and after the reduction, which implies that the electronic configuration of the La$^{3+}$ ion remains essentially unchanged. Nevertheless, we note a small reduction of the splitting from $\Delta E=4.1(3)$ eV to $3.60(5)$ eV from LaNiO$_3$ to LaNiO$_2$. We attribute this change to the presence of a subtle amount of LaOH in the reduced sample \cite{JLi2019}, which can be due to fragments in the broken crystals that were possibly already decomposed (see also Fig.~\ref{overred}). In this context, we also note that the sample presumably also contains insufficiently reduced regions with insulating LaNiO$_{2.5}$ \cite{Misra2016,Shin2022}, which can explain a slight charging effect that was observed during the XPS acquisition from the reduced sample but not for the LaNiO$_3$ sample.   

Compared to the La $3d_{5/2}$ peaks, the La $3d_{3/2}$ multiplet experiences a more distinct change upon reduction. While the shape of these peaks might also be influenced by the presence of LaOH, the multiplet overlaps with the Ni $2p_{3/2}$ binding energies. Thus, the changes in the region between 850 and 870 eV are likely mostly due to the transformation of the electronic configuration of Ni$^{3+}$ towards a monovalent state, although between 855 and 865 eV additional contributions from the La MNN Auger lines are also present. As a consequence of the various overlapping spectral contributions in the range between 850 and 870 eV, disentangling the evolution of the peaks associated with Ni $2p_{3/2}$ is problematic and we therefore refrain from carrying out peak profile fits and direct comparisons with the $2p_{3/2}$ lines of cuprates \cite{Marel1988}. A fit of the Ni $2p_{1/2}$ region is also challenging, which is due to the small intensity of these peaks and the intense background originating from inelastically scattered electrons from the La 3$d$ orbitals. Nevertheless, it can be recognized that the low binding energy edge of LaNiO$_3$ is steeper than that of LaNiO$_2$ (see inset in Fig.~\ref{XPS}a)), which is consistent with a reduced oxidation state of Ni in the latter case that emerges at lower binding energies.  


A multiplet of Ni that is free from overlap with other features is the Ni $3p$ region between 60 and 75 eV (Fig.~\ref{XPS}b)). The clear shift of the peak maximum upon reduction can be indicative of the reduced screening of the core hole potential of monovalent Ni after the reduction. In addition to the shift of the Ni $3p$ peak of LaNiO$_2$, we find an emerging shoulder at binding energies above the main peak (red arrow in Fig.~\ref{XPS}b)). In a previous study, the Ni $3p$ multiplet of reduced LaNiO$_3$ was fitted with two doublets for the $3p_{3/2}$ and $3p_{1/2}$ lines and the Ni$^{3+}$/Ni$^{2+}$ ratio was determined \cite{Qiao2011}. Recently, however, it was pointed out that the $3p$ peak shape is convoluted by a more complex multiplet splitting and an accurate modeling requires cluster-model calculations that take the local crystal field symmetry into account \cite{Baeumer2021}. While such calculations for Ni$^{1+}$ in square-planar coordination are an interesting topic for future studies, we emphasize the qualitative similarity to the $3p$ line of divalent cuprates, featuring a dominant main peak and one or more high binding energy satellite peaks that correspond to $3p^5 3d^{10}\underline  L$ and $3p^53d^{9}$ final states, respectively, with $\underline  L$ denoting a ligand hole \cite{Laan1981}.

\section{Conclusion}

In summary, we have synthesized millimeter-sized LaNiO$_2$ crystals using a direct contact CaH$_2$ treatment, whereas in previous studies on thin films and polycrystalline powders an indirect contact method was sufficient. Our characterization of the crystals revealed the hallmarks of a self-doped correlated metal with strong antiferromagnetic fluctuations. The observed non-Curie Weiss behavior of the intrinsic magnetic susceptibility corroborates the previously proposed analogy between parent IL nickelates and doped cuprates. Remarkable similarities to cuprates were also derived from the response of the lattice to hydrostatic pressure. Specifically, we unveiled a close similarity between the bulk moduli of IL nickelates and cuprates, and we anticipate that our report of the lattice parameters at 12 GPa will serve as an important input for theoretical calculations examining the $T_c$ enhancement in nickelates under hydrostatic pressure. Furthermore, the application of pressure can provide a platform to tune the hybridization between the orbitals that contribute to the low-energy electronic structure of IL nickelates, including the Ni 3$d$, O 2$p$, and rare-earth 5$d$ orbitals. Our comparison of the electronic transport of a pulverized crystal and topotactically reduced powder demonstrated the need for single crystals to determine the intrinsic behavior of IL nickelates and provides an explanation for previously observed insulating behavior of IL powder, as well as a hint about the origin of the absence of superconductivity in powder samples. Moreover, we demonstrated that hydrogen in reduced nickelates is predominantly incorporated in secondary and decomposed phases, such as LaOH, whereas  oxyhydride nickelates were not formed for our employed experimental conditions. Finally, our XPS measurements reveal a possible  fingerprint of the highly unusual Ni$^{1+}$  oxidation state in the electronic structure of IL nickelates. These results strongly encourage future investigations of the LaNiO$_2$ crystals with advanced spectroscopic methods, which might provide additional insights into the magnetic and electronic properties that remain elusive in thin film studies.
    
\section*{Acknowledgements}
We acknowledge technical support from Jeroen Jacobs and Stany Bauchau for the high-pressure experiments. We thank Valentin Zimmermann, Vignesh Sundaramurthy, Hangoo Lee, and Shohei Hayashida for helpful discussions, and Samir Hammoud for carrying out the hydrogen gas extraction studies.

\bibliographystyle{apsrev4-2}
\bibliography{nickelates}

\begin{thebibliography}{99}%
\makeatletter
\providecommand \@ifxundefined [1]{%
 \@ifx{#1\undefined}
}%
\providecommand \@ifnum [1]{%
 \ifnum #1\expandafter \@firstoftwo
 \else \expandafter \@secondoftwo
 \fi
}%
\providecommand \@ifx [1]{%
 \ifx #1\expandafter \@firstoftwo
 \else \expandafter \@secondoftwo
 \fi
}%
\providecommand \natexlab [1]{#1}%
\providecommand \enquote  [1]{``#1''}%
\providecommand \bibnamefont  [1]{#1}%
\providecommand \bibfnamefont [1]{#1}%
\providecommand \citenamefont [1]{#1}%
\providecommand \href@noop [0]{\@secondoftwo}%
\providecommand \href [0]{\begingroup \@sanitize@url \@href}%
\providecommand \@href[1]{\@@startlink{#1}\@@href}%
\providecommand \@@href[1]{\endgroup#1\@@endlink}%
\providecommand \@sanitize@url [0]{\catcode `\\12\catcode `\$12\catcode
  `\&12\catcode `\#12\catcode `\^12\catcode `\_12\catcode `\%12\relax}%
\providecommand \@@startlink[1]{}%
\providecommand \@@endlink[0]{}%
\providecommand \url  [0]{\begingroup\@sanitize@url \@url }%
\providecommand \@url [1]{\endgroup\@href {#1}{\urlprefix }}%
\providecommand \urlprefix  [0]{URL }%
\providecommand \Eprint [0]{\href }%
\providecommand \doibase [0]{https://doi.org/}%
\providecommand \selectlanguage [0]{\@gobble}%
\providecommand \bibinfo  [0]{\@secondoftwo}%
\providecommand \bibfield  [0]{\@secondoftwo}%
\providecommand \translation [1]{[#1]}%
\providecommand \BibitemOpen [0]{}%
\providecommand \bibitemStop [0]{}%
\providecommand \bibitemNoStop [0]{.\EOS\space}%
\providecommand \EOS [0]{\spacefactor3000\relax}%
\providecommand \BibitemShut  [1]{\csname bibitem#1\endcsname}%
\let\auto@bib@innerbib\@empty
\bibitem [{\citenamefont {Li}\ \emph {et~al.}(2019{\natexlab{a}})\citenamefont
  {Li}, \citenamefont {Lee}, \citenamefont {Wang}, \citenamefont {Osada},
  \citenamefont {Crossley}, \citenamefont {Lee}, \citenamefont {Cui},
  \citenamefont {Hikita},\ and\ \citenamefont {Hwang}}]{Li2019}%
  \BibitemOpen
  \bibfield  {author} {\bibinfo {author} {\bibfnamefont {D.}~\bibnamefont
  {Li}}, \bibinfo {author} {\bibfnamefont {K.}~\bibnamefont {Lee}}, \bibinfo
  {author} {\bibfnamefont {B.~Y.}\ \bibnamefont {Wang}}, \bibinfo {author}
  {\bibfnamefont {M.}~\bibnamefont {Osada}}, \bibinfo {author} {\bibfnamefont
  {S.}~\bibnamefont {Crossley}}, \bibinfo {author} {\bibfnamefont {H.~R.}\
  \bibnamefont {Lee}}, \bibinfo {author} {\bibfnamefont {Y.}~\bibnamefont
  {Cui}}, \bibinfo {author} {\bibfnamefont {Y.}~\bibnamefont {Hikita}},\ and\
  \bibinfo {author} {\bibfnamefont {H.~Y.}\ \bibnamefont {Hwang}},\ }\href
  {https://doi.org/10.1038/s41586-019-1496-5} {\bibfield  {journal} {\bibinfo
  {journal} {Nature}\ }\textbf {\bibinfo {volume} {572}},\ \bibinfo {pages}
  {624} (\bibinfo {year} {2019}{\natexlab{a}})}\BibitemShut {NoStop}%
\bibitem [{\citenamefont {Zeng}\ \emph {et~al.}(2020)\citenamefont {Zeng},
  \citenamefont {Tang}, \citenamefont {Yin}, \citenamefont {Li}, \citenamefont
  {Li}, \citenamefont {Huang}, \citenamefont {Hu}, \citenamefont {Liu},
  \citenamefont {Omar}, \citenamefont {Jani}, \citenamefont {Lim},
  \citenamefont {Han}, \citenamefont {Wan}, \citenamefont {Yang}, \citenamefont
  {Pennycook}, \citenamefont {Wee},\ and\ \citenamefont {Ariando}}]{Zeng2020}%
  \BibitemOpen
  \bibfield  {author} {\bibinfo {author} {\bibfnamefont {S.}~\bibnamefont
  {Zeng}}, \bibinfo {author} {\bibfnamefont {C.~S.}\ \bibnamefont {Tang}},
  \bibinfo {author} {\bibfnamefont {X.}~\bibnamefont {Yin}}, \bibinfo {author}
  {\bibfnamefont {C.}~\bibnamefont {Li}}, \bibinfo {author} {\bibfnamefont
  {M.}~\bibnamefont {Li}}, \bibinfo {author} {\bibfnamefont {Z.}~\bibnamefont
  {Huang}}, \bibinfo {author} {\bibfnamefont {J.}~\bibnamefont {Hu}}, \bibinfo
  {author} {\bibfnamefont {W.}~\bibnamefont {Liu}}, \bibinfo {author}
  {\bibfnamefont {G.~J.}\ \bibnamefont {Omar}}, \bibinfo {author}
  {\bibfnamefont {H.}~\bibnamefont {Jani}}, \bibinfo {author} {\bibfnamefont
  {Z.~S.}\ \bibnamefont {Lim}}, \bibinfo {author} {\bibfnamefont
  {K.}~\bibnamefont {Han}}, \bibinfo {author} {\bibfnamefont {D.}~\bibnamefont
  {Wan}}, \bibinfo {author} {\bibfnamefont {P.}~\bibnamefont {Yang}}, \bibinfo
  {author} {\bibfnamefont {S.~J.}\ \bibnamefont {Pennycook}}, \bibinfo {author}
  {\bibfnamefont {A.~T.~S.}\ \bibnamefont {Wee}},\ and\ \bibinfo {author}
  {\bibfnamefont {A.}~\bibnamefont {Ariando}},\ }\href
  {https://doi.org/10.1103/PhysRevLett.125.147003} {\bibfield  {journal}
  {\bibinfo  {journal} {Phys. Rev. Lett.}\ }\textbf {\bibinfo {volume} {125}},\
  \bibinfo {pages} {147003} (\bibinfo {year} {2020})}\BibitemShut {NoStop}%
\bibitem [{\citenamefont {Li}\ \emph {et~al.}(2021)\citenamefont {Li},
  \citenamefont {Sun}, \citenamefont {Yang}, \citenamefont {Cai}, \citenamefont
  {Guo}, \citenamefont {Gu}, \citenamefont {Zhu},\ and\ \citenamefont
  {Nie}}]{Li2021MBE}%
  \BibitemOpen
  \bibfield  {author} {\bibinfo {author} {\bibfnamefont {Y.}~\bibnamefont
  {Li}}, \bibinfo {author} {\bibfnamefont {W.}~\bibnamefont {Sun}}, \bibinfo
  {author} {\bibfnamefont {J.}~\bibnamefont {Yang}}, \bibinfo {author}
  {\bibfnamefont {X.}~\bibnamefont {Cai}}, \bibinfo {author} {\bibfnamefont
  {W.}~\bibnamefont {Guo}}, \bibinfo {author} {\bibfnamefont {Z.}~\bibnamefont
  {Gu}}, \bibinfo {author} {\bibfnamefont {Y.}~\bibnamefont {Zhu}},\ and\
  \bibinfo {author} {\bibfnamefont {Y.}~\bibnamefont {Nie}},\ }\href
  {https://www.frontiersin.org/articles/10.3389/fphy.2021.719534} {\bibfield
  {journal} {\bibinfo  {journal} {Front. Phys.}\ }\textbf {\bibinfo {volume}
  {9:719534}} (\bibinfo {year} {2021})}\BibitemShut {NoStop}%
\bibitem [{\citenamefont {Gao}\ \emph {et~al.}(2021)\citenamefont {Gao},
  \citenamefont {Zhao}, \citenamefont {Zhou},\ and\ \citenamefont
  {Zhu}}]{Gao2021}%
  \BibitemOpen
  \bibfield  {author} {\bibinfo {author} {\bibfnamefont {Q.}~\bibnamefont
  {Gao}}, \bibinfo {author} {\bibfnamefont {Y.}~\bibnamefont {Zhao}}, \bibinfo
  {author} {\bibfnamefont {X.-J.}\ \bibnamefont {Zhou}},\ and\ \bibinfo
  {author} {\bibfnamefont {Z.}~\bibnamefont {Zhu}},\ }\href
  {https://doi.org/10.1088/0256-307x/38/7/077401} {\bibfield  {journal}
  {\bibinfo  {journal} {Chin Phys. Lett.}\ }\textbf {\bibinfo {volume} {38}},\
  \bibinfo {pages} {077401} (\bibinfo {year} {2021})}\BibitemShut {NoStop}%
\bibitem [{\citenamefont {Li}\ \emph {et~al.}(2020{\natexlab{a}})\citenamefont
  {Li}, \citenamefont {Wang}, \citenamefont {Lee}, \citenamefont {Harvey},
  \citenamefont {Osada}, \citenamefont {Goodge}, \citenamefont {Kourkoutis},\
  and\ \citenamefont {Hwang}}]{Li20201}%
  \BibitemOpen
  \bibfield  {author} {\bibinfo {author} {\bibfnamefont {D.}~\bibnamefont
  {Li}}, \bibinfo {author} {\bibfnamefont {B.~Y.}\ \bibnamefont {Wang}},
  \bibinfo {author} {\bibfnamefont {K.}~\bibnamefont {Lee}}, \bibinfo {author}
  {\bibfnamefont {S.~P.}\ \bibnamefont {Harvey}}, \bibinfo {author}
  {\bibfnamefont {M.}~\bibnamefont {Osada}}, \bibinfo {author} {\bibfnamefont
  {B.~H.}\ \bibnamefont {Goodge}}, \bibinfo {author} {\bibfnamefont {L.~F.}\
  \bibnamefont {Kourkoutis}},\ and\ \bibinfo {author} {\bibfnamefont {H.~Y.}\
  \bibnamefont {Hwang}},\ }\href
  {https://doi.org/10.1103/PhysRevLett.125.027001} {\bibfield  {journal}
  {\bibinfo  {journal} {Phys. Rev. Lett.}\ }\textbf {\bibinfo {volume} {125}},\
  \bibinfo {pages} {027001} (\bibinfo {year} {2020}{\natexlab{a}})}\BibitemShut
  {NoStop}%
\bibitem [{\citenamefont {Osada}\ \emph {et~al.}(2020)\citenamefont {Osada},
  \citenamefont {Wang}, \citenamefont {Goodge}, \citenamefont {Lee},
  \citenamefont {Yoon}, \citenamefont {Sakuma}, \citenamefont {Li},
  \citenamefont {Miura}, \citenamefont {Kourkoutis},\ and\ \citenamefont
  {Hwang}}]{Osada2020}%
  \BibitemOpen
  \bibfield  {author} {\bibinfo {author} {\bibfnamefont {M.}~\bibnamefont
  {Osada}}, \bibinfo {author} {\bibfnamefont {B.~Y.}\ \bibnamefont {Wang}},
  \bibinfo {author} {\bibfnamefont {B.~H.}\ \bibnamefont {Goodge}}, \bibinfo
  {author} {\bibfnamefont {K.}~\bibnamefont {Lee}}, \bibinfo {author}
  {\bibfnamefont {H.}~\bibnamefont {Yoon}}, \bibinfo {author} {\bibfnamefont
  {K.}~\bibnamefont {Sakuma}}, \bibinfo {author} {\bibfnamefont
  {D.}~\bibnamefont {Li}}, \bibinfo {author} {\bibfnamefont {M.}~\bibnamefont
  {Miura}}, \bibinfo {author} {\bibfnamefont {L.~F.}\ \bibnamefont
  {Kourkoutis}},\ and\ \bibinfo {author} {\bibfnamefont {H.~Y.}\ \bibnamefont
  {Hwang}},\ }\href {https://doi.org/10.1021/acs.nanolett.0c01392} {\bibfield
  {journal} {\bibinfo  {journal} {Nano Lett.}\ }\textbf {\bibinfo {volume}
  {20}},\ \bibinfo {pages} {5735} (\bibinfo {year} {2020})}\BibitemShut
  {NoStop}%
\bibitem [{\citenamefont {Osada}\ \emph {et~al.}(2021)\citenamefont {Osada},
  \citenamefont {Wang}, \citenamefont {Goodge}, \citenamefont {Harvey},
  \citenamefont {Lee}, \citenamefont {Li}, \citenamefont {Kourkoutis},\ and\
  \citenamefont {Hwang}}]{osada2021}%
  \BibitemOpen
  \bibfield  {author} {\bibinfo {author} {\bibfnamefont {M.}~\bibnamefont
  {Osada}}, \bibinfo {author} {\bibfnamefont {B.~Y.}\ \bibnamefont {Wang}},
  \bibinfo {author} {\bibfnamefont {B.~H.}\ \bibnamefont {Goodge}}, \bibinfo
  {author} {\bibfnamefont {S.~P.}\ \bibnamefont {Harvey}}, \bibinfo {author}
  {\bibfnamefont {K.}~\bibnamefont {Lee}}, \bibinfo {author} {\bibfnamefont
  {D.}~\bibnamefont {Li}}, \bibinfo {author} {\bibfnamefont {L.~F.}\
  \bibnamefont {Kourkoutis}},\ and\ \bibinfo {author} {\bibfnamefont {H.~Y.}\
  \bibnamefont {Hwang}},\ }\href
  {https://doi.org/https://doi.org/10.1002/adma.202104083} {\bibfield
  {journal} {\bibinfo  {journal} {Adv. Mater.}\ }\textbf {\bibinfo {volume}
  {33}},\ \bibinfo {pages} {2104083} (\bibinfo {year} {2021})}\BibitemShut
  {NoStop}%
\bibitem [{\citenamefont {Wang}\ \emph
  {et~al.}(2022{\natexlab{a}})\citenamefont {Wang}, \citenamefont {Wang1},
  \citenamefont {Hsu}, \citenamefont {Osada}, \citenamefont {Lee},
  \citenamefont {Jia}, \citenamefont {Duffy}, \citenamefont {Li}, \citenamefont
  {Fowlie}, \citenamefont {Beasley}, \citenamefont {Devereaux}, \citenamefont
  {Fisher}, \citenamefont {Hussey},\ and\ \citenamefont {Hwang}}]{Wang2022}%
  \BibitemOpen
  \bibfield  {author} {\bibinfo {author} {\bibfnamefont {B.~Y.}\ \bibnamefont
  {Wang}}, \bibinfo {author} {\bibfnamefont {T.~C.}\ \bibnamefont {Wang1}},
  \bibinfo {author} {\bibfnamefont {Y.-T.}\ \bibnamefont {Hsu}}, \bibinfo
  {author} {\bibfnamefont {M.}~\bibnamefont {Osada}}, \bibinfo {author}
  {\bibfnamefont {K.}~\bibnamefont {Lee}}, \bibinfo {author} {\bibfnamefont
  {C.}~\bibnamefont {Jia}}, \bibinfo {author} {\bibfnamefont {C.}~\bibnamefont
  {Duffy}}, \bibinfo {author} {\bibfnamefont {D.}~\bibnamefont {Li}}, \bibinfo
  {author} {\bibfnamefont {J.}~\bibnamefont {Fowlie}}, \bibinfo {author}
  {\bibfnamefont {M.~R.}\ \bibnamefont {Beasley}}, \bibinfo {author}
  {\bibfnamefont {T.~P.}\ \bibnamefont {Devereaux}}, \bibinfo {author}
  {\bibfnamefont {I.~R.}\ \bibnamefont {Fisher}}, \bibinfo {author}
  {\bibfnamefont {N.~E.}\ \bibnamefont {Hussey}},\ and\ \bibinfo {author}
  {\bibfnamefont {H.~Y.}\ \bibnamefont {Hwang}},\ }\href@noop {} {\bibinfo
  {title} {Rare-earth control of the superconducting upper critical field in
  infinite-layer nickelates}} (\bibinfo {year} {2022}{\natexlab{a}}),\ \Eprint
  {https://arxiv.org/abs/2205.15355} {arXiv:2205.15355 [cond-mat.supr-con]}
  \BibitemShut {NoStop}%
\bibitem [{\citenamefont {Zeng}\ \emph
  {et~al.}(2022{\natexlab{a}})\citenamefont {Zeng}, \citenamefont {Li},
  \citenamefont {Chow}, \citenamefont {Cao}, \citenamefont {Zhang},
  \citenamefont {Tang}, \citenamefont {Yin}, \citenamefont {Lim}, \citenamefont
  {Hu}, \citenamefont {Yang},\ and\ \citenamefont {Ariando}}]{zeng2021}%
  \BibitemOpen
  \bibfield  {author} {\bibinfo {author} {\bibfnamefont {S.}~\bibnamefont
  {Zeng}}, \bibinfo {author} {\bibfnamefont {C.}~\bibnamefont {Li}}, \bibinfo
  {author} {\bibfnamefont {L.~E.}\ \bibnamefont {Chow}}, \bibinfo {author}
  {\bibfnamefont {Y.}~\bibnamefont {Cao}}, \bibinfo {author} {\bibfnamefont
  {Z.}~\bibnamefont {Zhang}}, \bibinfo {author} {\bibfnamefont {C.~S.}\
  \bibnamefont {Tang}}, \bibinfo {author} {\bibfnamefont {X.}~\bibnamefont
  {Yin}}, \bibinfo {author} {\bibfnamefont {Z.~S.}\ \bibnamefont {Lim}},
  \bibinfo {author} {\bibfnamefont {J.}~\bibnamefont {Hu}}, \bibinfo {author}
  {\bibfnamefont {P.}~\bibnamefont {Yang}},\ and\ \bibinfo {author}
  {\bibfnamefont {A.}~\bibnamefont {Ariando}},\ }\href
  {https://doi.org/10.1126/sciadv.abl9927} {\bibfield  {journal} {\bibinfo
  {journal} {Sci. Adv.}\ }\textbf {\bibinfo {volume} {8}},\ \bibinfo {pages}
  {eabl9927} (\bibinfo {year} {2022}{\natexlab{a}})}\BibitemShut {NoStop}%
\bibitem [{\citenamefont {Ren}\ \emph {et~al.}(2021)\citenamefont {Ren},
  \citenamefont {Gao}, \citenamefont {Zhao}, \citenamefont {Luo}, \citenamefont
  {Zhou},\ and\ \citenamefont {Zhu}}]{Ren2021}%
  \BibitemOpen
  \bibfield  {author} {\bibinfo {author} {\bibfnamefont {X.}~\bibnamefont
  {Ren}}, \bibinfo {author} {\bibfnamefont {Q.}~\bibnamefont {Gao}}, \bibinfo
  {author} {\bibfnamefont {Y.}~\bibnamefont {Zhao}}, \bibinfo {author}
  {\bibfnamefont {H.}~\bibnamefont {Luo}}, \bibinfo {author} {\bibfnamefont
  {X.}~\bibnamefont {Zhou}},\ and\ \bibinfo {author} {\bibfnamefont
  {Z.}~\bibnamefont {Zhu}},\ }\href@noop {} {\bibinfo {title}
  {Superconductivity in infinite-layer pr$_{0.8}$sr$_{0.2}$nio$_2$ films on
  different substrates}} (\bibinfo {year} {2021}),\ \Eprint
  {https://arxiv.org/abs/2109.05761} {arXiv:2109.05761 [cond-mat.supr-con]}
  \BibitemShut {NoStop}%
\bibitem [{\citenamefont {Lee}\ \emph {et~al.}(2022)\citenamefont {Lee},
  \citenamefont {Wang}, \citenamefont {Osada}, \citenamefont {Goodge},
  \citenamefont {Wang}, \citenamefont {Lee}, \citenamefont {Harvey},
  \citenamefont {Kim}, \citenamefont {Yu}, \citenamefont {Murthy},
  \citenamefont {Raghu}, \citenamefont {Kourkoutis},\ and\ \citenamefont
  {Hwang}}]{Lee2022}%
  \BibitemOpen
  \bibfield  {author} {\bibinfo {author} {\bibfnamefont {K.}~\bibnamefont
  {Lee}}, \bibinfo {author} {\bibfnamefont {B.~Y.}\ \bibnamefont {Wang}},
  \bibinfo {author} {\bibfnamefont {M.}~\bibnamefont {Osada}}, \bibinfo
  {author} {\bibfnamefont {B.~H.}\ \bibnamefont {Goodge}}, \bibinfo {author}
  {\bibfnamefont {T.~C.}\ \bibnamefont {Wang}}, \bibinfo {author}
  {\bibfnamefont {Y.}~\bibnamefont {Lee}}, \bibinfo {author} {\bibfnamefont
  {S.}~\bibnamefont {Harvey}}, \bibinfo {author} {\bibfnamefont {W.~J.}\
  \bibnamefont {Kim}}, \bibinfo {author} {\bibfnamefont {Y.}~\bibnamefont
  {Yu}}, \bibinfo {author} {\bibfnamefont {C.}~\bibnamefont {Murthy}}, \bibinfo
  {author} {\bibfnamefont {S.}~\bibnamefont {Raghu}}, \bibinfo {author}
  {\bibfnamefont {L.~F.}\ \bibnamefont {Kourkoutis}},\ and\ \bibinfo {author}
  {\bibfnamefont {H.~Y.}\ \bibnamefont {Hwang}},\ }\href@noop {} {\bibinfo
  {title} {Character of the "normal state" of the nickelate superconductors}}
  (\bibinfo {year} {2022}),\ \Eprint {https://arxiv.org/abs/2203.02580}
  {arXiv:2203.02580 [cond-mat.supr-con]} \BibitemShut {NoStop}%
\bibitem [{\citenamefont {Wang}\ \emph
  {et~al.}(2022{\natexlab{b}})\citenamefont {Wang}, \citenamefont {Yang},
  \citenamefont {Yang}, \citenamefont {Chen}, \citenamefont {Zhang},
  \citenamefont {Zhang}, \citenamefont {Zhu}, \citenamefont {Uwatoko},
  \citenamefont {Gu}, \citenamefont {Dong}, \citenamefont {Sun}, \citenamefont
  {Jin},\ and\ \citenamefont {Cheng}}]{Wang2022b}%
  \BibitemOpen
  \bibfield  {author} {\bibinfo {author} {\bibfnamefont {N.~N.}\ \bibnamefont
  {Wang}}, \bibinfo {author} {\bibfnamefont {M.~W.}\ \bibnamefont {Yang}},
  \bibinfo {author} {\bibfnamefont {Z.}~\bibnamefont {Yang}}, \bibinfo {author}
  {\bibfnamefont {K.~Y.}\ \bibnamefont {Chen}}, \bibinfo {author}
  {\bibfnamefont {H.}~\bibnamefont {Zhang}}, \bibinfo {author} {\bibfnamefont
  {Q.~H.}\ \bibnamefont {Zhang}}, \bibinfo {author} {\bibfnamefont {Z.~H.}\
  \bibnamefont {Zhu}}, \bibinfo {author} {\bibfnamefont {Y.}~\bibnamefont
  {Uwatoko}}, \bibinfo {author} {\bibfnamefont {L.}~\bibnamefont {Gu}},
  \bibinfo {author} {\bibfnamefont {X.~L.}\ \bibnamefont {Dong}}, \bibinfo
  {author} {\bibfnamefont {J.~P.}\ \bibnamefont {Sun}}, \bibinfo {author}
  {\bibfnamefont {K.~J.}\ \bibnamefont {Jin}},\ and\ \bibinfo {author}
  {\bibfnamefont {J.-G.}\ \bibnamefont {Cheng}},\ }\href
  {https://doi.org/10.1038/s41467-022-32065-x} {\bibfield  {journal} {\bibinfo
  {journal} {Nat. Commun.}\ }\textbf {\bibinfo {volume} {13}},\ \bibinfo
  {pages} {4367} (\bibinfo {year} {2022}{\natexlab{b}})}\BibitemShut {NoStop}%
\bibitem [{\citenamefont {Pan}\ \emph {et~al.}(2022)\citenamefont {Pan},
  \citenamefont {{Ferenc Segedin}}, \citenamefont {LaBollita}, \citenamefont
  {Song}, \citenamefont {Nica}, \citenamefont {Goodge}, \citenamefont {Pierce},
  \citenamefont {Doyle}, \citenamefont {Novakov}, \citenamefont {{C{\'{o}}rdova
  Carrizales}}, \citenamefont {N'Diaye}, \citenamefont {Shafer}, \citenamefont
  {Paik}, \citenamefont {Heron}, \citenamefont {Mason}, \citenamefont {Yacoby},
  \citenamefont {Kourkoutis}, \citenamefont {Erten}, \citenamefont {Brooks},
  \citenamefont {Botana},\ and\ \citenamefont {Mundy}}]{Pan2021}%
  \BibitemOpen
  \bibfield  {author} {\bibinfo {author} {\bibfnamefont {G.~A.}\ \bibnamefont
  {Pan}}, \bibinfo {author} {\bibfnamefont {D.}~\bibnamefont {{Ferenc
  Segedin}}}, \bibinfo {author} {\bibfnamefont {H.}~\bibnamefont {LaBollita}},
  \bibinfo {author} {\bibfnamefont {Q.}~\bibnamefont {Song}}, \bibinfo {author}
  {\bibfnamefont {E.~M.}\ \bibnamefont {Nica}}, \bibinfo {author}
  {\bibfnamefont {B.~H.}\ \bibnamefont {Goodge}}, \bibinfo {author}
  {\bibfnamefont {A.~T.}\ \bibnamefont {Pierce}}, \bibinfo {author}
  {\bibfnamefont {S.}~\bibnamefont {Doyle}}, \bibinfo {author} {\bibfnamefont
  {S.}~\bibnamefont {Novakov}}, \bibinfo {author} {\bibfnamefont
  {D.}~\bibnamefont {{C{\'{o}}rdova Carrizales}}}, \bibinfo {author}
  {\bibfnamefont {A.~T.}\ \bibnamefont {N'Diaye}}, \bibinfo {author}
  {\bibfnamefont {P.}~\bibnamefont {Shafer}}, \bibinfo {author} {\bibfnamefont
  {H.}~\bibnamefont {Paik}}, \bibinfo {author} {\bibfnamefont {J.~T.}\
  \bibnamefont {Heron}}, \bibinfo {author} {\bibfnamefont {J.~A.}\ \bibnamefont
  {Mason}}, \bibinfo {author} {\bibfnamefont {A.}~\bibnamefont {Yacoby}},
  \bibinfo {author} {\bibfnamefont {L.~F.}\ \bibnamefont {Kourkoutis}},
  \bibinfo {author} {\bibfnamefont {O.}~\bibnamefont {Erten}}, \bibinfo
  {author} {\bibfnamefont {C.~M.}\ \bibnamefont {Brooks}}, \bibinfo {author}
  {\bibfnamefont {A.~S.}\ \bibnamefont {Botana}},\ and\ \bibinfo {author}
  {\bibfnamefont {J.~A.}\ \bibnamefont {Mundy}},\ }\href
  {https://doi.org/10.1038/s41563-021-01142-9} {\bibfield  {journal} {\bibinfo
  {journal} {Nat. Mater.}\ }\textbf {\bibinfo {volume} {21}},\ \bibinfo {pages}
  {160} (\bibinfo {year} {2022})}\BibitemShut {NoStop}%
\bibitem [{\citenamefont {Hayward}\ \emph {et~al.}(1999)\citenamefont
  {Hayward}, \citenamefont {Green}, \citenamefont {Rosseinsky},\ and\
  \citenamefont {Sloan}}]{Hayward1999}%
  \BibitemOpen
  \bibfield  {author} {\bibinfo {author} {\bibfnamefont {M.~A.}\ \bibnamefont
  {Hayward}}, \bibinfo {author} {\bibfnamefont {M.~A.}\ \bibnamefont {Green}},
  \bibinfo {author} {\bibfnamefont {M.~J.}\ \bibnamefont {Rosseinsky}},\ and\
  \bibinfo {author} {\bibfnamefont {J.}~\bibnamefont {Sloan}},\ }\href
  {https://doi.org/10.1021/ja991573i} {\bibfield  {journal} {\bibinfo
  {journal} {J. Am. Chem. Soc.}\ }\textbf {\bibinfo {volume} {121}},\ \bibinfo
  {pages} {8843} (\bibinfo {year} {1999})}\BibitemShut {NoStop}%
\bibitem [{\citenamefont {Hayward}\ and\ \citenamefont
  {Rosseinsky}(2003)}]{Hayward2003}%
  \BibitemOpen
  \bibfield  {author} {\bibinfo {author} {\bibfnamefont {M.~A.}\ \bibnamefont
  {Hayward}}\ and\ \bibinfo {author} {\bibfnamefont {M.~J.}\ \bibnamefont
  {Rosseinsky}},\ }\href
  {https://doi.org/https://doi.org/10.1016/S1293-2558(03)00111-0} {\bibfield
  {journal} {\bibinfo  {journal} {Solid State Sci.}\ }\textbf {\bibinfo
  {volume} {5}},\ \bibinfo {pages} {839 } (\bibinfo {year} {2003})}\BibitemShut
  {NoStop}%
\bibitem [{\citenamefont {Lin}\ \emph {et~al.}(2022)\citenamefont {Lin},
  \citenamefont {Gawryluk}, \citenamefont {Klein}, \citenamefont {Huangfu},
  \citenamefont {Pomjakushina}, \citenamefont {von Rohr},\ and\ \citenamefont
  {Schilling}}]{Lin202101}%
  \BibitemOpen
  \bibfield  {author} {\bibinfo {author} {\bibfnamefont {H.}~\bibnamefont
  {Lin}}, \bibinfo {author} {\bibfnamefont {D.~J.}\ \bibnamefont {Gawryluk}},
  \bibinfo {author} {\bibfnamefont {Y.~M.}\ \bibnamefont {Klein}}, \bibinfo
  {author} {\bibfnamefont {S.}~\bibnamefont {Huangfu}}, \bibinfo {author}
  {\bibfnamefont {E.}~\bibnamefont {Pomjakushina}}, \bibinfo {author}
  {\bibfnamefont {F.}~\bibnamefont {von Rohr}},\ and\ \bibinfo {author}
  {\bibfnamefont {A.}~\bibnamefont {Schilling}},\ }\href
  {https://doi.org/10.1088/1367-2630/ac465e} {\bibfield  {journal} {\bibinfo
  {journal} {New J. Phys.}\ }\textbf {\bibinfo {volume} {24}},\ \bibinfo
  {pages} {013022} (\bibinfo {year} {2022})}\BibitemShut {NoStop}%
\bibitem [{\citenamefont {Fowlie}\ \emph {et~al.}(2022)\citenamefont {Fowlie},
  \citenamefont {Hadjimichael}, \citenamefont {Martins}, \citenamefont {Li},
  \citenamefont {Osada}, \citenamefont {Wang}, \citenamefont {Lee},
  \citenamefont {Lee}, \citenamefont {Salman}, \citenamefont {Prokscha},
  \citenamefont {Triscone}, \citenamefont {Hwang},\ and\ \citenamefont
  {Suter}}]{Fowlie2022}%
  \BibitemOpen
  \bibfield  {author} {\bibinfo {author} {\bibfnamefont {J.}~\bibnamefont
  {Fowlie}}, \bibinfo {author} {\bibfnamefont {M.}~\bibnamefont
  {Hadjimichael}}, \bibinfo {author} {\bibfnamefont {M.~M.}\ \bibnamefont
  {Martins}}, \bibinfo {author} {\bibfnamefont {D.}~\bibnamefont {Li}},
  \bibinfo {author} {\bibfnamefont {M.}~\bibnamefont {Osada}}, \bibinfo
  {author} {\bibfnamefont {B.~Y.}\ \bibnamefont {Wang}}, \bibinfo {author}
  {\bibfnamefont {K.}~\bibnamefont {Lee}}, \bibinfo {author} {\bibfnamefont
  {Y.}~\bibnamefont {Lee}}, \bibinfo {author} {\bibfnamefont {Z.}~\bibnamefont
  {Salman}}, \bibinfo {author} {\bibfnamefont {T.}~\bibnamefont {Prokscha}},
  \bibinfo {author} {\bibfnamefont {J.-M.}\ \bibnamefont {Triscone}}, \bibinfo
  {author} {\bibfnamefont {H.~Y.}\ \bibnamefont {Hwang}},\ and\ \bibinfo
  {author} {\bibfnamefont {A.}~\bibnamefont {Suter}},\ }\href
  {https://www.nature.com/articles/s41567-022-01684-y} {\bibfield  {journal}
  {\bibinfo  {journal} {Nat. Phys.}\ } (\bibinfo {year} {2022})}\BibitemShut
  {NoStop}%
\bibitem [{\citenamefont {Ortiz}\ \emph {et~al.}(2022)\citenamefont {Ortiz},
  \citenamefont {Puphal}, \citenamefont {Klett}, \citenamefont {Hotz},
  \citenamefont {Kremer}, \citenamefont {Trepka}, \citenamefont {Hemmida},
  \citenamefont {von Nidda}, \citenamefont {Isobe}, \citenamefont {Khasanov},
  \citenamefont {Luetkens}, \citenamefont {Hansmann}, \citenamefont {Keimer},
  \citenamefont {Sch\"afer},\ and\ \citenamefont {Hepting}}]{Ortiz2022}%
  \BibitemOpen
  \bibfield  {author} {\bibinfo {author} {\bibfnamefont {R.~A.}\ \bibnamefont
  {Ortiz}}, \bibinfo {author} {\bibfnamefont {P.}~\bibnamefont {Puphal}},
  \bibinfo {author} {\bibfnamefont {M.}~\bibnamefont {Klett}}, \bibinfo
  {author} {\bibfnamefont {F.}~\bibnamefont {Hotz}}, \bibinfo {author}
  {\bibfnamefont {R.~K.}\ \bibnamefont {Kremer}}, \bibinfo {author}
  {\bibfnamefont {H.}~\bibnamefont {Trepka}}, \bibinfo {author} {\bibfnamefont
  {M.}~\bibnamefont {Hemmida}}, \bibinfo {author} {\bibfnamefont {H.-A.~K.}\
  \bibnamefont {von Nidda}}, \bibinfo {author} {\bibfnamefont {M.}~\bibnamefont
  {Isobe}}, \bibinfo {author} {\bibfnamefont {R.}~\bibnamefont {Khasanov}},
  \bibinfo {author} {\bibfnamefont {H.}~\bibnamefont {Luetkens}}, \bibinfo
  {author} {\bibfnamefont {P.}~\bibnamefont {Hansmann}}, \bibinfo {author}
  {\bibfnamefont {B.}~\bibnamefont {Keimer}}, \bibinfo {author} {\bibfnamefont
  {T.}~\bibnamefont {Sch\"afer}},\ and\ \bibinfo {author} {\bibfnamefont
  {M.}~\bibnamefont {Hepting}},\ }\href
  {https://doi.org/10.1103/PhysRevResearch.4.023093} {\bibfield  {journal}
  {\bibinfo  {journal} {Phys. Rev. Research}\ }\textbf {\bibinfo {volume}
  {4}},\ \bibinfo {pages} {023093} (\bibinfo {year} {2022})}\BibitemShut
  {NoStop}%
\bibitem [{\citenamefont {Scalapino}(2012)}]{Scalapino2012}%
  \BibitemOpen
  \bibfield  {author} {\bibinfo {author} {\bibfnamefont {D.~J.}\ \bibnamefont
  {Scalapino}},\ }\href {https://doi.org/10.1103/RevModPhys.84.1383} {\bibfield
   {journal} {\bibinfo  {journal} {Rev. Mod. Phys.}\ }\textbf {\bibinfo
  {volume} {84}},\ \bibinfo {pages} {1383} (\bibinfo {year}
  {2012})}\BibitemShut {NoStop}%
\bibitem [{\citenamefont {Hepting}\ \emph {et~al.}(2020)\citenamefont
  {Hepting}, \citenamefont {Li}, \citenamefont {Jia}, \citenamefont {Lu},
  \citenamefont {Paris}, \citenamefont {Tseng}, \citenamefont {Feng},
  \citenamefont {Osada}, \citenamefont {Been}, \citenamefont {Hikita},
  \citenamefont {Chuang}, \citenamefont {Hussain}, \citenamefont {Zhou},
  \citenamefont {Nag}, \citenamefont {Garcia-Fernandez}, \citenamefont {Rossi},
  \citenamefont {Huang}, \citenamefont {Huang}, \citenamefont {Shen},
  \citenamefont {Schmitt}, \citenamefont {Hwang}, \citenamefont {Moritz},
  \citenamefont {Zaanen}, \citenamefont {Devereaux},\ and\ \citenamefont
  {Lee}}]{Hepting2020}%
  \BibitemOpen
  \bibfield  {author} {\bibinfo {author} {\bibfnamefont {M.}~\bibnamefont
  {Hepting}}, \bibinfo {author} {\bibfnamefont {D.}~\bibnamefont {Li}},
  \bibinfo {author} {\bibfnamefont {C.~J.}\ \bibnamefont {Jia}}, \bibinfo
  {author} {\bibfnamefont {H.}~\bibnamefont {Lu}}, \bibinfo {author}
  {\bibfnamefont {E.}~\bibnamefont {Paris}}, \bibinfo {author} {\bibfnamefont
  {Y.}~\bibnamefont {Tseng}}, \bibinfo {author} {\bibfnamefont
  {X.}~\bibnamefont {Feng}}, \bibinfo {author} {\bibfnamefont {M.}~\bibnamefont
  {Osada}}, \bibinfo {author} {\bibfnamefont {E.}~\bibnamefont {Been}},
  \bibinfo {author} {\bibfnamefont {Y.}~\bibnamefont {Hikita}}, \bibinfo
  {author} {\bibfnamefont {Y.-D.}\ \bibnamefont {Chuang}}, \bibinfo {author}
  {\bibfnamefont {Z.}~\bibnamefont {Hussain}}, \bibinfo {author} {\bibfnamefont
  {K.~J.}\ \bibnamefont {Zhou}}, \bibinfo {author} {\bibfnamefont
  {A.}~\bibnamefont {Nag}}, \bibinfo {author} {\bibfnamefont {M.}~\bibnamefont
  {Garcia-Fernandez}}, \bibinfo {author} {\bibfnamefont {M.}~\bibnamefont
  {Rossi}}, \bibinfo {author} {\bibfnamefont {H.~Y.}\ \bibnamefont {Huang}},
  \bibinfo {author} {\bibfnamefont {D.~J.}\ \bibnamefont {Huang}}, \bibinfo
  {author} {\bibfnamefont {Z.~X.}\ \bibnamefont {Shen}}, \bibinfo {author}
  {\bibfnamefont {T.}~\bibnamefont {Schmitt}}, \bibinfo {author} {\bibfnamefont
  {H.~Y.}\ \bibnamefont {Hwang}}, \bibinfo {author} {\bibfnamefont
  {B.}~\bibnamefont {Moritz}}, \bibinfo {author} {\bibfnamefont
  {J.}~\bibnamefont {Zaanen}}, \bibinfo {author} {\bibfnamefont {T.~P.}\
  \bibnamefont {Devereaux}},\ and\ \bibinfo {author} {\bibfnamefont {W.~S.}\
  \bibnamefont {Lee}},\ }\href {https://doi.org/10.1038/s41563-019-0585-z}
  {\bibfield  {journal} {\bibinfo  {journal} {Nat. Mater.}\ }\textbf {\bibinfo
  {volume} {19}},\ \bibinfo {pages} {381} (\bibinfo {year} {2020})}\BibitemShut
  {NoStop}%
\bibitem [{\citenamefont {Goodge}\ \emph {et~al.}(2021)\citenamefont {Goodge},
  \citenamefont {Li}, \citenamefont {Lee}, \citenamefont {Osada}, \citenamefont
  {Wang}, \citenamefont {Sawatzky}, \citenamefont {Hwang},\ and\ \citenamefont
  {Kourkoutis}}]{Goodge2021}%
  \BibitemOpen
  \bibfield  {author} {\bibinfo {author} {\bibfnamefont {B.~H.}\ \bibnamefont
  {Goodge}}, \bibinfo {author} {\bibfnamefont {D.}~\bibnamefont {Li}}, \bibinfo
  {author} {\bibfnamefont {K.}~\bibnamefont {Lee}}, \bibinfo {author}
  {\bibfnamefont {M.}~\bibnamefont {Osada}}, \bibinfo {author} {\bibfnamefont
  {B.~Y.}\ \bibnamefont {Wang}}, \bibinfo {author} {\bibfnamefont {G.~A.}\
  \bibnamefont {Sawatzky}}, \bibinfo {author} {\bibfnamefont {H.~Y.}\
  \bibnamefont {Hwang}},\ and\ \bibinfo {author} {\bibfnamefont {L.~F.}\
  \bibnamefont {Kourkoutis}},\ }\href {https://doi.org/10.1073/pnas.2007683118}
  {\bibfield  {journal} {\bibinfo  {journal} {Proc. Natl. Acad. Sci. U.S.A. 1}\
  }\textbf {\bibinfo {volume} {118}},\ \bibinfo {pages} {e2007683118} (\bibinfo
  {year} {2021})}\BibitemShut {NoStop}%
\bibitem [{\citenamefont {Bernardini}\ \emph
  {et~al.}(2022{\natexlab{a}})\citenamefont {Bernardini}, \citenamefont
  {Iglesias}, \citenamefont {Bibes},\ and\ \citenamefont
  {Cano}}]{Bernardini2022a}%
  \BibitemOpen
  \bibfield  {author} {\bibinfo {author} {\bibfnamefont {F.}~\bibnamefont
  {Bernardini}}, \bibinfo {author} {\bibfnamefont {L.}~\bibnamefont
  {Iglesias}}, \bibinfo {author} {\bibfnamefont {M.}~\bibnamefont {Bibes}},\
  and\ \bibinfo {author} {\bibfnamefont {A.}~\bibnamefont {Cano}},\ }\href
  {https://www.frontiersin.org/articles/10.3389/fphy.2022.828007} {\bibfield
  {journal} {\bibinfo  {journal} {Front. Phys.}\ }\textbf {\bibinfo {volume}
  {10:828007}} (\bibinfo {year} {2022}{\natexlab{a}})}\BibitemShut {NoStop}%
\bibitem [{\citenamefont {Li}\ \emph {et~al.}(2020{\natexlab{b}})\citenamefont
  {Li}, \citenamefont {He}, \citenamefont {Si}, \citenamefont {Zhu},
  \citenamefont {Zhang},\ and\ \citenamefont {Wen}}]{Li2020}%
  \BibitemOpen
  \bibfield  {author} {\bibinfo {author} {\bibfnamefont {Q.}~\bibnamefont
  {Li}}, \bibinfo {author} {\bibfnamefont {C.}~\bibnamefont {He}}, \bibinfo
  {author} {\bibfnamefont {J.}~\bibnamefont {Si}}, \bibinfo {author}
  {\bibfnamefont {X.}~\bibnamefont {Zhu}}, \bibinfo {author} {\bibfnamefont
  {Y.}~\bibnamefont {Zhang}},\ and\ \bibinfo {author} {\bibfnamefont {H.-H.}\
  \bibnamefont {Wen}},\ }\href {https://doi.org/10.1038/s43246-020-0018-1}
  {\bibfield  {journal} {\bibinfo  {journal} {Commun. Mater.}\ }\textbf
  {\bibinfo {volume} {1}},\ \bibinfo {pages} {16} (\bibinfo {year}
  {2020}{\natexlab{b}})}\BibitemShut {NoStop}%
\bibitem [{\citenamefont {Wang}\ \emph {et~al.}(2020)\citenamefont {Wang},
  \citenamefont {Zheng}, \citenamefont {Krivyakina}, \citenamefont {Chmaissem},
  \citenamefont {Lopes}, \citenamefont {Lynn}, \citenamefont {Gallington},
  \citenamefont {Ren}, \citenamefont {Rosenkranz}, \citenamefont {Mitchell},\
  and\ \citenamefont {Phelan}}]{Wang20201}%
  \BibitemOpen
  \bibfield  {author} {\bibinfo {author} {\bibfnamefont {B.-X.}\ \bibnamefont
  {Wang}}, \bibinfo {author} {\bibfnamefont {H.}~\bibnamefont {Zheng}},
  \bibinfo {author} {\bibfnamefont {E.}~\bibnamefont {Krivyakina}}, \bibinfo
  {author} {\bibfnamefont {O.}~\bibnamefont {Chmaissem}}, \bibinfo {author}
  {\bibfnamefont {P.~P.}\ \bibnamefont {Lopes}}, \bibinfo {author}
  {\bibfnamefont {J.~W.}\ \bibnamefont {Lynn}}, \bibinfo {author}
  {\bibfnamefont {L.~C.}\ \bibnamefont {Gallington}}, \bibinfo {author}
  {\bibfnamefont {Y.}~\bibnamefont {Ren}}, \bibinfo {author} {\bibfnamefont
  {S.}~\bibnamefont {Rosenkranz}}, \bibinfo {author} {\bibfnamefont {J.~F.}\
  \bibnamefont {Mitchell}},\ and\ \bibinfo {author} {\bibfnamefont
  {D.}~\bibnamefont {Phelan}},\ }\href
  {https://doi.org/10.1103/PhysRevMaterials.4.084409} {\bibfield  {journal}
  {\bibinfo  {journal} {Phys. Rev. Mater.}\ }\textbf {\bibinfo {volume} {4}},\
  \bibinfo {pages} {084409} (\bibinfo {year} {2020})}\BibitemShut {NoStop}%
\bibitem [{\citenamefont {He}\ \emph {et~al.}(2021)\citenamefont {He},
  \citenamefont {Ming}, \citenamefont {Li}, \citenamefont {Zhu}, \citenamefont
  {Si},\ and\ \citenamefont {Wen}}]{He2021}%
  \BibitemOpen
  \bibfield  {author} {\bibinfo {author} {\bibfnamefont {C.}~\bibnamefont
  {He}}, \bibinfo {author} {\bibfnamefont {X.}~\bibnamefont {Ming}}, \bibinfo
  {author} {\bibfnamefont {Q.}~\bibnamefont {Li}}, \bibinfo {author}
  {\bibfnamefont {X.}~\bibnamefont {Zhu}}, \bibinfo {author} {\bibfnamefont
  {J.}~\bibnamefont {Si}},\ and\ \bibinfo {author} {\bibfnamefont {H.-H.}\
  \bibnamefont {Wen}},\ }\href {https://doi.org/10.1088/1361-648x/abfb90}
  {\bibfield  {journal} {\bibinfo  {journal} {J. Phys. Condens. Matter}\
  }\textbf {\bibinfo {volume} {33}},\ \bibinfo {pages} {265701} (\bibinfo
  {year} {2021})}\BibitemShut {NoStop}%
\bibitem [{\citenamefont {Rao}\ \emph {et~al.}(1993)\citenamefont {Rao},
  \citenamefont {Nagarajan},\ and\ \citenamefont {Vijayaraghaven}}]{Rao1993}%
  \BibitemOpen
  \bibfield  {author} {\bibinfo {author} {\bibfnamefont {C.~N.~R.}\
  \bibnamefont {Rao}}, \bibinfo {author} {\bibfnamefont {R.}~\bibnamefont
  {Nagarajan}},\ and\ \bibinfo {author} {\bibfnamefont {R.}~\bibnamefont
  {Vijayaraghaven}},\ }\href {https://doi.org/10.1088/0953-2048/6/1/001}
  {\bibfield  {journal} {\bibinfo  {journal} {Supercond. Sci. Technol.}\
  }\textbf {\bibinfo {volume} {6}},\ \bibinfo {pages} {1} (\bibinfo {year}
  {1993})}\BibitemShut {NoStop}%
\bibitem [{\citenamefont {Geisler}\ and\ \citenamefont
  {Pentcheva}(2020)}]{Geisler2020}%
  \BibitemOpen
  \bibfield  {author} {\bibinfo {author} {\bibfnamefont {B.}~\bibnamefont
  {Geisler}}\ and\ \bibinfo {author} {\bibfnamefont {R.}~\bibnamefont
  {Pentcheva}},\ }\href {https://doi.org/10.1103/PhysRevB.102.020502}
  {\bibfield  {journal} {\bibinfo  {journal} {Phys. Rev. B}\ }\textbf {\bibinfo
  {volume} {102}},\ \bibinfo {pages} {020502} (\bibinfo {year}
  {2020})}\BibitemShut {NoStop}%
\bibitem [{\citenamefont {He}\ \emph {et~al.}(2020)\citenamefont {He},
  \citenamefont {Jiang}, \citenamefont {Lu}, \citenamefont {Song},
  \citenamefont {Chen}, \citenamefont {Jin}, \citenamefont {Shui},\ and\
  \citenamefont {Zhong}}]{He2020}%
  \BibitemOpen
  \bibfield  {author} {\bibinfo {author} {\bibfnamefont {R.}~\bibnamefont
  {He}}, \bibinfo {author} {\bibfnamefont {P.}~\bibnamefont {Jiang}}, \bibinfo
  {author} {\bibfnamefont {Y.}~\bibnamefont {Lu}}, \bibinfo {author}
  {\bibfnamefont {Y.}~\bibnamefont {Song}}, \bibinfo {author} {\bibfnamefont
  {M.}~\bibnamefont {Chen}}, \bibinfo {author} {\bibfnamefont {M.}~\bibnamefont
  {Jin}}, \bibinfo {author} {\bibfnamefont {L.}~\bibnamefont {Shui}},\ and\
  \bibinfo {author} {\bibfnamefont {Z.}~\bibnamefont {Zhong}},\ }\href
  {https://doi.org/10.1103/PhysRevB.102.035118} {\bibfield  {journal} {\bibinfo
   {journal} {Phys. Rev. B}\ }\textbf {\bibinfo {volume} {102}},\ \bibinfo
  {pages} {035118} (\bibinfo {year} {2020})}\BibitemShut {NoStop}%
\bibitem [{\citenamefont {Crespin}\ \emph {et~al.}(2005)\citenamefont
  {Crespin}, \citenamefont {Isnard}, \citenamefont {Dubois}, \citenamefont
  {Choisnet},\ and\ \citenamefont {Odier}}]{Crespin2005}%
  \BibitemOpen
  \bibfield  {author} {\bibinfo {author} {\bibfnamefont {M.}~\bibnamefont
  {Crespin}}, \bibinfo {author} {\bibfnamefont {O.}~\bibnamefont {Isnard}},
  \bibinfo {author} {\bibfnamefont {F.}~\bibnamefont {Dubois}}, \bibinfo
  {author} {\bibfnamefont {J.}~\bibnamefont {Choisnet}},\ and\ \bibinfo
  {author} {\bibfnamefont {P.}~\bibnamefont {Odier}},\ }\href
  {https://doi.org/https://doi.org/10.1016/j.jssc.2005.01.023} {\bibfield
  {journal} {\bibinfo  {journal} {J. Solid State Chem.}\ }\textbf {\bibinfo
  {volume} {178}},\ \bibinfo {pages} {1326 } (\bibinfo {year}
  {2005})}\BibitemShut {NoStop}%
\bibitem [{\citenamefont {Puphal}\ \emph {et~al.}(2022)\citenamefont {Puphal},
  \citenamefont {Pomjakushin}, \citenamefont {Ortiz}, \citenamefont {Hammoud},
  \citenamefont {Isobe}, \citenamefont {Keimer},\ and\ \citenamefont
  {Hepting}}]{Puphal2022}%
  \BibitemOpen
  \bibfield  {author} {\bibinfo {author} {\bibfnamefont {P.}~\bibnamefont
  {Puphal}}, \bibinfo {author} {\bibfnamefont {V.}~\bibnamefont {Pomjakushin}},
  \bibinfo {author} {\bibfnamefont {R.~A.}\ \bibnamefont {Ortiz}}, \bibinfo
  {author} {\bibfnamefont {S.}~\bibnamefont {Hammoud}}, \bibinfo {author}
  {\bibfnamefont {M.}~\bibnamefont {Isobe}}, \bibinfo {author} {\bibfnamefont
  {B.}~\bibnamefont {Keimer}},\ and\ \bibinfo {author} {\bibfnamefont
  {M.}~\bibnamefont {Hepting}},\ }\href
  {https://www.frontiersin.org/articles/10.3389/fphy.2022.842578} {\bibfield
  {journal} {\bibinfo  {journal} {Front. Phys.}\ }\textbf {\bibinfo {volume}
  {10:842578}} (\bibinfo {year} {2022})}\BibitemShut {NoStop}%
\bibitem [{\citenamefont {Puphal}\ \emph {et~al.}(2021)\citenamefont {Puphal},
  \citenamefont {Wu}, \citenamefont {F{\"{u}}rsich}, \citenamefont {Lee},
  \citenamefont {Pakdaman}, \citenamefont {Bruin}, \citenamefont {Nuss},
  \citenamefont {Suyolcu}, \citenamefont {van Aken}, \citenamefont {Keimer},
  \citenamefont {Isobe},\ and\ \citenamefont {Hepting}}]{Puphal2021}%
  \BibitemOpen
  \bibfield  {author} {\bibinfo {author} {\bibfnamefont {P.}~\bibnamefont
  {Puphal}}, \bibinfo {author} {\bibfnamefont {Y.-M.}\ \bibnamefont {Wu}},
  \bibinfo {author} {\bibfnamefont {K.}~\bibnamefont {F{\"{u}}rsich}}, \bibinfo
  {author} {\bibfnamefont {H.}~\bibnamefont {Lee}}, \bibinfo {author}
  {\bibfnamefont {M.}~\bibnamefont {Pakdaman}}, \bibinfo {author}
  {\bibfnamefont {J.~A.~N.}\ \bibnamefont {Bruin}}, \bibinfo {author}
  {\bibfnamefont {J.}~\bibnamefont {Nuss}}, \bibinfo {author} {\bibfnamefont
  {Y.~E.}\ \bibnamefont {Suyolcu}}, \bibinfo {author} {\bibfnamefont {P.~A.}\
  \bibnamefont {van Aken}}, \bibinfo {author} {\bibfnamefont {B.}~\bibnamefont
  {Keimer}}, \bibinfo {author} {\bibfnamefont {M.}~\bibnamefont {Isobe}},\ and\
  \bibinfo {author} {\bibfnamefont {M.}~\bibnamefont {Hepting}},\ }\href
  {https://doi.org/10.1126/sciadv.abl8091} {\bibfield  {journal} {\bibinfo
  {journal} {Sci. Adv.}\ }\textbf {\bibinfo {volume} {7}},\ \bibinfo {pages}
  {eabl8091} (\bibinfo {year} {2021})}\BibitemShut {NoStop}%
\bibitem [{\citenamefont {Damascelli}\ \emph {et~al.}(2003)\citenamefont
  {Damascelli}, \citenamefont {Hussain},\ and\ \citenamefont
  {Shen}}]{Damascelli2003}%
  \BibitemOpen
  \bibfield  {author} {\bibinfo {author} {\bibfnamefont {A.}~\bibnamefont
  {Damascelli}}, \bibinfo {author} {\bibfnamefont {Z.}~\bibnamefont
  {Hussain}},\ and\ \bibinfo {author} {\bibfnamefont {Z.-X.}\ \bibnamefont
  {Shen}},\ }\href {https://doi.org/10.1103/RevModPhys.75.473} {\bibfield
  {journal} {\bibinfo  {journal} {Rev. Mod. Phys.}\ }\textbf {\bibinfo {volume}
  {75}},\ \bibinfo {pages} {473} (\bibinfo {year} {2003})}\BibitemShut
  {NoStop}%
\bibitem [{\citenamefont {Fink}\ \emph {et~al.}(2001)\citenamefont {Fink},
  \citenamefont {Knupfer}, \citenamefont {Atzkern},\ and\ \citenamefont
  {Golden}}]{Fink2001}%
  \BibitemOpen
  \bibfield  {author} {\bibinfo {author} {\bibfnamefont {J.}~\bibnamefont
  {Fink}}, \bibinfo {author} {\bibfnamefont {M.}~\bibnamefont {Knupfer}},
  \bibinfo {author} {\bibfnamefont {S.}~\bibnamefont {Atzkern}},\ and\ \bibinfo
  {author} {\bibfnamefont {M.}~\bibnamefont {Golden}},\ }\href
  {https://doi.org/10.1016/S0368-2048(01)00254-7} {\bibfield  {journal}
  {\bibinfo  {journal} {J. Electron Spectros. Relat. Phenomena}\ }\textbf
  {\bibinfo {volume} {117-118}},\ \bibinfo {pages} {287} (\bibinfo {year}
  {2001})}\BibitemShut {NoStop}%
\bibitem [{\citenamefont {Fischer}\ \emph {et~al.}(2007)\citenamefont
  {Fischer}, \citenamefont {Kugler}, \citenamefont {Maggio-Aprile},
  \citenamefont {Berthod},\ and\ \citenamefont {Renner}}]{Fischer2007}%
  \BibitemOpen
  \bibfield  {author} {\bibinfo {author} {\bibfnamefont {O.}~\bibnamefont
  {Fischer}}, \bibinfo {author} {\bibfnamefont {M.}~\bibnamefont {Kugler}},
  \bibinfo {author} {\bibfnamefont {I.}~\bibnamefont {Maggio-Aprile}}, \bibinfo
  {author} {\bibfnamefont {C.}~\bibnamefont {Berthod}},\ and\ \bibinfo {author}
  {\bibfnamefont {C.}~\bibnamefont {Renner}},\ }\href
  {https://doi.org/10.1103/RevModPhys.79.353} {\bibfield  {journal} {\bibinfo
  {journal} {Rev. Mod. Phys.}\ }\textbf {\bibinfo {volume} {79}},\ \bibinfo
  {pages} {353} (\bibinfo {year} {2007})}\BibitemShut {NoStop}%
\bibitem [{\citenamefont {Basov}\ and\ \citenamefont
  {Timusk}(2005)}]{Basov2005}%
  \BibitemOpen
  \bibfield  {author} {\bibinfo {author} {\bibfnamefont {D.~N.}\ \bibnamefont
  {Basov}}\ and\ \bibinfo {author} {\bibfnamefont {T.}~\bibnamefont {Timusk}},\
  }\href {https://doi.org/10.1103/RevModPhys.77.721} {\bibfield  {journal}
  {\bibinfo  {journal} {Rev. Mod. Phys.}\ }\textbf {\bibinfo {volume} {77}},\
  \bibinfo {pages} {721} (\bibinfo {year} {2005})}\BibitemShut {NoStop}%
\bibitem [{\citenamefont {Devereaux}\ and\ \citenamefont
  {Hackl}(2007)}]{Devereaux2007}%
  \BibitemOpen
  \bibfield  {author} {\bibinfo {author} {\bibfnamefont {T.~P.}\ \bibnamefont
  {Devereaux}}\ and\ \bibinfo {author} {\bibfnamefont {R.}~\bibnamefont
  {Hackl}},\ }\href {https://doi.org/10.1103/RevModPhys.79.175} {\bibfield
  {journal} {\bibinfo  {journal} {Rev. Mod. Phys.}\ }\textbf {\bibinfo {volume}
  {79}},\ \bibinfo {pages} {175} (\bibinfo {year} {2007})}\BibitemShut
  {NoStop}%
\bibitem [{\citenamefont {Ament}\ \emph {et~al.}(2011)\citenamefont {Ament},
  \citenamefont {van Veenendaal}, \citenamefont {Devereaux}, \citenamefont
  {Hill},\ and\ \citenamefont {van~den Brink}}]{Ament2011}%
  \BibitemOpen
  \bibfield  {author} {\bibinfo {author} {\bibfnamefont {L.~J.~P.}\
  \bibnamefont {Ament}}, \bibinfo {author} {\bibfnamefont {M.}~\bibnamefont
  {van Veenendaal}}, \bibinfo {author} {\bibfnamefont {T.~P.}\ \bibnamefont
  {Devereaux}}, \bibinfo {author} {\bibfnamefont {J.~P.}\ \bibnamefont
  {Hill}},\ and\ \bibinfo {author} {\bibfnamefont {J.}~\bibnamefont {van~den
  Brink}},\ }\href {https://doi.org/10.1103/RevModPhys.83.705} {\bibfield
  {journal} {\bibinfo  {journal} {Rev. Mod. Phys.}\ }\textbf {\bibinfo {volume}
  {83}},\ \bibinfo {pages} {705} (\bibinfo {year} {2011})}\BibitemShut
  {NoStop}%
\bibitem [{\citenamefont {Fujita}\ \emph {et~al.}(2012)\citenamefont {Fujita},
  \citenamefont {Hiraka}, \citenamefont {Matsuda}, \citenamefont {Matsuura},
  \citenamefont {M.~Tranquada}, \citenamefont {Wakimoto}, \citenamefont {Xu},\
  and\ \citenamefont {Yamada}}]{Fujita2012}%
  \BibitemOpen
  \bibfield  {author} {\bibinfo {author} {\bibfnamefont {M.}~\bibnamefont
  {Fujita}}, \bibinfo {author} {\bibfnamefont {H.}~\bibnamefont {Hiraka}},
  \bibinfo {author} {\bibfnamefont {M.}~\bibnamefont {Matsuda}}, \bibinfo
  {author} {\bibfnamefont {M.}~\bibnamefont {Matsuura}}, \bibinfo {author}
  {\bibfnamefont {J.}~\bibnamefont {M.~Tranquada}}, \bibinfo {author}
  {\bibfnamefont {S.}~\bibnamefont {Wakimoto}}, \bibinfo {author}
  {\bibfnamefont {G.}~\bibnamefont {Xu}},\ and\ \bibinfo {author}
  {\bibfnamefont {K.}~\bibnamefont {Yamada}},\ }\href
  {https://doi.org/10.1143/JPSJ.81.011007} {\bibfield  {journal} {\bibinfo
  {journal} {J. Phys. Soc. Japan}\ }\textbf {\bibinfo {volume} {81}},\ \bibinfo
  {pages} {011007} (\bibinfo {year} {2012})}\BibitemShut {NoStop}%
\bibitem [{\citenamefont {Hepting}\ \emph {et~al.}(2021)\citenamefont
  {Hepting}, \citenamefont {Dean},\ and\ \citenamefont {Lee}}]{Hepting2021}%
  \BibitemOpen
  \bibfield  {author} {\bibinfo {author} {\bibfnamefont {M.}~\bibnamefont
  {Hepting}}, \bibinfo {author} {\bibfnamefont {M.~P.~M.}\ \bibnamefont
  {Dean}},\ and\ \bibinfo {author} {\bibfnamefont {W.-S.}\ \bibnamefont
  {Lee}},\ }\href
  {https://www.frontiersin.org/article/10.3389/fphy.2021.808683} {\bibfield
  {journal} {\bibinfo  {journal} {Front. Phys.}\ }\textbf {\bibinfo {volume}
  {9:808683}} (\bibinfo {year} {2021})}\BibitemShut {NoStop}%
\bibitem [{\citenamefont {Rossi}\ \emph {et~al.}(2021)\citenamefont {Rossi},
  \citenamefont {Lu}, \citenamefont {Nag}, \citenamefont {Li}, \citenamefont
  {Osada}, \citenamefont {Lee}, \citenamefont {Wang}, \citenamefont
  {Agrestini}, \citenamefont {Garcia-Fernandez}, \citenamefont {Kas},
  \citenamefont {Chuang}, \citenamefont {Shen}, \citenamefont {Hwang},
  \citenamefont {Moritz}, \citenamefont {Zhou}, \citenamefont {Devereaux},\
  and\ \citenamefont {Lee}}]{Rossi2020}%
  \BibitemOpen
  \bibfield  {author} {\bibinfo {author} {\bibfnamefont {M.}~\bibnamefont
  {Rossi}}, \bibinfo {author} {\bibfnamefont {H.}~\bibnamefont {Lu}}, \bibinfo
  {author} {\bibfnamefont {A.}~\bibnamefont {Nag}}, \bibinfo {author}
  {\bibfnamefont {D.}~\bibnamefont {Li}}, \bibinfo {author} {\bibfnamefont
  {M.}~\bibnamefont {Osada}}, \bibinfo {author} {\bibfnamefont
  {K.}~\bibnamefont {Lee}}, \bibinfo {author} {\bibfnamefont {B.~Y.}\
  \bibnamefont {Wang}}, \bibinfo {author} {\bibfnamefont {S.}~\bibnamefont
  {Agrestini}}, \bibinfo {author} {\bibfnamefont {M.}~\bibnamefont
  {Garcia-Fernandez}}, \bibinfo {author} {\bibfnamefont {J.~J.}\ \bibnamefont
  {Kas}}, \bibinfo {author} {\bibfnamefont {Y.-D.}\ \bibnamefont {Chuang}},
  \bibinfo {author} {\bibfnamefont {Z.~X.}\ \bibnamefont {Shen}}, \bibinfo
  {author} {\bibfnamefont {H.~Y.}\ \bibnamefont {Hwang}}, \bibinfo {author}
  {\bibfnamefont {B.}~\bibnamefont {Moritz}}, \bibinfo {author} {\bibfnamefont
  {K.-J.}\ \bibnamefont {Zhou}}, \bibinfo {author} {\bibfnamefont {T.~P.}\
  \bibnamefont {Devereaux}},\ and\ \bibinfo {author} {\bibfnamefont {W.~S.}\
  \bibnamefont {Lee}},\ }\href {https://doi.org/10.1103/PhysRevB.104.L220505}
  {\bibfield  {journal} {\bibinfo  {journal} {Phys. Rev. B}\ }\textbf {\bibinfo
  {volume} {104}},\ \bibinfo {pages} {L220505} (\bibinfo {year}
  {2021})}\BibitemShut {NoStop}%
\bibitem [{\citenamefont {Gu}\ \emph {et~al.}(2020)\citenamefont {Gu},
  \citenamefont {Li}, \citenamefont {Wan}, \citenamefont {Li}, \citenamefont
  {Guo}, \citenamefont {Yang}, \citenamefont {Li}, \citenamefont {Zhu},
  \citenamefont {Pan}, \citenamefont {Nie},\ and\ \citenamefont
  {Wen}}]{Gu2020}%
  \BibitemOpen
  \bibfield  {author} {\bibinfo {author} {\bibfnamefont {Q.}~\bibnamefont
  {Gu}}, \bibinfo {author} {\bibfnamefont {Y.}~\bibnamefont {Li}}, \bibinfo
  {author} {\bibfnamefont {S.}~\bibnamefont {Wan}}, \bibinfo {author}
  {\bibfnamefont {H.}~\bibnamefont {Li}}, \bibinfo {author} {\bibfnamefont
  {W.}~\bibnamefont {Guo}}, \bibinfo {author} {\bibfnamefont {H.}~\bibnamefont
  {Yang}}, \bibinfo {author} {\bibfnamefont {Q.}~\bibnamefont {Li}}, \bibinfo
  {author} {\bibfnamefont {X.}~\bibnamefont {Zhu}}, \bibinfo {author}
  {\bibfnamefont {X.}~\bibnamefont {Pan}}, \bibinfo {author} {\bibfnamefont
  {Y.}~\bibnamefont {Nie}},\ and\ \bibinfo {author} {\bibfnamefont {H.-H.}\
  \bibnamefont {Wen}},\ }\href {https://doi.org/10.1038/s41467-020-19908-1}
  {\bibfield  {journal} {\bibinfo  {journal} {Nat. Commun.}\ }\textbf {\bibinfo
  {volume} {11}},\ \bibinfo {pages} {6027} (\bibinfo {year}
  {2020})}\BibitemShut {NoStop}%
\bibitem [{\citenamefont {Ortiz}\ \emph {et~al.}(2021)\citenamefont {Ortiz},
  \citenamefont {Menke}, \citenamefont {Misj\'ak}, \citenamefont {Mantadakis},
  \citenamefont {F\"ursich}, \citenamefont {Schierle}, \citenamefont
  {Logvenov}, \citenamefont {Kaiser}, \citenamefont {Keimer}, \citenamefont
  {Hansmann},\ and\ \citenamefont {Benckiser}}]{Ortiz2021}%
  \BibitemOpen
  \bibfield  {author} {\bibinfo {author} {\bibfnamefont {R.~A.}\ \bibnamefont
  {Ortiz}}, \bibinfo {author} {\bibfnamefont {H.}~\bibnamefont {Menke}},
  \bibinfo {author} {\bibfnamefont {F.}~\bibnamefont {Misj\'ak}}, \bibinfo
  {author} {\bibfnamefont {D.~T.}\ \bibnamefont {Mantadakis}}, \bibinfo
  {author} {\bibfnamefont {K.}~\bibnamefont {F\"ursich}}, \bibinfo {author}
  {\bibfnamefont {E.}~\bibnamefont {Schierle}}, \bibinfo {author}
  {\bibfnamefont {G.}~\bibnamefont {Logvenov}}, \bibinfo {author}
  {\bibfnamefont {U.}~\bibnamefont {Kaiser}}, \bibinfo {author} {\bibfnamefont
  {B.}~\bibnamefont {Keimer}}, \bibinfo {author} {\bibfnamefont
  {P.}~\bibnamefont {Hansmann}},\ and\ \bibinfo {author} {\bibfnamefont
  {E.}~\bibnamefont {Benckiser}},\ }\href
  {https://doi.org/10.1103/PhysRevB.104.165137} {\bibfield  {journal} {\bibinfo
   {journal} {Phys. Rev. B}\ }\textbf {\bibinfo {volume} {104}},\ \bibinfo
  {pages} {165137} (\bibinfo {year} {2021})}\BibitemShut {NoStop}%
\bibitem [{\citenamefont {Rossi}\ \emph {et~al.}(2022)\citenamefont {Rossi},
  \citenamefont {Osada}, \citenamefont {Choi}, \citenamefont {Agrestini},
  \citenamefont {Jost}, \citenamefont {Lee}, \citenamefont {Lu}, \citenamefont
  {Wang}, \citenamefont {Lee}, \citenamefont {Nag}, \citenamefont {Chuang},
  \citenamefont {Kuo}, \citenamefont {Lee}, \citenamefont {Moritz},
  \citenamefont {Devereaux}, \citenamefont {Shen}, \citenamefont {Lee},
  \citenamefont {Zhou}, \citenamefont {Hwang},\ and\ \citenamefont
  {Lee}}]{Rossi2022}%
  \BibitemOpen
  \bibfield  {author} {\bibinfo {author} {\bibfnamefont {M.}~\bibnamefont
  {Rossi}}, \bibinfo {author} {\bibfnamefont {M.}~\bibnamefont {Osada}},
  \bibinfo {author} {\bibfnamefont {J.}~\bibnamefont {Choi}}, \bibinfo {author}
  {\bibfnamefont {S.}~\bibnamefont {Agrestini}}, \bibinfo {author}
  {\bibfnamefont {D.}~\bibnamefont {Jost}}, \bibinfo {author} {\bibfnamefont
  {Y.}~\bibnamefont {Lee}}, \bibinfo {author} {\bibfnamefont {H.}~\bibnamefont
  {Lu}}, \bibinfo {author} {\bibfnamefont {B.~Y.}\ \bibnamefont {Wang}},
  \bibinfo {author} {\bibfnamefont {K.}~\bibnamefont {Lee}}, \bibinfo {author}
  {\bibfnamefont {A.}~\bibnamefont {Nag}}, \bibinfo {author} {\bibfnamefont
  {Y.-D.}\ \bibnamefont {Chuang}}, \bibinfo {author} {\bibfnamefont {C.-T.}\
  \bibnamefont {Kuo}}, \bibinfo {author} {\bibfnamefont {S.-J.}\ \bibnamefont
  {Lee}}, \bibinfo {author} {\bibfnamefont {B.}~\bibnamefont {Moritz}},
  \bibinfo {author} {\bibfnamefont {T.~P.}\ \bibnamefont {Devereaux}}, \bibinfo
  {author} {\bibfnamefont {Z.-X.}\ \bibnamefont {Shen}}, \bibinfo {author}
  {\bibfnamefont {J.-S.}\ \bibnamefont {Lee}}, \bibinfo {author} {\bibfnamefont
  {K.-J.}\ \bibnamefont {Zhou}}, \bibinfo {author} {\bibfnamefont {H.~Y.}\
  \bibnamefont {Hwang}},\ and\ \bibinfo {author} {\bibfnamefont {W.-S.}\
  \bibnamefont {Lee}},\ }\href
  {https://www.nature.com/articles/s41567-022-01660-6} {\bibfield  {journal}
  {\bibinfo  {journal} {Nat. Phys.}\ }\textbf {\bibinfo {volume} {18}},\
  \bibinfo {pages} {869} (\bibinfo {year} {2022})}\BibitemShut {NoStop}%
\bibitem [{\citenamefont {Lu}\ \emph {et~al.}(2021)\citenamefont {Lu},
  \citenamefont {Rossi}, \citenamefont {Nag}, \citenamefont {Osada},
  \citenamefont {Li}, \citenamefont {Lee}, \citenamefont {Wang}, \citenamefont
  {Garcia-Fernandez}, \citenamefont {Agrestini}, \citenamefont {Shen},
  \citenamefont {Been}, \citenamefont {Moritz}, \citenamefont {Devereaux},
  \citenamefont {Zaanen}, \citenamefont {Hwang}, \citenamefont {Zhou},\ and\
  \citenamefont {Lee}}]{Lu2021}%
  \BibitemOpen
  \bibfield  {author} {\bibinfo {author} {\bibfnamefont {H.}~\bibnamefont
  {Lu}}, \bibinfo {author} {\bibfnamefont {M.}~\bibnamefont {Rossi}}, \bibinfo
  {author} {\bibfnamefont {A.}~\bibnamefont {Nag}}, \bibinfo {author}
  {\bibfnamefont {M.}~\bibnamefont {Osada}}, \bibinfo {author} {\bibfnamefont
  {D.~F.}\ \bibnamefont {Li}}, \bibinfo {author} {\bibfnamefont
  {K.}~\bibnamefont {Lee}}, \bibinfo {author} {\bibfnamefont {B.~Y.}\
  \bibnamefont {Wang}}, \bibinfo {author} {\bibfnamefont {M.}~\bibnamefont
  {Garcia-Fernandez}}, \bibinfo {author} {\bibfnamefont {S.}~\bibnamefont
  {Agrestini}}, \bibinfo {author} {\bibfnamefont {Z.~X.}\ \bibnamefont {Shen}},
  \bibinfo {author} {\bibfnamefont {E.~M.}\ \bibnamefont {Been}}, \bibinfo
  {author} {\bibfnamefont {B.}~\bibnamefont {Moritz}}, \bibinfo {author}
  {\bibfnamefont {T.~P.}\ \bibnamefont {Devereaux}}, \bibinfo {author}
  {\bibfnamefont {J.}~\bibnamefont {Zaanen}}, \bibinfo {author} {\bibfnamefont
  {H.~Y.}\ \bibnamefont {Hwang}}, \bibinfo {author} {\bibfnamefont {K.-J.}\
  \bibnamefont {Zhou}},\ and\ \bibinfo {author} {\bibfnamefont {W.~S.}\
  \bibnamefont {Lee}},\ }\href {https://doi.org/10.1126/science.abd7726}
  {\bibfield  {journal} {\bibinfo  {journal} {Science}\ }\textbf {\bibinfo
  {volume} {373}},\ \bibinfo {pages} {213} (\bibinfo {year}
  {2021})}\BibitemShut {NoStop}%
\bibitem [{\citenamefont {Zeng}\ \emph
  {et~al.}(2022{\natexlab{b}})\citenamefont {Zeng}, \citenamefont {Yin},
  \citenamefont {Li}, \citenamefont {Chow}, \citenamefont {Tang}, \citenamefont
  {Han}, \citenamefont {Huang}, \citenamefont {Cao}, \citenamefont {Wan},
  \citenamefont {Zhang}, \citenamefont {Lim}, \citenamefont {Diao},
  \citenamefont {Yang}, \citenamefont {Wee}, \citenamefont {Pennycook},\ and\
  \citenamefont {Ariando}}]{Zeng2022n}%
  \BibitemOpen
  \bibfield  {author} {\bibinfo {author} {\bibfnamefont {S.~W.}\ \bibnamefont
  {Zeng}}, \bibinfo {author} {\bibfnamefont {X.~M.}\ \bibnamefont {Yin}},
  \bibinfo {author} {\bibfnamefont {C.~J.}\ \bibnamefont {Li}}, \bibinfo
  {author} {\bibfnamefont {L.~E.}\ \bibnamefont {Chow}}, \bibinfo {author}
  {\bibfnamefont {C.~S.}\ \bibnamefont {Tang}}, \bibinfo {author}
  {\bibfnamefont {K.}~\bibnamefont {Han}}, \bibinfo {author} {\bibfnamefont
  {Z.}~\bibnamefont {Huang}}, \bibinfo {author} {\bibfnamefont
  {Y.}~\bibnamefont {Cao}}, \bibinfo {author} {\bibfnamefont {D.~Y.}\
  \bibnamefont {Wan}}, \bibinfo {author} {\bibfnamefont {Z.~T.}\ \bibnamefont
  {Zhang}}, \bibinfo {author} {\bibfnamefont {Z.~S.}\ \bibnamefont {Lim}},
  \bibinfo {author} {\bibfnamefont {C.~Z.}\ \bibnamefont {Diao}}, \bibinfo
  {author} {\bibfnamefont {P.}~\bibnamefont {Yang}}, \bibinfo {author}
  {\bibfnamefont {A.~T.~S.}\ \bibnamefont {Wee}}, \bibinfo {author}
  {\bibfnamefont {S.~J.}\ \bibnamefont {Pennycook}},\ and\ \bibinfo {author}
  {\bibfnamefont {A.}~\bibnamefont {Ariando}},\ }\href
  {https://doi.org/10.1038/s41467-022-28390-w} {\bibfield  {journal} {\bibinfo
  {journal} {Nat. Commun.}\ }\textbf {\bibinfo {volume} {13}},\ \bibinfo
  {pages} {743} (\bibinfo {year} {2022}{\natexlab{b}})}\BibitemShut {NoStop}%
\bibitem [{\citenamefont {Cervasio}\ \emph {et~al.}(2022)\citenamefont
  {Cervasio}, \citenamefont {Tomarchio}, \citenamefont {Verseils},
  \citenamefont {Brubach}, \citenamefont {Macis}, \citenamefont {Zeng},
  \citenamefont {Ariando}, \citenamefont {Roy},\ and\ \citenamefont
  {Lupi}}]{Cervasio2022}%
  \BibitemOpen
  \bibfield  {author} {\bibinfo {author} {\bibfnamefont {R.}~\bibnamefont
  {Cervasio}}, \bibinfo {author} {\bibfnamefont {L.}~\bibnamefont {Tomarchio}},
  \bibinfo {author} {\bibfnamefont {M.}~\bibnamefont {Verseils}}, \bibinfo
  {author} {\bibfnamefont {J.-B.}\ \bibnamefont {Brubach}}, \bibinfo {author}
  {\bibfnamefont {S.}~\bibnamefont {Macis}}, \bibinfo {author} {\bibfnamefont
  {S.}~\bibnamefont {Zeng}}, \bibinfo {author} {\bibfnamefont {A.}~\bibnamefont
  {Ariando}}, \bibinfo {author} {\bibfnamefont {P.}~\bibnamefont {Roy}},\ and\
  \bibinfo {author} {\bibfnamefont {S.}~\bibnamefont {Lupi}},\ }\href@noop {}
  {\bibinfo {title} {Optical properties of superconducting nd0.8sr0.2nio2
  nickelate}} (\bibinfo {year} {2022}),\ \Eprint
  {https://arxiv.org/abs/2203.16986} {arXiv:2203.16986 [cond-mat.supr-con]}
  \BibitemShut {NoStop}%
\bibitem [{\citenamefont {Shen}\ \emph {et~al.}(2022)\citenamefont {Shen},
  \citenamefont {Sears}, \citenamefont {Fabbris}, \citenamefont {Li},
  \citenamefont {Pelliciari}, \citenamefont {Jarrige}, \citenamefont {He},
  \citenamefont {Bo\ifmmode \check{z}\else \v{z}\fi{}ovi\ifmmode~\acute{c}\else
  \'{c}\fi{}}, \citenamefont {Mitrano}, \citenamefont {Zhang}, \citenamefont
  {Mitchell}, \citenamefont {Botana}, \citenamefont {Bisogni}, \citenamefont
  {Norman}, \citenamefont {Johnston},\ and\ \citenamefont {Dean}}]{Shen2022}%
  \BibitemOpen
  \bibfield  {author} {\bibinfo {author} {\bibfnamefont {Y.}~\bibnamefont
  {Shen}}, \bibinfo {author} {\bibfnamefont {J.}~\bibnamefont {Sears}},
  \bibinfo {author} {\bibfnamefont {G.}~\bibnamefont {Fabbris}}, \bibinfo
  {author} {\bibfnamefont {J.}~\bibnamefont {Li}}, \bibinfo {author}
  {\bibfnamefont {J.}~\bibnamefont {Pelliciari}}, \bibinfo {author}
  {\bibfnamefont {I.}~\bibnamefont {Jarrige}}, \bibinfo {author} {\bibfnamefont
  {X.}~\bibnamefont {He}}, \bibinfo {author} {\bibfnamefont {I.}~\bibnamefont
  {Bo\ifmmode \check{z}\else \v{z}\fi{}ovi\ifmmode~\acute{c}\else \'{c}\fi{}}},
  \bibinfo {author} {\bibfnamefont {M.}~\bibnamefont {Mitrano}}, \bibinfo
  {author} {\bibfnamefont {J.}~\bibnamefont {Zhang}}, \bibinfo {author}
  {\bibfnamefont {J.~F.}\ \bibnamefont {Mitchell}}, \bibinfo {author}
  {\bibfnamefont {A.~S.}\ \bibnamefont {Botana}}, \bibinfo {author}
  {\bibfnamefont {V.}~\bibnamefont {Bisogni}}, \bibinfo {author} {\bibfnamefont
  {M.~R.}\ \bibnamefont {Norman}}, \bibinfo {author} {\bibfnamefont
  {S.}~\bibnamefont {Johnston}},\ and\ \bibinfo {author} {\bibfnamefont
  {M.~P.~M.}\ \bibnamefont {Dean}},\ }\href
  {https://doi.org/10.1103/PhysRevX.12.011055} {\bibfield  {journal} {\bibinfo
  {journal} {Phys. Rev. X}\ }\textbf {\bibinfo {volume} {12}},\ \bibinfo
  {pages} {011055} (\bibinfo {year} {2022})}\BibitemShut {NoStop}%
\bibitem [{\citenamefont {F\"ursich}\ \emph {et~al.}(2022)\citenamefont
  {F\"ursich}, \citenamefont {Pons}, \citenamefont {Bluschke}, \citenamefont
  {Ortiz}, \citenamefont {Wintz}, \citenamefont {Schierle}, \citenamefont
  {Weigand}, \citenamefont {Logvenov}, \citenamefont {Sch\"utz}, \citenamefont
  {Keimer},\ and\ \citenamefont {Benckiser}}]{Fursich2022}%
  \BibitemOpen
  \bibfield  {author} {\bibinfo {author} {\bibfnamefont {K.}~\bibnamefont
  {F\"ursich}}, \bibinfo {author} {\bibfnamefont {R.}~\bibnamefont {Pons}},
  \bibinfo {author} {\bibfnamefont {M.}~\bibnamefont {Bluschke}}, \bibinfo
  {author} {\bibfnamefont {R.~A.}\ \bibnamefont {Ortiz}}, \bibinfo {author}
  {\bibfnamefont {S.}~\bibnamefont {Wintz}}, \bibinfo {author} {\bibfnamefont
  {E.}~\bibnamefont {Schierle}}, \bibinfo {author} {\bibfnamefont
  {M.}~\bibnamefont {Weigand}}, \bibinfo {author} {\bibfnamefont
  {G.}~\bibnamefont {Logvenov}}, \bibinfo {author} {\bibfnamefont
  {G.}~\bibnamefont {Sch\"utz}}, \bibinfo {author} {\bibfnamefont
  {B.}~\bibnamefont {Keimer}},\ and\ \bibinfo {author} {\bibfnamefont
  {E.}~\bibnamefont {Benckiser}},\ }\href
  {https://www.frontiersin.org/articles/10.3389/fphy.2021.810220} {\bibfield
  {journal} {\bibinfo  {journal} {Front. Phys.}\ }\textbf {\bibinfo {volume}
  {9:810220}} (\bibinfo {year} {2022})}\BibitemShut {NoStop}%
\bibitem [{\citenamefont {Krieger}\ \emph {et~al.}(2022)\citenamefont
  {Krieger}, \citenamefont {Martinelli}, \citenamefont {Zeng}, \citenamefont
  {Chow}, \citenamefont {Kummer}, \citenamefont {Arpaia}, \citenamefont
  {Moretti~Sala}, \citenamefont {Brookes}, \citenamefont {Ariando},
  \citenamefont {Viart}, \citenamefont {Salluzzo}, \citenamefont
  {Ghiringhelli},\ and\ \citenamefont {Preziosi}}]{Krieger2022}%
  \BibitemOpen
  \bibfield  {author} {\bibinfo {author} {\bibfnamefont {G.}~\bibnamefont
  {Krieger}}, \bibinfo {author} {\bibfnamefont {L.}~\bibnamefont {Martinelli}},
  \bibinfo {author} {\bibfnamefont {S.}~\bibnamefont {Zeng}}, \bibinfo {author}
  {\bibfnamefont {L.~E.}\ \bibnamefont {Chow}}, \bibinfo {author}
  {\bibfnamefont {K.}~\bibnamefont {Kummer}}, \bibinfo {author} {\bibfnamefont
  {R.}~\bibnamefont {Arpaia}}, \bibinfo {author} {\bibfnamefont
  {M.}~\bibnamefont {Moretti~Sala}}, \bibinfo {author} {\bibfnamefont {N.~B.}\
  \bibnamefont {Brookes}}, \bibinfo {author} {\bibfnamefont {A.}~\bibnamefont
  {Ariando}}, \bibinfo {author} {\bibfnamefont {N.}~\bibnamefont {Viart}},
  \bibinfo {author} {\bibfnamefont {M.}~\bibnamefont {Salluzzo}}, \bibinfo
  {author} {\bibfnamefont {G.}~\bibnamefont {Ghiringhelli}},\ and\ \bibinfo
  {author} {\bibfnamefont {D.}~\bibnamefont {Preziosi}},\ }\href
  {https://doi.org/10.1103/PhysRevLett.129.027002} {\bibfield  {journal}
  {\bibinfo  {journal} {Phys. Rev. Lett.}\ }\textbf {\bibinfo {volume} {129}},\
  \bibinfo {pages} {027002} (\bibinfo {year} {2022})}\BibitemShut {NoStop}%
\bibitem [{\citenamefont {Tam}\ \emph {et~al.}(2022)\citenamefont {Tam},
  \citenamefont {Choi}, \citenamefont {Ding}, \citenamefont {Agrestini},
  \citenamefont {Nag}, \citenamefont {Wu}, \citenamefont {Huang}, \citenamefont
  {Luo}, \citenamefont {Gao}, \citenamefont {Garc{\'{i}}a-Fern{\'{a}}ndez},
  \citenamefont {Qiao},\ and\ \citenamefont {Zhou}}]{Tam2021}%
  \BibitemOpen
  \bibfield  {author} {\bibinfo {author} {\bibfnamefont {C.~C.}\ \bibnamefont
  {Tam}}, \bibinfo {author} {\bibfnamefont {J.}~\bibnamefont {Choi}}, \bibinfo
  {author} {\bibfnamefont {X.}~\bibnamefont {Ding}}, \bibinfo {author}
  {\bibfnamefont {S.}~\bibnamefont {Agrestini}}, \bibinfo {author}
  {\bibfnamefont {A.}~\bibnamefont {Nag}}, \bibinfo {author} {\bibfnamefont
  {M.}~\bibnamefont {Wu}}, \bibinfo {author} {\bibfnamefont {B.}~\bibnamefont
  {Huang}}, \bibinfo {author} {\bibfnamefont {H.}~\bibnamefont {Luo}}, \bibinfo
  {author} {\bibfnamefont {P.}~\bibnamefont {Gao}}, \bibinfo {author}
  {\bibfnamefont {M.}~\bibnamefont {Garc{\'{i}}a-Fern{\'{a}}ndez}}, \bibinfo
  {author} {\bibfnamefont {L.}~\bibnamefont {Qiao}},\ and\ \bibinfo {author}
  {\bibfnamefont {K.-J.}\ \bibnamefont {Zhou}},\ }\href
  {https://www.nature.com/articles/s41563-022-01330-1} {\bibfield  {journal}
  {\bibinfo  {journal} {Nat. Mater.}\ } (\bibinfo {year} {2022})}\BibitemShut
  {NoStop}%
\bibitem [{\citenamefont {Tomioka}\ \emph {et~al.}(2021)\citenamefont
  {Tomioka}, \citenamefont {Ito}, \citenamefont {Maruyama}, \citenamefont
  {Kimura},\ and\ \citenamefont {Shindo}}]{Tomioka2021}%
  \BibitemOpen
  \bibfield  {author} {\bibinfo {author} {\bibfnamefont {Y.}~\bibnamefont
  {Tomioka}}, \bibinfo {author} {\bibfnamefont {T.}~\bibnamefont {Ito}},
  \bibinfo {author} {\bibfnamefont {E.}~\bibnamefont {Maruyama}}, \bibinfo
  {author} {\bibfnamefont {S.}~\bibnamefont {Kimura}},\ and\ \bibinfo {author}
  {\bibfnamefont {I.}~\bibnamefont {Shindo}},\ }\href
  {https://doi.org/10.7566/JPSJ.90.034704} {\bibfield  {journal} {\bibinfo
  {journal} {J. Phys. Soc. Japan}\ }\textbf {\bibinfo {volume} {90}},\ \bibinfo
  {pages} {034704} (\bibinfo {year} {2021})}\BibitemShut {NoStop}%
\bibitem [{\citenamefont {Phelan}\ \emph {et~al.}(2019)\citenamefont {Phelan},
  \citenamefont {Zahn}, \citenamefont {Kennedy},\ and\ \citenamefont
  {McQueen}}]{Phelan2019}%
  \BibitemOpen
  \bibfield  {author} {\bibinfo {author} {\bibfnamefont {W.~A.}\ \bibnamefont
  {Phelan}}, \bibinfo {author} {\bibfnamefont {J.}~\bibnamefont {Zahn}},
  \bibinfo {author} {\bibfnamefont {Z.}~\bibnamefont {Kennedy}},\ and\ \bibinfo
  {author} {\bibfnamefont {T.~M.}\ \bibnamefont {McQueen}},\ }\href
  {https://doi.org/10.1016/j.jssc.2018.12.013} {\bibfield  {journal} {\bibinfo
  {journal} {J. Solid State Chem.}\ }\textbf {\bibinfo {volume} {270}},\
  \bibinfo {pages} {705} (\bibinfo {year} {2019})}\BibitemShut {NoStop}%
\bibitem [{\citenamefont {Zhang}\ \emph {et~al.}(2017)\citenamefont {Zhang},
  \citenamefont {Zheng}, \citenamefont {Ren},\ and\ \citenamefont
  {Mitchell}}]{Zhang2017}%
  \BibitemOpen
  \bibfield  {author} {\bibinfo {author} {\bibfnamefont {J.}~\bibnamefont
  {Zhang}}, \bibinfo {author} {\bibfnamefont {H.}~\bibnamefont {Zheng}},
  \bibinfo {author} {\bibfnamefont {Y.}~\bibnamefont {Ren}},\ and\ \bibinfo
  {author} {\bibfnamefont {J.~F.}\ \bibnamefont {Mitchell}},\ }\href
  {https://doi.org/10.1021/acs.cgd.7b00205} {\bibfield  {journal} {\bibinfo
  {journal} {Cryst. Growth Des.}\ }\textbf {\bibinfo {volume} {17}},\ \bibinfo
  {pages} {2730} (\bibinfo {year} {2017})}\BibitemShut {NoStop}%
\bibitem [{\citenamefont {Guo}\ \emph {et~al.}(2018)\citenamefont {Guo},
  \citenamefont {Li}, \citenamefont {Zhao}, \citenamefont {Hu}, \citenamefont
  {Chang}, \citenamefont {Kuo}, \citenamefont {Schmidt}, \citenamefont
  {Piovano}, \citenamefont {Pi}, \citenamefont {Sobolev}, \citenamefont
  {Khomskii}, \citenamefont {Tjeng},\ and\ \citenamefont {Komarek}}]{Guo2018}%
  \BibitemOpen
  \bibfield  {author} {\bibinfo {author} {\bibfnamefont {H.}~\bibnamefont
  {Guo}}, \bibinfo {author} {\bibfnamefont {Z.~W.}\ \bibnamefont {Li}},
  \bibinfo {author} {\bibfnamefont {L.}~\bibnamefont {Zhao}}, \bibinfo {author}
  {\bibfnamefont {Z.}~\bibnamefont {Hu}}, \bibinfo {author} {\bibfnamefont
  {C.~F.}\ \bibnamefont {Chang}}, \bibinfo {author} {\bibfnamefont {C.-Y.}\
  \bibnamefont {Kuo}}, \bibinfo {author} {\bibfnamefont {W.}~\bibnamefont
  {Schmidt}}, \bibinfo {author} {\bibfnamefont {A.}~\bibnamefont {Piovano}},
  \bibinfo {author} {\bibfnamefont {T.~W.}\ \bibnamefont {Pi}}, \bibinfo
  {author} {\bibfnamefont {O.}~\bibnamefont {Sobolev}}, \bibinfo {author}
  {\bibfnamefont {D.~I.}\ \bibnamefont {Khomskii}}, \bibinfo {author}
  {\bibfnamefont {L.~H.}\ \bibnamefont {Tjeng}},\ and\ \bibinfo {author}
  {\bibfnamefont {A.~C.}\ \bibnamefont {Komarek}},\ }\href
  {https://doi.org/10.1038/s41467-017-02524-x} {\bibfield  {journal} {\bibinfo
  {journal} {Nat. Commun.}\ }\textbf {\bibinfo {volume} {9}},\ \bibinfo {pages}
  {49} (\bibinfo {year} {2018})}\BibitemShut {NoStop}%
\bibitem [{\citenamefont {Wang}\ \emph {et~al.}(2018)\citenamefont {Wang},
  \citenamefont {Rosenkranz}, \citenamefont {Rui}, \citenamefont {Zhang},
  \citenamefont {Ye}, \citenamefont {Zheng}, \citenamefont {Klie},
  \citenamefont {Mitchell},\ and\ \citenamefont {Phelan}}]{Wang2018}%
  \BibitemOpen
  \bibfield  {author} {\bibinfo {author} {\bibfnamefont {B.-X.}\ \bibnamefont
  {Wang}}, \bibinfo {author} {\bibfnamefont {S.}~\bibnamefont {Rosenkranz}},
  \bibinfo {author} {\bibfnamefont {X.}~\bibnamefont {Rui}}, \bibinfo {author}
  {\bibfnamefont {J.}~\bibnamefont {Zhang}}, \bibinfo {author} {\bibfnamefont
  {F.}~\bibnamefont {Ye}}, \bibinfo {author} {\bibfnamefont {H.}~\bibnamefont
  {Zheng}}, \bibinfo {author} {\bibfnamefont {R.~F.}\ \bibnamefont {Klie}},
  \bibinfo {author} {\bibfnamefont {J.~F.}\ \bibnamefont {Mitchell}},\ and\
  \bibinfo {author} {\bibfnamefont {D.}~\bibnamefont {Phelan}},\ }\href
  {https://doi.org/10.1103/physrevmaterials.2.064404} {\bibfield  {journal}
  {\bibinfo  {journal} {Phys. Rev. Materials}\ }\textbf {\bibinfo {volume}
  {2}},\ \bibinfo {pages} {064404} (\bibinfo {year} {2018})}\BibitemShut
  {NoStop}%
\bibitem [{\citenamefont {Dey}\ \emph {et~al.}(2019)\citenamefont {Dey},
  \citenamefont {Hergett}, \citenamefont {Telang}, \citenamefont
  {Abdel-Hafiez},\ and\ \citenamefont {Klingeler}}]{Dey2019}%
  \BibitemOpen
  \bibfield  {author} {\bibinfo {author} {\bibfnamefont {K.}~\bibnamefont
  {Dey}}, \bibinfo {author} {\bibfnamefont {W.}~\bibnamefont {Hergett}},
  \bibinfo {author} {\bibfnamefont {P.}~\bibnamefont {Telang}}, \bibinfo
  {author} {\bibfnamefont {M.~M.}\ \bibnamefont {Abdel-Hafiez}},\ and\ \bibinfo
  {author} {\bibfnamefont {R.}~\bibnamefont {Klingeler}},\ }\href
  {https://doi.org/10.1016/j.jcrysgro.2019.125157} {\bibfield  {journal}
  {\bibinfo  {journal} {J. Cryst. Growth}\ }\textbf {\bibinfo {volume} {524}},\
  \bibinfo {pages} {125157} (\bibinfo {year} {2019})}\BibitemShut {NoStop}%
\bibitem [{\citenamefont {Zheng}\ \emph {et~al.}(2020)\citenamefont {Zheng},
  \citenamefont {Wang}, \citenamefont {Phelan}, \citenamefont {Zhang},
  \citenamefont {Ren}, \citenamefont {Krogstad}, \citenamefont {Rosenkranz},
  \citenamefont {Osborn},\ and\ \citenamefont {Mitchell}}]{Zheng2020}%
  \BibitemOpen
  \bibfield  {author} {\bibinfo {author} {\bibfnamefont {H.}~\bibnamefont
  {Zheng}}, \bibinfo {author} {\bibfnamefont {B.-X.}\ \bibnamefont {Wang}},
  \bibinfo {author} {\bibfnamefont {D.}~\bibnamefont {Phelan}}, \bibinfo
  {author} {\bibfnamefont {J.}~\bibnamefont {Zhang}}, \bibinfo {author}
  {\bibfnamefont {Y.}~\bibnamefont {Ren}}, \bibinfo {author} {\bibfnamefont
  {M.}~\bibnamefont {Krogstad}}, \bibinfo {author} {\bibfnamefont
  {S.}~\bibnamefont {Rosenkranz}}, \bibinfo {author} {\bibfnamefont
  {R.}~\bibnamefont {Osborn}},\ and\ \bibinfo {author} {\bibfnamefont
  {J.}~\bibnamefont {Mitchell}},\ }\href
  {https://doi.org/10.3390/cryst10070557} {\bibfield  {journal} {\bibinfo
  {journal} {Crystals}\ }\textbf {\bibinfo {volume} {10}},\ \bibinfo {pages}
  {557} (\bibinfo {year} {2020})}\BibitemShut {NoStop}%
\bibitem [{\citenamefont {Zheng}\ \emph {et~al.}(2019)\citenamefont {Zheng},
  \citenamefont {Zhang}, \citenamefont {Wang}, \citenamefont {Phelan},
  \citenamefont {Krogstad}, \citenamefont {Ren}, \citenamefont {Phelan},
  \citenamefont {Chmaissem}, \citenamefont {Poudel},\ and\ \citenamefont
  {Mitchell}}]{Zheng2019}%
  \BibitemOpen
  \bibfield  {author} {\bibinfo {author} {\bibfnamefont {H.}~\bibnamefont
  {Zheng}}, \bibinfo {author} {\bibfnamefont {J.}~\bibnamefont {Zhang}},
  \bibinfo {author} {\bibfnamefont {B.}~\bibnamefont {Wang}}, \bibinfo {author}
  {\bibfnamefont {D.}~\bibnamefont {Phelan}}, \bibinfo {author} {\bibfnamefont
  {M.~J.}\ \bibnamefont {Krogstad}}, \bibinfo {author} {\bibfnamefont
  {Y.}~\bibnamefont {Ren}}, \bibinfo {author} {\bibfnamefont {W.~A.}\
  \bibnamefont {Phelan}}, \bibinfo {author} {\bibfnamefont {O.}~\bibnamefont
  {Chmaissem}}, \bibinfo {author} {\bibfnamefont {B.}~\bibnamefont {Poudel}},\
  and\ \bibinfo {author} {\bibfnamefont {J.~F.}\ \bibnamefont {Mitchell}},\
  }\href {https://doi.org/10.3390/cryst9070324} {\bibfield  {journal} {\bibinfo
   {journal} {Crystals}\ }\textbf {\bibinfo {volume} {9}},\ \bibinfo {pages}
  {324} (\bibinfo {year} {2019})}\BibitemShut {NoStop}%
\bibitem [{\citenamefont {Alonso}\ \emph {et~al.}(1997)\citenamefont {Alonso},
  \citenamefont {Mart{\'{\i}}nez-Lope}, \citenamefont
  {Garc{\'{\i}}a-Mu{\~{n}}oz},\ and\ \citenamefont
  {Fern{\'{a}}ndez-D{\'{\i}}az}}]{Alonso1997}%
  \BibitemOpen
  \bibfield  {author} {\bibinfo {author} {\bibfnamefont {J.~A.}\ \bibnamefont
  {Alonso}}, \bibinfo {author} {\bibfnamefont {M.~J.}\ \bibnamefont
  {Mart{\'{\i}}nez-Lope}}, \bibinfo {author} {\bibfnamefont {J.~L.}\
  \bibnamefont {Garc{\'{\i}}a-Mu{\~{n}}oz}},\ and\ \bibinfo {author}
  {\bibfnamefont {M.~T.}\ \bibnamefont {Fern{\'{a}}ndez-D{\'{\i}}az}},\ }\href
  {https://doi.org/10.1088/0953-8984/9/30/010} {\bibfield  {journal} {\bibinfo
  {journal} {J. Phys.: Condens. Matter}\ }\textbf {\bibinfo {volume} {9}},\
  \bibinfo {pages} {6417} (\bibinfo {year} {1997})}\BibitemShut {NoStop}%
\bibitem [{\citenamefont {Malaman}\ and\ \citenamefont
  {Brice}(1984)}]{Malaman1984}%
  \BibitemOpen
  \bibfield  {author} {\bibinfo {author} {\bibfnamefont {B.}~\bibnamefont
  {Malaman}}\ and\ \bibinfo {author} {\bibfnamefont {J.}~\bibnamefont
  {Brice}},\ }\href {https://doi.org/10.1016/0022-4596(84)90226-3} {\bibfield
  {journal} {\bibinfo  {journal} {J. Solid State Chem.}\ }\textbf {\bibinfo
  {volume} {53}},\ \bibinfo {pages} {44} (\bibinfo {year} {1984})}\BibitemShut
  {NoStop}%
\bibitem [{\citenamefont {Onozuka}\ \emph {et~al.}(2016)\citenamefont
  {Onozuka}, \citenamefont {Chikamatsu}, \citenamefont {Katayama},
  \citenamefont {Fukumura},\ and\ \citenamefont {Hasegawa}}]{Onozuka2016}%
  \BibitemOpen
  \bibfield  {author} {\bibinfo {author} {\bibfnamefont {T.}~\bibnamefont
  {Onozuka}}, \bibinfo {author} {\bibfnamefont {A.}~\bibnamefont {Chikamatsu}},
  \bibinfo {author} {\bibfnamefont {T.}~\bibnamefont {Katayama}}, \bibinfo
  {author} {\bibfnamefont {T.}~\bibnamefont {Fukumura}},\ and\ \bibinfo
  {author} {\bibfnamefont {T.}~\bibnamefont {Hasegawa}},\ }\href
  {https://doi.org/10.1039/C6DT01737A} {\bibfield  {journal} {\bibinfo
  {journal} {Dalton Trans.}\ }\textbf {\bibinfo {volume} {45}},\ \bibinfo
  {pages} {12114} (\bibinfo {year} {2016})}\BibitemShut {NoStop}%
\bibitem [{\citenamefont {Si}\ \emph {et~al.}(2020)\citenamefont {Si},
  \citenamefont {Xiao}, \citenamefont {Kaufmann}, \citenamefont {Tomczak},
  \citenamefont {Lu}, \citenamefont {Zhong},\ and\ \citenamefont
  {Held}}]{Si2020}%
  \BibitemOpen
  \bibfield  {author} {\bibinfo {author} {\bibfnamefont {L.}~\bibnamefont
  {Si}}, \bibinfo {author} {\bibfnamefont {W.}~\bibnamefont {Xiao}}, \bibinfo
  {author} {\bibfnamefont {J.}~\bibnamefont {Kaufmann}}, \bibinfo {author}
  {\bibfnamefont {J.~M.}\ \bibnamefont {Tomczak}}, \bibinfo {author}
  {\bibfnamefont {Y.}~\bibnamefont {Lu}}, \bibinfo {author} {\bibfnamefont
  {Z.}~\bibnamefont {Zhong}},\ and\ \bibinfo {author} {\bibfnamefont
  {K.}~\bibnamefont {Held}},\ }\href
  {https://doi.org/10.1103/PhysRevLett.124.166402} {\bibfield  {journal}
  {\bibinfo  {journal} {Phys. Rev. Lett.}\ }\textbf {\bibinfo {volume} {124}},\
  \bibinfo {pages} {166402} (\bibinfo {year} {2020})}\BibitemShut {NoStop}%
\bibitem [{\citenamefont {Malyi}\ \emph {et~al.}(2022)\citenamefont {Malyi},
  \citenamefont {Varignon},\ and\ \citenamefont {Zunger}}]{Malyi2022}%
  \BibitemOpen
  \bibfield  {author} {\bibinfo {author} {\bibfnamefont {O.~I.}\ \bibnamefont
  {Malyi}}, \bibinfo {author} {\bibfnamefont {J.}~\bibnamefont {Varignon}},\
  and\ \bibinfo {author} {\bibfnamefont {A.}~\bibnamefont {Zunger}},\ }\href
  {https://doi.org/10.1103/PhysRevB.105.014106} {\bibfield  {journal} {\bibinfo
   {journal} {Phys. Rev. B}\ }\textbf {\bibinfo {volume} {105}},\ \bibinfo
  {pages} {014106} (\bibinfo {year} {2022})}\BibitemShut {NoStop}%
\bibitem [{\citenamefont {Carrasco~\'Alvarez}\ \emph
  {et~al.}(2022)\citenamefont {Carrasco~\'Alvarez}, \citenamefont {Petit},
  \citenamefont {Iglesias}, \citenamefont {Prellier}, \citenamefont {Bibes},\
  and\ \citenamefont {Varignon}}]{Alvarez2022}%
  \BibitemOpen
  \bibfield  {author} {\bibinfo {author} {\bibfnamefont {A.~A.}\ \bibnamefont
  {Carrasco~\'Alvarez}}, \bibinfo {author} {\bibfnamefont {S.}~\bibnamefont
  {Petit}}, \bibinfo {author} {\bibfnamefont {L.}~\bibnamefont {Iglesias}},
  \bibinfo {author} {\bibfnamefont {W.}~\bibnamefont {Prellier}}, \bibinfo
  {author} {\bibfnamefont {M.}~\bibnamefont {Bibes}},\ and\ \bibinfo {author}
  {\bibfnamefont {J.}~\bibnamefont {Varignon}},\ }\href
  {https://doi.org/10.1103/PhysRevResearch.4.023064} {\bibfield  {journal}
  {\bibinfo  {journal} {Phys. Rev. Research}\ }\textbf {\bibinfo {volume}
  {4}},\ \bibinfo {pages} {023064} (\bibinfo {year} {2022})}\BibitemShut
  {NoStop}%
\bibitem [{\citenamefont {Bernardini}\ \emph
  {et~al.}(2022{\natexlab{b}})\citenamefont {Bernardini}, \citenamefont
  {Bosin},\ and\ \citenamefont {Cano}}]{Bernardini2022}%
  \BibitemOpen
  \bibfield  {author} {\bibinfo {author} {\bibfnamefont {F.}~\bibnamefont
  {Bernardini}}, \bibinfo {author} {\bibfnamefont {A.}~\bibnamefont {Bosin}},\
  and\ \bibinfo {author} {\bibfnamefont {A.}~\bibnamefont {Cano}},\ }\href
  {https://doi.org/10.1103/PhysRevMaterials.6.044807} {\bibfield  {journal}
  {\bibinfo  {journal} {Phys. Rev. Materials}\ }\textbf {\bibinfo {volume}
  {6}},\ \bibinfo {pages} {044807} (\bibinfo {year}
  {2022}{\natexlab{b}})}\BibitemShut {NoStop}%
\bibitem [{\citenamefont {Si}\ \emph {et~al.}(2022)\citenamefont {Si},
  \citenamefont {Worm},\ and\ \citenamefont {Held}}]{Si2022}%
  \BibitemOpen
  \bibfield  {author} {\bibinfo {author} {\bibfnamefont {L.}~\bibnamefont
  {Si}}, \bibinfo {author} {\bibfnamefont {P.}~\bibnamefont {Worm}},\ and\
  \bibinfo {author} {\bibfnamefont {K.}~\bibnamefont {Held}},\ }\href
  {https://doi.org/10.3390/cryst12050656} {\bibfield  {journal} {\bibinfo
  {journal} {Crystals}\ }\textbf {\bibinfo {volume} {12}},\ \bibinfo {pages}
  {656} (\bibinfo {year} {2022})}\BibitemShut {NoStop}%
\bibitem [{\citenamefont {Denis~Romero}\ \emph {et~al.}(2014)\citenamefont
  {Denis~Romero}, \citenamefont {Leach}, \citenamefont {M\"oller},
  \citenamefont {Foronda}, \citenamefont {Blundell},\ and\ \citenamefont
  {Hayward}}]{DenisRomero2014}%
  \BibitemOpen
  \bibfield  {author} {\bibinfo {author} {\bibfnamefont {F.}~\bibnamefont
  {Denis~Romero}}, \bibinfo {author} {\bibfnamefont {A.}~\bibnamefont {Leach}},
  \bibinfo {author} {\bibfnamefont {J.~S.}\ \bibnamefont {M\"oller}}, \bibinfo
  {author} {\bibfnamefont {F.}~\bibnamefont {Foronda}}, \bibinfo {author}
  {\bibfnamefont {S.~J.}\ \bibnamefont {Blundell}},\ and\ \bibinfo {author}
  {\bibfnamefont {M.~A.}\ \bibnamefont {Hayward}},\ }\href
  {https://doi.org/https://doi.org/10.1002/anie.201403536} {\bibfield
  {journal} {\bibinfo  {journal} {Angew. Chem. Int. Ed.}\ }\textbf {\bibinfo
  {volume} {53}},\ \bibinfo {pages} {7556} (\bibinfo {year}
  {2014})}\BibitemShut {NoStop}%
\bibitem [{\citenamefont {Ikeda}\ \emph {et~al.}(2014)\citenamefont {Ikeda},
  \citenamefont {Manabe},\ and\ \citenamefont {Naito}}]{Ikeda2014}%
  \BibitemOpen
  \bibfield  {author} {\bibinfo {author} {\bibfnamefont {A.}~\bibnamefont
  {Ikeda}}, \bibinfo {author} {\bibfnamefont {T.}~\bibnamefont {Manabe}},\ and\
  \bibinfo {author} {\bibfnamefont {M.}~\bibnamefont {Naito}},\ }\href
  {https://doi.org/10.1016/j.physc.2014.09.002} {\bibfield  {journal} {\bibinfo
   {journal} {Physica C Supercond.}\ }\textbf {\bibinfo {volume} {506}},\
  \bibinfo {pages} {83} (\bibinfo {year} {2014})}\BibitemShut {NoStop}%
\bibitem [{\citenamefont {Qin}\ \emph {et~al.}(2005)\citenamefont {Qin},
  \citenamefont {Liu}, \citenamefont {Yu}, \citenamefont {Bao}, \citenamefont
  {Li}, \citenamefont {Yu}, \citenamefont {Liu},\ and\ \citenamefont
  {Jin}}]{Qin2005}%
  \BibitemOpen
  \bibfield  {author} {\bibinfo {author} {\bibfnamefont {X.}~\bibnamefont
  {Qin}}, \bibinfo {author} {\bibfnamefont {Q.}~\bibnamefont {Liu}}, \bibinfo
  {author} {\bibfnamefont {Y.}~\bibnamefont {Yu}}, \bibinfo {author}
  {\bibfnamefont {Z.}~\bibnamefont {Bao}}, \bibinfo {author} {\bibfnamefont
  {F.}~\bibnamefont {Li}}, \bibinfo {author} {\bibfnamefont {R.}~\bibnamefont
  {Yu}}, \bibinfo {author} {\bibfnamefont {J.}~\bibnamefont {Liu}},\ and\
  \bibinfo {author} {\bibfnamefont {C.}~\bibnamefont {Jin}},\ }\href
  {https://doi.org/10.1016/j.stam.2005.06.012} {\bibfield  {journal} {\bibinfo
  {journal} {Sci. Technol. Adv. Mater.}\ }\textbf {\bibinfo {volume} {6}},\
  \bibinfo {pages} {828} (\bibinfo {year} {2005})}\BibitemShut {NoStop}%
\bibitem [{\citenamefont {Hyatt}\ \emph {et~al.}(2001)\citenamefont {Hyatt},
  \citenamefont {Hriljac}, \citenamefont {Miyazaki}, \citenamefont {Gameson},
  \citenamefont {Edwards},\ and\ \citenamefont {Jephcoat}}]{Hyatt2001}%
  \BibitemOpen
  \bibfield  {author} {\bibinfo {author} {\bibfnamefont {N.~C.}\ \bibnamefont
  {Hyatt}}, \bibinfo {author} {\bibfnamefont {J.~A.}\ \bibnamefont {Hriljac}},
  \bibinfo {author} {\bibfnamefont {Y.}~\bibnamefont {Miyazaki}}, \bibinfo
  {author} {\bibfnamefont {I.}~\bibnamefont {Gameson}}, \bibinfo {author}
  {\bibfnamefont {P.~P.}\ \bibnamefont {Edwards}},\ and\ \bibinfo {author}
  {\bibfnamefont {A.~P.}\ \bibnamefont {Jephcoat}},\ }\href
  {https://doi.org/10.1103/physrevb.65.014507} {\bibfield  {journal} {\bibinfo
  {journal} {Phys. Rev. B}\ }\textbf {\bibinfo {volume} {65}},\ \bibinfo
  {pages} {014507} (\bibinfo {year} {2001})}\BibitemShut {NoStop}%
\bibitem [{\citenamefont {Mark}\ \emph {et~al.}(2022)\citenamefont {Mark},
  \citenamefont {Campuzano},\ and\ \citenamefont {Hemley}}]{Mark2022}%
  \BibitemOpen
  \bibfield  {author} {\bibinfo {author} {\bibfnamefont {A.~C.}\ \bibnamefont
  {Mark}}, \bibinfo {author} {\bibfnamefont {J.~C.}\ \bibnamefont
  {Campuzano}},\ and\ \bibinfo {author} {\bibfnamefont {R.~J.}\ \bibnamefont
  {Hemley}},\ }\href {https://doi.org/10.1080/08957959.2022.2059366} {\bibfield
   {journal} {\bibinfo  {journal} {High Press. Res.}\ }\textbf {\bibinfo
  {volume} {42}},\ \bibinfo {pages} {137} (\bibinfo {year} {2022})}\BibitemShut
  {NoStop}%
\bibitem [{\citenamefont {Chu}\ \emph {et~al.}(1993)\citenamefont {Chu},
  \citenamefont {Gao}, \citenamefont {Chen}, \citenamefont {Huang},
  \citenamefont {Meng},\ and\ \citenamefont {Xue}}]{Chu1993}%
  \BibitemOpen
  \bibfield  {author} {\bibinfo {author} {\bibfnamefont {C.~W.}\ \bibnamefont
  {Chu}}, \bibinfo {author} {\bibfnamefont {L.}~\bibnamefont {Gao}}, \bibinfo
  {author} {\bibfnamefont {F.}~\bibnamefont {Chen}}, \bibinfo {author}
  {\bibfnamefont {Z.~J.}\ \bibnamefont {Huang}}, \bibinfo {author}
  {\bibfnamefont {R.~L.}\ \bibnamefont {Meng}},\ and\ \bibinfo {author}
  {\bibfnamefont {Y.~Y.}\ \bibnamefont {Xue}},\ }\href
  {https://doi.org/10.1038/365323a0} {\bibfield  {journal} {\bibinfo  {journal}
  {Nature}\ }\textbf {\bibinfo {volume} {365}},\ \bibinfo {pages} {323}
  (\bibinfo {year} {1993})}\BibitemShut {NoStop}%
\bibitem [{\citenamefont {Botana}\ and\ \citenamefont
  {Norman}(2020)}]{Botana2020}%
  \BibitemOpen
  \bibfield  {author} {\bibinfo {author} {\bibfnamefont {A.~S.}\ \bibnamefont
  {Botana}}\ and\ \bibinfo {author} {\bibfnamefont {M.~R.}\ \bibnamefont
  {Norman}},\ }\href {https://doi.org/10.1103/PhysRevX.10.011024} {\bibfield
  {journal} {\bibinfo  {journal} {Phys. Rev. X}\ }\textbf {\bibinfo {volume}
  {10}},\ \bibinfo {pages} {011024} (\bibinfo {year} {2020})}\BibitemShut
  {NoStop}%
\bibitem [{\citenamefont {Kang}\ \emph {et~al.}(2022)\citenamefont {Kang},
  \citenamefont {Kim},\ and\ \citenamefont {Zhu}}]{Kang2022}%
  \BibitemOpen
  \bibfield  {author} {\bibinfo {author} {\bibfnamefont {B.}~\bibnamefont
  {Kang}}, \bibinfo {author} {\bibfnamefont {H.}~\bibnamefont {Kim}},\ and\
  \bibinfo {author} {\bibfnamefont {Q.}~\bibnamefont {Zhu}},\ }\href@noop {}
  {\bibinfo {title} {Tunable kondo screening and interlayer hybridization
  effects in infinite-layer nickelates}} (\bibinfo {year} {2022}),\ \Eprint
  {https://arxiv.org/abs/2207.04388} {arXiv:2207.04388 [cond-mat.supr-con]}
  \BibitemShut {NoStop}%
\bibitem [{\citenamefont {Lee}\ and\ \citenamefont {Pickett}(2004)}]{Lee2004}%
  \BibitemOpen
  \bibfield  {author} {\bibinfo {author} {\bibfnamefont {K.-W.}\ \bibnamefont
  {Lee}}\ and\ \bibinfo {author} {\bibfnamefont {W.~E.}\ \bibnamefont
  {Pickett}},\ }\href {https://doi.org/10.1103/PhysRevB.70.165109} {\bibfield
  {journal} {\bibinfo  {journal} {Phys. Rev. B}\ }\textbf {\bibinfo {volume}
  {70}},\ \bibinfo {pages} {165109} (\bibinfo {year} {2004})}\BibitemShut
  {NoStop}%
\bibitem [{\citenamefont {Huangfu}\ \emph {et~al.}(2020)\citenamefont
  {Huangfu}, \citenamefont {Guguchia}, \citenamefont {Cheptiakov},
  \citenamefont {Zhang}, \citenamefont {Luetkens}, \citenamefont {Gawryluk},
  \citenamefont {Shang}, \citenamefont {von Rohr},\ and\ \citenamefont
  {Schilling}}]{Huangfu2020}%
  \BibitemOpen
  \bibfield  {author} {\bibinfo {author} {\bibfnamefont {S.}~\bibnamefont
  {Huangfu}}, \bibinfo {author} {\bibfnamefont {Z.}~\bibnamefont {Guguchia}},
  \bibinfo {author} {\bibfnamefont {D.}~\bibnamefont {Cheptiakov}}, \bibinfo
  {author} {\bibfnamefont {X.}~\bibnamefont {Zhang}}, \bibinfo {author}
  {\bibfnamefont {H.}~\bibnamefont {Luetkens}}, \bibinfo {author}
  {\bibfnamefont {D.~J.}\ \bibnamefont {Gawryluk}}, \bibinfo {author}
  {\bibfnamefont {T.}~\bibnamefont {Shang}}, \bibinfo {author} {\bibfnamefont
  {F.~O.}\ \bibnamefont {von Rohr}},\ and\ \bibinfo {author} {\bibfnamefont
  {A.}~\bibnamefont {Schilling}},\ }\href
  {https://doi.org/10.1103/PhysRevB.102.054423} {\bibfield  {journal} {\bibinfo
   {journal} {Phys. Rev. B}\ }\textbf {\bibinfo {volume} {102}},\ \bibinfo
  {pages} {054423} (\bibinfo {year} {2020})}\BibitemShut {NoStop}%
\bibitem [{\citenamefont {Wissel}\ \emph {et~al.}(2022)\citenamefont {Wissel},
  \citenamefont {Bernardini}, \citenamefont {Oh}, \citenamefont {Vasala},
  \citenamefont {Schoch}, \citenamefont {Blaschkowski}, \citenamefont
  {Glatzel}, \citenamefont {Bauer}, \citenamefont {Clemens},\ and\
  \citenamefont {Cano}}]{Wissel2022}%
  \BibitemOpen
  \bibfield  {author} {\bibinfo {author} {\bibfnamefont {K.}~\bibnamefont
  {Wissel}}, \bibinfo {author} {\bibfnamefont {F.}~\bibnamefont {Bernardini}},
  \bibinfo {author} {\bibfnamefont {H.}~\bibnamefont {Oh}}, \bibinfo {author}
  {\bibfnamefont {S.}~\bibnamefont {Vasala}}, \bibinfo {author} {\bibfnamefont
  {R.}~\bibnamefont {Schoch}}, \bibinfo {author} {\bibfnamefont
  {B.}~\bibnamefont {Blaschkowski}}, \bibinfo {author} {\bibfnamefont
  {P.}~\bibnamefont {Glatzel}}, \bibinfo {author} {\bibfnamefont
  {M.}~\bibnamefont {Bauer}}, \bibinfo {author} {\bibfnamefont
  {O.}~\bibnamefont {Clemens}},\ and\ \bibinfo {author} {\bibfnamefont
  {A.}~\bibnamefont {Cano}},\ }\href
  {https://doi.org/10.1021/acs.chemmater.2c00726} {\bibfield  {journal}
  {\bibinfo  {journal} {Chem. Mater.}\ }\textbf {\bibinfo {volume} {34}},\
  \bibinfo {pages} {7201} (\bibinfo {year} {2022})}\BibitemShut {NoStop}%
\bibitem [{\citenamefont {Honda}(1910)}]{Honda1910}%
  \BibitemOpen
  \bibfield  {author} {\bibinfo {author} {\bibfnamefont {K.}~\bibnamefont
  {Honda}},\ }\href {https://doi.org/https://doi.org/10.1002/andp.19103371006}
  {\bibfield  {journal} {\bibinfo  {journal} {Ann. Phys.}\ }\textbf {\bibinfo
  {volume} {337}},\ \bibinfo {pages} {1027} (\bibinfo {year}
  {1910})}\BibitemShut {NoStop}%
\bibitem [{\citenamefont {Owen}(1912)}]{Owen1912}%
  \BibitemOpen
  \bibfield  {author} {\bibinfo {author} {\bibfnamefont {M.}~\bibnamefont
  {Owen}},\ }\href {https://doi.org/https://doi.org/10.1002/andp.19123420404}
  {\bibfield  {journal} {\bibinfo  {journal} {Ann. Phys.}\ }\textbf {\bibinfo
  {volume} {342}},\ \bibinfo {pages} {657} (\bibinfo {year}
  {1912})}\BibitemShut {NoStop}%
\bibitem [{\citenamefont {Nakano}\ \emph {et~al.}(1994)\citenamefont {Nakano},
  \citenamefont {Oda}, \citenamefont {Manabe}, \citenamefont {Momono},
  \citenamefont {Miura},\ and\ \citenamefont {Ido}}]{Nakano1994}%
  \BibitemOpen
  \bibfield  {author} {\bibinfo {author} {\bibfnamefont {T.}~\bibnamefont
  {Nakano}}, \bibinfo {author} {\bibfnamefont {M.}~\bibnamefont {Oda}},
  \bibinfo {author} {\bibfnamefont {C.}~\bibnamefont {Manabe}}, \bibinfo
  {author} {\bibfnamefont {N.}~\bibnamefont {Momono}}, \bibinfo {author}
  {\bibfnamefont {Y.}~\bibnamefont {Miura}},\ and\ \bibinfo {author}
  {\bibfnamefont {M.}~\bibnamefont {Ido}},\ }\href
  {https://doi.org/10.1103/PhysRevB.49.16000} {\bibfield  {journal} {\bibinfo
  {journal} {Phys. Rev. B}\ }\textbf {\bibinfo {volume} {49}},\ \bibinfo
  {pages} {16000} (\bibinfo {year} {1994})}\BibitemShut {NoStop}%
\bibitem [{\citenamefont {Takagi}\ \emph {et~al.}(1992)\citenamefont {Takagi},
  \citenamefont {Batlogg}, \citenamefont {Kao}, \citenamefont {Kwo},
  \citenamefont {Cava}, \citenamefont {Krajewski},\ and\ \citenamefont
  {Peck}}]{Takagi1992}%
  \BibitemOpen
  \bibfield  {author} {\bibinfo {author} {\bibfnamefont {H.}~\bibnamefont
  {Takagi}}, \bibinfo {author} {\bibfnamefont {B.}~\bibnamefont {Batlogg}},
  \bibinfo {author} {\bibfnamefont {H.~L.}\ \bibnamefont {Kao}}, \bibinfo
  {author} {\bibfnamefont {J.}~\bibnamefont {Kwo}}, \bibinfo {author}
  {\bibfnamefont {R.~J.}\ \bibnamefont {Cava}}, \bibinfo {author}
  {\bibfnamefont {J.~J.}\ \bibnamefont {Krajewski}},\ and\ \bibinfo {author}
  {\bibfnamefont {W.~F.}\ \bibnamefont {Peck}},\ }\href
  {https://doi.org/10.1103/PhysRevLett.69.2975} {\bibfield  {journal} {\bibinfo
   {journal} {Phys. Rev. Lett.}\ }\textbf {\bibinfo {volume} {69}},\ \bibinfo
  {pages} {2975} (\bibinfo {year} {1992})}\BibitemShut {NoStop}%
\bibitem [{\citenamefont {Kawai}\ \emph {et~al.}(2009)\citenamefont {Kawai},
  \citenamefont {Inoue}, \citenamefont {Mizumaki}, \citenamefont {Kawamura},
  \citenamefont {Ichikawa},\ and\ \citenamefont {Shimakawa}}]{Kawai2009}%
  \BibitemOpen
  \bibfield  {author} {\bibinfo {author} {\bibfnamefont {M.}~\bibnamefont
  {Kawai}}, \bibinfo {author} {\bibfnamefont {S.}~\bibnamefont {Inoue}},
  \bibinfo {author} {\bibfnamefont {M.}~\bibnamefont {Mizumaki}}, \bibinfo
  {author} {\bibfnamefont {N.}~\bibnamefont {Kawamura}}, \bibinfo {author}
  {\bibfnamefont {N.}~\bibnamefont {Ichikawa}},\ and\ \bibinfo {author}
  {\bibfnamefont {Y.}~\bibnamefont {Shimakawa}},\ }\href
  {https://doi.org/10.1063/1.3078276} {\bibfield  {journal} {\bibinfo
  {journal} {Appl. Phys. Lett.}\ }\textbf {\bibinfo {volume} {94}},\ \bibinfo
  {pages} {082102} (\bibinfo {year} {2009})}\BibitemShut {NoStop}%
\bibitem [{\citenamefont {Kaneko}\ \emph {et~al.}(2009)\citenamefont {Kaneko},
  \citenamefont {Yamagishi}, \citenamefont {Tsukada}, \citenamefont {Manabe},\
  and\ \citenamefont {Naito}}]{Kaneko2009}%
  \BibitemOpen
  \bibfield  {author} {\bibinfo {author} {\bibfnamefont {D.}~\bibnamefont
  {Kaneko}}, \bibinfo {author} {\bibfnamefont {K.}~\bibnamefont {Yamagishi}},
  \bibinfo {author} {\bibfnamefont {A.}~\bibnamefont {Tsukada}}, \bibinfo
  {author} {\bibfnamefont {T.}~\bibnamefont {Manabe}},\ and\ \bibinfo {author}
  {\bibfnamefont {M.}~\bibnamefont {Naito}},\ }\href
  {https://doi.org/https://doi.org/10.1016/j.physc.2009.05.104} {\bibfield
  {journal} {\bibinfo  {journal} {Physica C Supercond.}\ }\textbf {\bibinfo
  {volume} {469}},\ \bibinfo {pages} {936} (\bibinfo {year}
  {2009})}\BibitemShut {NoStop}%
\bibitem [{\citenamefont {Ikeda}\ \emph {et~al.}(2016)\citenamefont {Ikeda},
  \citenamefont {Krockenberger}, \citenamefont {Irie}, \citenamefont {Naito},\
  and\ \citenamefont {Yamamoto}}]{Ikeda2016}%
  \BibitemOpen
  \bibfield  {author} {\bibinfo {author} {\bibfnamefont {A.}~\bibnamefont
  {Ikeda}}, \bibinfo {author} {\bibfnamefont {Y.}~\bibnamefont
  {Krockenberger}}, \bibinfo {author} {\bibfnamefont {H.}~\bibnamefont {Irie}},
  \bibinfo {author} {\bibfnamefont {M.}~\bibnamefont {Naito}},\ and\ \bibinfo
  {author} {\bibfnamefont {H.}~\bibnamefont {Yamamoto}},\ }\href
  {https://doi.org/10.7567/apex.9.061101} {\bibfield  {journal} {\bibinfo
  {journal} {Appl. Phys. Express}\ }\textbf {\bibinfo {volume} {9}},\ \bibinfo
  {pages} {061101} (\bibinfo {year} {2016})}\BibitemShut {NoStop}%
\bibitem [{\citenamefont {Hsu}\ \emph {et~al.}(2022)\citenamefont {Hsu},
  \citenamefont {Osada}, \citenamefont {Wang}, \citenamefont {Berben},
  \citenamefont {Duffy}, \citenamefont {Harvey}, \citenamefont {Lee},
  \citenamefont {Li}, \citenamefont {Wiedmann}, \citenamefont {Hwang},\ and\
  \citenamefont {Hussey}}]{Hsu2022}%
  \BibitemOpen
  \bibfield  {author} {\bibinfo {author} {\bibfnamefont {Y.-T.}\ \bibnamefont
  {Hsu}}, \bibinfo {author} {\bibfnamefont {M.}~\bibnamefont {Osada}}, \bibinfo
  {author} {\bibfnamefont {B.~Y.}\ \bibnamefont {Wang}}, \bibinfo {author}
  {\bibfnamefont {M.}~\bibnamefont {Berben}}, \bibinfo {author} {\bibfnamefont
  {C.}~\bibnamefont {Duffy}}, \bibinfo {author} {\bibfnamefont {S.~P.}\
  \bibnamefont {Harvey}}, \bibinfo {author} {\bibfnamefont {K.}~\bibnamefont
  {Lee}}, \bibinfo {author} {\bibfnamefont {D.}~\bibnamefont {Li}}, \bibinfo
  {author} {\bibfnamefont {S.}~\bibnamefont {Wiedmann}}, \bibinfo {author}
  {\bibfnamefont {H.~Y.}\ \bibnamefont {Hwang}},\ and\ \bibinfo {author}
  {\bibfnamefont {N.~E.}\ \bibnamefont {Hussey}},\ }\href
  {https://www.frontiersin.org/articles/10.3389/fphy.2022.846639/full}
  {\bibfield  {journal} {\bibinfo  {journal} {Front. Phys.}\ }\textbf {\bibinfo
  {volume} {10:846639}} (\bibinfo {year} {2022})}\BibitemShut {NoStop}%
\bibitem [{\citenamefont {Zhang}\ \emph {et~al.}(2020)\citenamefont {Zhang},
  \citenamefont {Yang},\ and\ \citenamefont {Zhang}}]{Zhang20201}%
  \BibitemOpen
  \bibfield  {author} {\bibinfo {author} {\bibfnamefont {G.-M.}\ \bibnamefont
  {Zhang}}, \bibinfo {author} {\bibfnamefont {Y.-f.}\ \bibnamefont {Yang}},\
  and\ \bibinfo {author} {\bibfnamefont {F.-C.}\ \bibnamefont {Zhang}},\ }\href
  {https://doi.org/10.1103/PhysRevB.101.020501} {\bibfield  {journal} {\bibinfo
   {journal} {Phys. Rev. B}\ }\textbf {\bibinfo {volume} {101}},\ \bibinfo
  {pages} {020501(R)} (\bibinfo {year} {2020})}\BibitemShut {NoStop}%
\bibitem [{\citenamefont {Yang}\ and\ \citenamefont {Zhang}(2022)}]{Yang2022}%
  \BibitemOpen
  \bibfield  {author} {\bibinfo {author} {\bibfnamefont {Y.-F.}\ \bibnamefont
  {Yang}}\ and\ \bibinfo {author} {\bibfnamefont {G.-M.}\ \bibnamefont
  {Zhang}},\ }\href
  {https://www.frontiersin.org/articles/10.3389/fphy.2021.801236/full}
  {\bibfield  {journal} {\bibinfo  {journal} {Front. Phys.}\ }\textbf {\bibinfo
  {volume} {9:801236}} (\bibinfo {year} {2022})}\BibitemShut {NoStop}%
\bibitem [{\citenamefont {Hsu}\ \emph {et~al.}(2021)\citenamefont {Hsu},
  \citenamefont {Wang}, \citenamefont {Berben}, \citenamefont {Li},
  \citenamefont {Lee}, \citenamefont {Duffy}, \citenamefont {Ottenbros},
  \citenamefont {Kim}, \citenamefont {Osada}, \citenamefont {Wiedmann},
  \citenamefont {Hwang},\ and\ \citenamefont {Hussey}}]{Hsu2021}%
  \BibitemOpen
  \bibfield  {author} {\bibinfo {author} {\bibfnamefont {Y.-T.}\ \bibnamefont
  {Hsu}}, \bibinfo {author} {\bibfnamefont {B.~Y.}\ \bibnamefont {Wang}},
  \bibinfo {author} {\bibfnamefont {M.}~\bibnamefont {Berben}}, \bibinfo
  {author} {\bibfnamefont {D.}~\bibnamefont {Li}}, \bibinfo {author}
  {\bibfnamefont {K.}~\bibnamefont {Lee}}, \bibinfo {author} {\bibfnamefont
  {C.}~\bibnamefont {Duffy}}, \bibinfo {author} {\bibfnamefont
  {T.}~\bibnamefont {Ottenbros}}, \bibinfo {author} {\bibfnamefont {W.~J.}\
  \bibnamefont {Kim}}, \bibinfo {author} {\bibfnamefont {M.}~\bibnamefont
  {Osada}}, \bibinfo {author} {\bibfnamefont {S.}~\bibnamefont {Wiedmann}},
  \bibinfo {author} {\bibfnamefont {H.~Y.}\ \bibnamefont {Hwang}},\ and\
  \bibinfo {author} {\bibfnamefont {N.~E.}\ \bibnamefont {Hussey}},\ }\href
  {https://doi.org/10.1103/PhysRevResearch.3.L042015} {\bibfield  {journal}
  {\bibinfo  {journal} {Phys. Rev. Research}\ }\textbf {\bibinfo {volume}
  {3}},\ \bibinfo {pages} {L042015} (\bibinfo {year} {2021})}\BibitemShut
  {NoStop}%
\bibitem [{\citenamefont {Chen}\ \emph {et~al.}(2022)\citenamefont {Chen},
  \citenamefont {Osada}, \citenamefont {Li}, \citenamefont {Been},
  \citenamefont {Chen}, \citenamefont {Hashimoto}, \citenamefont {Lu},
  \citenamefont {Mo}, \citenamefont {Lee}, \citenamefont {Wang}, \citenamefont
  {Rodolakis}, \citenamefont {McChesney}, \citenamefont {Jia}, \citenamefont
  {Moritz}, \citenamefont {Devereaux}, \citenamefont {Hwang},\ and\
  \citenamefont {Shen}}]{Chen2022}%
  \BibitemOpen
  \bibfield  {author} {\bibinfo {author} {\bibfnamefont {Z.}~\bibnamefont
  {Chen}}, \bibinfo {author} {\bibfnamefont {M.}~\bibnamefont {Osada}},
  \bibinfo {author} {\bibfnamefont {D.}~\bibnamefont {Li}}, \bibinfo {author}
  {\bibfnamefont {E.~M.}\ \bibnamefont {Been}}, \bibinfo {author}
  {\bibfnamefont {S.-D.}\ \bibnamefont {Chen}}, \bibinfo {author}
  {\bibfnamefont {M.}~\bibnamefont {Hashimoto}}, \bibinfo {author}
  {\bibfnamefont {D.}~\bibnamefont {Lu}}, \bibinfo {author} {\bibfnamefont
  {S.-K.}\ \bibnamefont {Mo}}, \bibinfo {author} {\bibfnamefont
  {K.}~\bibnamefont {Lee}}, \bibinfo {author} {\bibfnamefont {B.~Y.}\
  \bibnamefont {Wang}}, \bibinfo {author} {\bibfnamefont {F.}~\bibnamefont
  {Rodolakis}}, \bibinfo {author} {\bibfnamefont {J.~L.}\ \bibnamefont
  {McChesney}}, \bibinfo {author} {\bibfnamefont {C.}~\bibnamefont {Jia}},
  \bibinfo {author} {\bibfnamefont {B.}~\bibnamefont {Moritz}}, \bibinfo
  {author} {\bibfnamefont {T.~P.}\ \bibnamefont {Devereaux}}, \bibinfo {author}
  {\bibfnamefont {H.~Y.}\ \bibnamefont {Hwang}},\ and\ \bibinfo {author}
  {\bibfnamefont {Z.-X.}\ \bibnamefont {Shen}},\ }\href
  {https://doi.org/10.1016/j.matt.2022.01.020} {\bibfield  {journal} {\bibinfo
  {journal} {Matter}\ }\textbf {\bibinfo {volume} {5}},\ \bibinfo {pages}
  {1806} (\bibinfo {year} {2022})}\BibitemShut {NoStop}%
\bibitem [{\citenamefont {Fu}\ \emph {et~al.}(2020)\citenamefont {Fu},
  \citenamefont {Wang}, \citenamefont {Cheng}, \citenamefont {Pei},
  \citenamefont {Zhou}, \citenamefont {Chen}, \citenamefont {Wang},
  \citenamefont {Zhao}, \citenamefont {Jiang}, \citenamefont {Liu},
  \citenamefont {Huang}, \citenamefont {Wang}, \citenamefont {Zhao},
  \citenamefont {Yu}, \citenamefont {Ye}, \citenamefont {Wang},\ and\
  \citenamefont {Mei}}]{Fu2020}%
  \BibitemOpen
  \bibfield  {author} {\bibinfo {author} {\bibfnamefont {Y.}~\bibnamefont
  {Fu}}, \bibinfo {author} {\bibfnamefont {L.}~\bibnamefont {Wang}}, \bibinfo
  {author} {\bibfnamefont {H.}~\bibnamefont {Cheng}}, \bibinfo {author}
  {\bibfnamefont {S.}~\bibnamefont {Pei}}, \bibinfo {author} {\bibfnamefont
  {X.}~\bibnamefont {Zhou}}, \bibinfo {author} {\bibfnamefont {J.}~\bibnamefont
  {Chen}}, \bibinfo {author} {\bibfnamefont {S.}~\bibnamefont {Wang}}, \bibinfo
  {author} {\bibfnamefont {R.}~\bibnamefont {Zhao}}, \bibinfo {author}
  {\bibfnamefont {W.}~\bibnamefont {Jiang}}, \bibinfo {author} {\bibfnamefont
  {C.}~\bibnamefont {Liu}}, \bibinfo {author} {\bibfnamefont {M.}~\bibnamefont
  {Huang}}, \bibinfo {author} {\bibfnamefont {X.}~\bibnamefont {Wang}},
  \bibinfo {author} {\bibfnamefont {Y.}~\bibnamefont {Zhao}}, \bibinfo {author}
  {\bibfnamefont {D.}~\bibnamefont {Yu}}, \bibinfo {author} {\bibfnamefont
  {F.}~\bibnamefont {Ye}}, \bibinfo {author} {\bibfnamefont {S.}~\bibnamefont
  {Wang}},\ and\ \bibinfo {author} {\bibfnamefont {J.-W.}\ \bibnamefont
  {Mei}},\ }\href@noop {} {\bibinfo {title} {{Core-level x-ray photoemission
  and Raman spectroscopy studies on electronic structures in Mott-Hubbard type
  nickelate oxide NdNiO$_2$}}} (\bibinfo {year} {2020}),\ \Eprint
  {https://arxiv.org/abs/1911.03177v2} {arXiv:1911.03177v2 [cond-mat.supr-con]}
  \BibitemShut {NoStop}%
\bibitem [{\citenamefont {Higashi}\ \emph {et~al.}(2021)\citenamefont
  {Higashi}, \citenamefont {Winder}, \citenamefont {Kune\ifmmode~\check{s}\else
  \v{s}\fi{}},\ and\ \citenamefont {Hariki}}]{Higashi2021}%
  \BibitemOpen
  \bibfield  {author} {\bibinfo {author} {\bibfnamefont {K.}~\bibnamefont
  {Higashi}}, \bibinfo {author} {\bibfnamefont {M.}~\bibnamefont {Winder}},
  \bibinfo {author} {\bibfnamefont {J.}~\bibnamefont
  {Kune\ifmmode~\check{s}\else \v{s}\fi{}}},\ and\ \bibinfo {author}
  {\bibfnamefont {A.}~\bibnamefont {Hariki}},\ }\href
  {https://doi.org/10.1103/PhysRevX.11.041009} {\bibfield  {journal} {\bibinfo
  {journal} {Phys. Rev. X}\ }\textbf {\bibinfo {volume} {11}},\ \bibinfo
  {pages} {041009} (\bibinfo {year} {2021})}\BibitemShut {NoStop}%
\bibitem [{\citenamefont {Mickevi{\v{c}}ius}\ \emph {et~al.}(2006)\citenamefont
  {Mickevi{\v{c}}ius}, \citenamefont {Grebinskij}, \citenamefont {Bondarenka},
  \citenamefont {Vengalis}, \citenamefont {{\v{S}}liu{\v{z}}iene},
  \citenamefont {Orlowski}, \citenamefont {Osinniy},\ and\ \citenamefont
  {Drube}}]{Mickevicius2006}%
  \BibitemOpen
  \bibfield  {author} {\bibinfo {author} {\bibfnamefont {S.}~\bibnamefont
  {Mickevi{\v{c}}ius}}, \bibinfo {author} {\bibfnamefont {S.}~\bibnamefont
  {Grebinskij}}, \bibinfo {author} {\bibfnamefont {V.}~\bibnamefont
  {Bondarenka}}, \bibinfo {author} {\bibfnamefont {B.}~\bibnamefont
  {Vengalis}}, \bibinfo {author} {\bibfnamefont {K.}~\bibnamefont
  {{\v{S}}liu{\v{z}}iene}}, \bibinfo {author} {\bibfnamefont {B.}~\bibnamefont
  {Orlowski}}, \bibinfo {author} {\bibfnamefont {V.}~\bibnamefont {Osinniy}},\
  and\ \bibinfo {author} {\bibfnamefont {W.}~\bibnamefont {Drube}},\ }\href
  {https://doi.org/10.1016/j.jallcom.2005.12.038} {\bibfield  {journal}
  {\bibinfo  {journal} {J. Alloys Compd.}\ }\textbf {\bibinfo {volume} {423}},\
  \bibinfo {pages} {107} (\bibinfo {year} {2006})}\BibitemShut {NoStop}%
\bibitem [{\citenamefont {Qiao}\ and\ \citenamefont {Bi}(2011)}]{Qiao2011}%
  \BibitemOpen
  \bibfield  {author} {\bibinfo {author} {\bibfnamefont {L.}~\bibnamefont
  {Qiao}}\ and\ \bibinfo {author} {\bibfnamefont {X.}~\bibnamefont {Bi}},\
  }\href {https://doi.org/10.1209/0295-5075/93/57002} {\bibfield  {journal}
  {\bibinfo  {journal} {EPL}\ }\textbf {\bibinfo {volume} {93}},\ \bibinfo
  {pages} {57002} (\bibinfo {year} {2011})}\BibitemShut {NoStop}%
\bibitem [{\citenamefont {Li}\ \emph {et~al.}(2019{\natexlab{b}})\citenamefont
  {Li}, \citenamefont {Zhou}, \citenamefont {Pang}, \citenamefont {Zhu},
  \citenamefont {Vovk}, \citenamefont {Cong}, \citenamefont {van Bavel},
  \citenamefont {Li},\ and\ \citenamefont {Yang}}]{JLi2019}%
  \BibitemOpen
  \bibfield  {author} {\bibinfo {author} {\bibfnamefont {J.~P.~H.}\
  \bibnamefont {Li}}, \bibinfo {author} {\bibfnamefont {X.}~\bibnamefont
  {Zhou}}, \bibinfo {author} {\bibfnamefont {Y.}~\bibnamefont {Pang}}, \bibinfo
  {author} {\bibfnamefont {L.}~\bibnamefont {Zhu}}, \bibinfo {author}
  {\bibfnamefont {E.~I.}\ \bibnamefont {Vovk}}, \bibinfo {author}
  {\bibfnamefont {L.}~\bibnamefont {Cong}}, \bibinfo {author} {\bibfnamefont
  {A.~P.}\ \bibnamefont {van Bavel}}, \bibinfo {author} {\bibfnamefont
  {S.}~\bibnamefont {Li}},\ and\ \bibinfo {author} {\bibfnamefont
  {Y.}~\bibnamefont {Yang}},\ }\href {https://doi.org/10.1039/c9cp04187g}
  {\bibfield  {journal} {\bibinfo  {journal} {Phys. Chem. Chem. Phys.}\
  }\textbf {\bibinfo {volume} {21}},\ \bibinfo {pages} {22351} (\bibinfo {year}
  {2019}{\natexlab{b}})}\BibitemShut {NoStop}%
\bibitem [{\citenamefont {Misra}\ and\ \citenamefont
  {Kundu}(2016)}]{Misra2016}%
  \BibitemOpen
  \bibfield  {author} {\bibinfo {author} {\bibfnamefont {D.}~\bibnamefont
  {Misra}}\ and\ \bibinfo {author} {\bibfnamefont {T.~K.}\ \bibnamefont
  {Kundu}},\ }\href {https://doi.org/10.1088/2053-1591/3/9/095701} {\bibfield
  {journal} {\bibinfo  {journal} {Mater. Res. Express}\ }\textbf {\bibinfo
  {volume} {3}},\ \bibinfo {pages} {095701} (\bibinfo {year}
  {2016})}\BibitemShut {NoStop}%
\bibitem [{\citenamefont {Shin}\ and\ \citenamefont
  {Rondinelli}(2022)}]{Shin2022}%
  \BibitemOpen
  \bibfield  {author} {\bibinfo {author} {\bibfnamefont {Y.}~\bibnamefont
  {Shin}}\ and\ \bibinfo {author} {\bibfnamefont {J.~M.}\ \bibnamefont
  {Rondinelli}},\ }\href {https://doi.org/10.1103/PhysRevResearch.4.L022069}
  {\bibfield  {journal} {\bibinfo  {journal} {Phys. Rev. Research}\ }\textbf
  {\bibinfo {volume} {4}},\ \bibinfo {pages} {L022069} (\bibinfo {year}
  {2022})}\BibitemShut {NoStop}%
\bibitem [{\citenamefont {van~der Marel}\ \emph {et~al.}(1988)\citenamefont
  {van~der Marel}, \citenamefont {van Elp}, \citenamefont {Sawatzky},\ and\
  \citenamefont {Heitmann}}]{Marel1988}%
  \BibitemOpen
  \bibfield  {author} {\bibinfo {author} {\bibfnamefont {D.}~\bibnamefont
  {van~der Marel}}, \bibinfo {author} {\bibfnamefont {J.}~\bibnamefont {van
  Elp}}, \bibinfo {author} {\bibfnamefont {G.~A.}\ \bibnamefont {Sawatzky}},\
  and\ \bibinfo {author} {\bibfnamefont {D.}~\bibnamefont {Heitmann}},\ }\href
  {https://doi.org/10.1103/PhysRevB.37.5136} {\bibfield  {journal} {\bibinfo
  {journal} {Phys. Rev. B}\ }\textbf {\bibinfo {volume} {37}},\ \bibinfo
  {pages} {5136} (\bibinfo {year} {1988})}\BibitemShut {NoStop}%
\bibitem [{\citenamefont {Baeumer}\ \emph {et~al.}(2021)\citenamefont
  {Baeumer}, \citenamefont {Li}, \citenamefont {Lu}, \citenamefont {Liang},
  \citenamefont {Jin}, \citenamefont {Martins}, \citenamefont {Duchon},
  \citenamefont {Gl{\"{o}}{\ss}}, \citenamefont {Gericke}, \citenamefont
  {Wohlgemuth}, \citenamefont {Giesen}, \citenamefont {Penn}, \citenamefont
  {Dittmann}, \citenamefont {Gunkel}, \citenamefont {Waser}, \citenamefont
  {Bajdich}, \citenamefont {Nem{\v{s}}{\'{a}}k}, \citenamefont {Mefford},\ and\
  \citenamefont {Chueh}}]{Baeumer2021}%
  \BibitemOpen
  \bibfield  {author} {\bibinfo {author} {\bibfnamefont {C.}~\bibnamefont
  {Baeumer}}, \bibinfo {author} {\bibfnamefont {J.}~\bibnamefont {Li}},
  \bibinfo {author} {\bibfnamefont {Q.}~\bibnamefont {Lu}}, \bibinfo {author}
  {\bibfnamefont {A.~Y.-L.}\ \bibnamefont {Liang}}, \bibinfo {author}
  {\bibfnamefont {L.}~\bibnamefont {Jin}}, \bibinfo {author} {\bibfnamefont
  {H.~P.}\ \bibnamefont {Martins}}, \bibinfo {author} {\bibfnamefont
  {T.}~\bibnamefont {Duchon}}, \bibinfo {author} {\bibfnamefont
  {M.}~\bibnamefont {Gl{\"{o}}{\ss}}}, \bibinfo {author} {\bibfnamefont
  {S.~M.}\ \bibnamefont {Gericke}}, \bibinfo {author} {\bibfnamefont {M.~A.}\
  \bibnamefont {Wohlgemuth}}, \bibinfo {author} {\bibfnamefont
  {M.}~\bibnamefont {Giesen}}, \bibinfo {author} {\bibfnamefont {E.~E.}\
  \bibnamefont {Penn}}, \bibinfo {author} {\bibfnamefont {R.}~\bibnamefont
  {Dittmann}}, \bibinfo {author} {\bibfnamefont {F.}~\bibnamefont {Gunkel}},
  \bibinfo {author} {\bibfnamefont {R.}~\bibnamefont {Waser}}, \bibinfo
  {author} {\bibfnamefont {M.}~\bibnamefont {Bajdich}}, \bibinfo {author}
  {\bibfnamefont {S.}~\bibnamefont {Nem{\v{s}}{\'{a}}k}}, \bibinfo {author}
  {\bibfnamefont {J.~T.}\ \bibnamefont {Mefford}},\ and\ \bibinfo {author}
  {\bibfnamefont {W.~C.}\ \bibnamefont {Chueh}},\ }\href
  {https://doi.org/10.1038/s41563-020-00877-1} {\bibfield  {journal} {\bibinfo
  {journal} {Nat. Mater.}\ }\textbf {\bibinfo {volume} {20}},\ \bibinfo {pages}
  {674} (\bibinfo {year} {2021})}\BibitemShut {NoStop}%
\bibitem [{\citenamefont {van~der Laan}\ \emph {et~al.}(1981)\citenamefont
  {van~der Laan}, \citenamefont {Westra}, \citenamefont {Haas},\ and\
  \citenamefont {Sawatzky}}]{Laan1981}%
  \BibitemOpen
  \bibfield  {author} {\bibinfo {author} {\bibfnamefont {G.}~\bibnamefont
  {van~der Laan}}, \bibinfo {author} {\bibfnamefont {C.}~\bibnamefont
  {Westra}}, \bibinfo {author} {\bibfnamefont {C.}~\bibnamefont {Haas}},\ and\
  \bibinfo {author} {\bibfnamefont {G.~A.}\ \bibnamefont {Sawatzky}},\ }\href
  {https://doi.org/10.1103/PhysRevB.23.4369} {\bibfield  {journal} {\bibinfo
  {journal} {Phys. Rev. B}\ }\textbf {\bibinfo {volume} {23}},\ \bibinfo
  {pages} {4369} (\bibinfo {year} {1981})}\BibitemShut {NoStop}%
\end{thebibliography}%
\end{document}